\documentclass{egpubl}
\usepackage{pg2026}

\SpecialIssuePaper         

\CGFStandardLicense

\usepackage[T1]{fontenc}
\usepackage{dfadobe}  

\usepackage{cite}  
\BibtexOrBiblatex
\printedVersion
\PrintedOrElectronic
\ifpdf \usepackage[pdftex]{graphicx} \pdfcompresslevel=9
\else \usepackage[dvips]{graphicx} \fi

\usepackage{egweblnk} 
\usepackage{booktabs}
\usepackage{array}
\usepackage{amsmath}
\usepackage{amssymb}
\usepackage{algorithm}
\usepackage{algpseudocode}
\usepackage{tikz}

\makeatletter
\let\ps@titlepage\ps@empty
\makeatother

\title[Construction of Simple Quadrangulations of the Sphere at Skeletal Joints]%
{Construction of Simple Quadrangulations\\of the Sphere at Skeletal Joints}

\author[S. Hwang \& Q. Y. Hong]
{\parbox{\textwidth}{\centering Suil Hwang\orcid{0009-0007-5172-8869} 
      and Q Youn Hong\orcid{0000-0002-3411-049X}}
      \\
{\parbox{\textwidth}{\centering Hanyang University ERICA, Department of Computer Science and Engineering, Korea}}
}

\begin{document}


\maketitle
\begin{abstract}
We present a method for constructing a simple quadrangulation of the sphere from an unordered set of $n$ direction vectors, with application to parametric volumetric models of skeletal joints. 
We first select a quadrangulation encoding the joint topology by exhaustively enumerating candidate dual graphs, prioritizing convexity and regularity of the resulting quads. 
We then realize this topology on the sphere without crossings by solving a conflict-constrained assignment problem with lazy cyclic-order separation, and recover the corresponding primal graph. 
Experiments on direction sets from spherical Fibonacci sampling of size $3$ to $20$ show that the method robustly produces valid, high-quality quadrangulations across a range of input configurations.
The resulting primal graph is directly converted into exact rational Bézier shell and column volumes, yielding a compact and precise volumetric representation of the joint. 
The approach extends naturally to multi-joint assemblies, enabling the construction of truss-based lattice structures within 3D domains.


\end{abstract}
\section{Introduction}
\label{sec:intro}
Selecting an appropriate representation is fundamental to geometric modeling, as it determines how effectively a model captures both the geometric and topological complexity of a shape. 
Among the many representations proposed for three-dimensional objects, skeletal representations provide a compact and informative abstraction. 
A \textit{skeleton}  is a thin medial structure that encodes the topology and geometry of a shape, typically represented as a one-dimensional network of curves, optionally augmented with surface elements~\cite{TDS*16}. 
Owing to their compactness and structural expressiveness, skeletons have been widely adopted in computer graphics, computer vision, and shape analysis, including applications such as character rigging, motion editing, pose estimation and tracking, and shape understanding. 

Beyond serving as an abstract representation, skeletons also play a central role in applications where the network itself defines the physical structure of an object. 
The process of constructing volumetric models around skeletal networks, commonly referred to as \textit{skeleton scaffolding}, has been adopted in engineering and design applications~\cite{PRW*18}. 
In these settings, skeletal networks form the backbone of truss-based lattice structures for architectural and additive manufacturing designs. 
Similarly, in biomedical applications, skeletal representations facilitate the modeling of vascular structures while reducing modeling and simulation complexity. 

Constructing volumetric models in skeleton scaffolding requires defining both the boundary geometry and the internal volumetric structure. 
Among the volumetric representations available, hexahedral meshes are particularly attractive due to their regular structure and superior numerical stability compared to tetrahedral meshes~\cite{PCS*22}. 
Furthermore, they provide a natural foundation for isogeometric analysis (IGA)~\cite{HCB05, CHB09}, which is based on volumetric tensor-product spline parameterizations. 

A key challenge in skeleton scaffolding arises in \textit{joint regions} where multiple tubular branches meet. 
These regions must be connected conformingly without introducing T-junctions or fragmented components, while simultaneously supporting high-quality parameterizations suitable for numerical analysis. 
In particular, the resulting parameterization should avoid singularities and maintain non-vanishing Jacobians. 
A common strategy is to partition the sphere according to the directions of the incident branches and construct a quadrilateral mesh that defines the interface between neighboring segments~\cite{PRW*18, VKB21, PMA21}. 
Such spherical quadrangulations naturally support tensor-product parameterizations and volumetric spline construction. 

Among these approaches, simple quadrangulations of the sphere (SQS)~\cite{BGG*05, PMA21} provide a principled combinatorial framework for generating minimal quadrilateral layouts consistent with prescribed branch directions. 
However, the space of valid quadrangulations of the sphere is highly combinatorial, making it challenging to identify configurations that are both topologically valid and geometrically well-conditioned. 
Existing methods typically rely on incremental constructions whose outcomes may vary with branch insertion order and offer limited guarantees on geometric quality, often producing irregular or degenerate quadrilateral elements. 

In this work, we address these limitations by formulating spherical joint construction as a global combinatorial-geometric problem. 
Rather than incrementally building quadrangulations, we enumerate candidate topologies and select one that admits a valid embedding for the given directions. 
This formulation enables the robust construction of high-quality quadrangulations and provides a reliable foundation for volumetric parameterization at skeletal joints. 

The main contributions of this work are as follows:
\begin{itemize}
  \item We formulate the construction of a simple quadrangulation of the sphere (SQS) at a skeletal joint as the realization of an enumerated dual plane graph on~$\mathbb{S}^2$, and this formulation uses all branch direction vectors jointly and no branch insertion order.
  \item We formulate this realization as a variant of the Assignment Problem with Conflicts (APC)~\cite{OSAA19} and solve the resulting model using the Google OR-Tools CP-SAT solver~\cite{PD26}.
  Crossing pairs of candidate minor arcs are precomputed as conflicts, while cyclic-order compatibility is enforced lazily through a solve--verify--separate loop.
  \item We convert the resulting SQS into a volumetric representation composed of exact rational B\'{e}zier shell and column volumes, demonstrating the suitability of the constructed quadrangulations for skeletal-joint modeling. 
  \item We demonstrate the practical utility of the proposed framework through the construction of multi-joint assemblies containing hundreds of skeletal joints. 
\end{itemize}

The remainder of the paper is organized as follows: 
Section~\ref{sec:related} reviews related work on skeletal scaffolding. 
Section~\ref{sec:method} presents our method for constructing quadrangulations of the sphere from a given set of branch directions. 
Section~\ref{sec:results} evaluates the proposed quadrangulation method both qualitatively and quantitatively and compares it with existing quadrangulation methods used in skeleton scaffolding. 
In Section~\ref{sec:sqs_to_volume}, we adopt quadrangulation results to construct parametric volumetric models of skeletal joints, and extend the framework to multi-joint assemblies. 
Finally, Section~\ref{sec:conclusion} concludes the paper and discusses directions for future work.
\section{Related work}
\label{sec:related}
Scaffolding skeletons requires handling both tubular segments and joint regions where multiple branches meet. 
Joints may be modeled explicitly or embedded in a global surface or volume. 
In both cases, the challenge is to determine suitable topological and geometric structures from given branch directions.

Early work of Srinivasan et al.~\cite{SMA05} derives joint meshes from convex hulls of cross-sectional profiles arranged along a wireframe. 
B{\ae}rentzen et al.~\cite{BMW12} generate quad-dominant meshes by obtaining polyhedra from the partitioning of the sphere at joints, which are subsequently refined to connect incident branches. 
This idea of spherical partitioning has been adopted by later methods. 

Panotopoulou et al.~\cite{PRW*18} generate minimal quadrangulations with the number of quads matching the number of branches using an incremental insertion scheme. 
Although topologically minimal, this approach may produce poor parameterization because adjacent edges can lie on the same geodesic. 
Fuentes Su{\'a}rez and Hubert~\cite{FSH18} model joint regions from spherical Voronoi diagrams. 
The boundaries of the Voronoi regions are subdivided to form edges of quads. 
Cross-sections of tubular branches follow the boundaries of the Voronoi regions, limiting the method's direct applicability to volumetric parameterization.

Extending beyond surface constructions, several methods address volumetric modeling of skeletons. 
Livesu et al.~\cite{LMPS16} generate hexahedral meshes by adaptively splitting the initial hexahedra at joints according to branch directions. 
Viville et al.~\cite{VKB21} construct hexahedral meshes by enforcing degree-4 dual graphs derived from convex hulls, though remeshing operations to reduce or elevate the degree of vertices may induce double edges that are undesirable for parameterization.

Peltier et al.~\cite{PMA21} propose a framework for piecewise polynomial volumetric models of tubular shapes where joint structures are obtained by incrementally modifying the dual graph through opening and closing operations. 
Although topological consistency is maintained, geometric validity is not guaranteed, and edge crossings may occur depending on input configurations. 
In contrast, we also exploit the primal--dual relationship between a quadrangulation and its dual graph, but impose stricter constraints to ensure both topological validity and geometric quality.

Related work in hexahedral meshing further explores volumetric decompositions. 
Verhetsel et al.~\cite{VPR19} construct shellable hexahedral meshes of the sphere by enumerating boundary quadrangulations and applying quad-flip operations to generate interior structures. 
Although fully volumetric, this method provides limited control over element quality, and concave elements may arise, hindering parameterization.

Joint modeling is also relevant in lattice and truss construction. Van Sosin et al.~\cite{VRS*21} generate truss structures from spheres packing the 3D domain, where neighboring spheres are connected using trimmed surfaces of revolution with fillets. 
Chen and Zhu~\cite{CZ22} derive branching structures of tubular surfaces based on spherical Voronoi diagrams, but focus on surface representations without explicit volumetric structure. 
Zhang et al.~\cite{ZXY*26} reconstruct truss CAD models from point clouds by fitting convex hulls to clustered data. 
However, the resulting primitives lack explicit topological connectivity.
\section{Method}
\label{sec:method}

\begin{figure*}[t]
  \centering
  \setlength{\tabcolsep}{2pt}
  {\begin{tabular}{cccc}
    \includegraphics[width=0.22\linewidth]{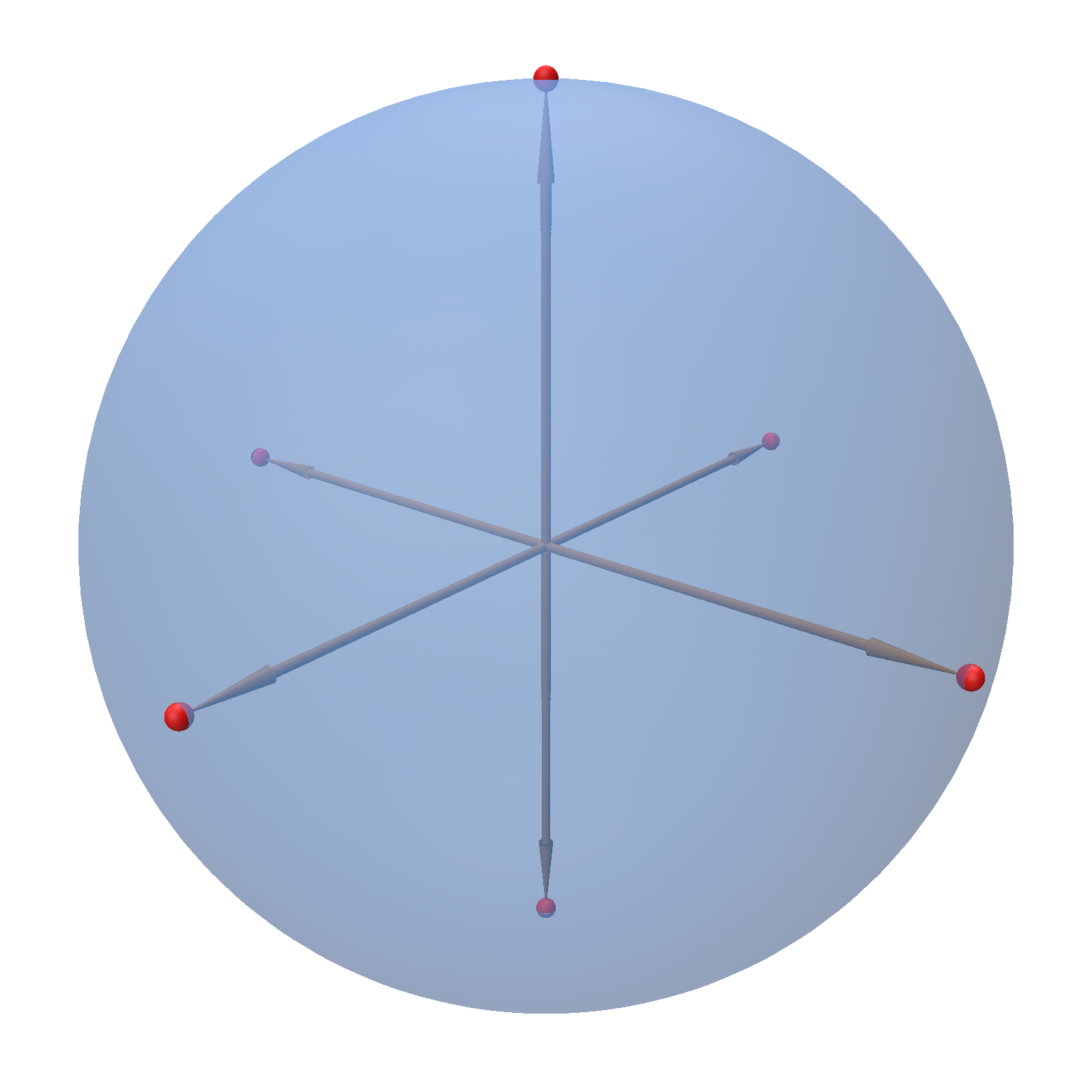} &
    \includegraphics[width=0.30\linewidth]{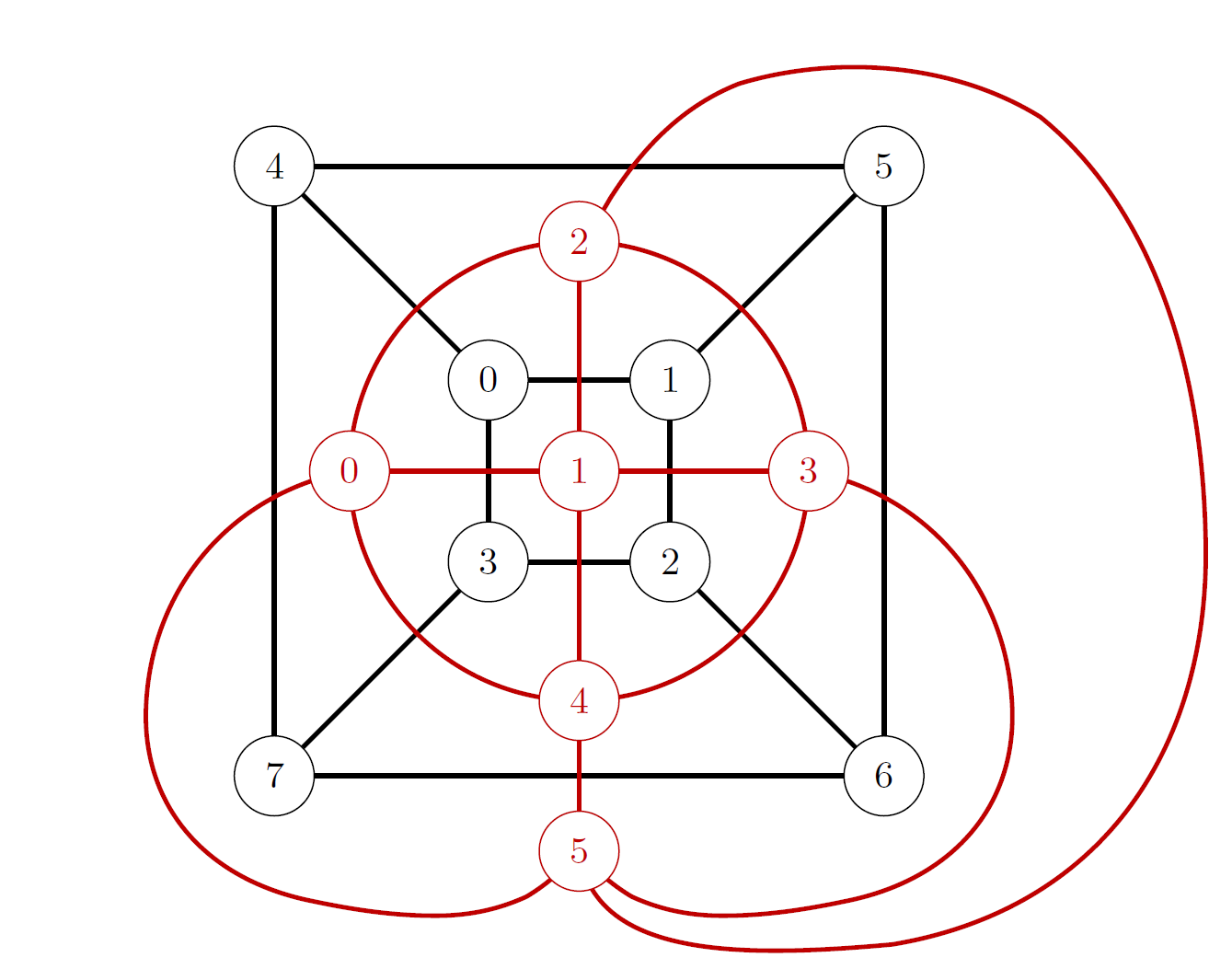} &
    \includegraphics[width=0.22\linewidth]{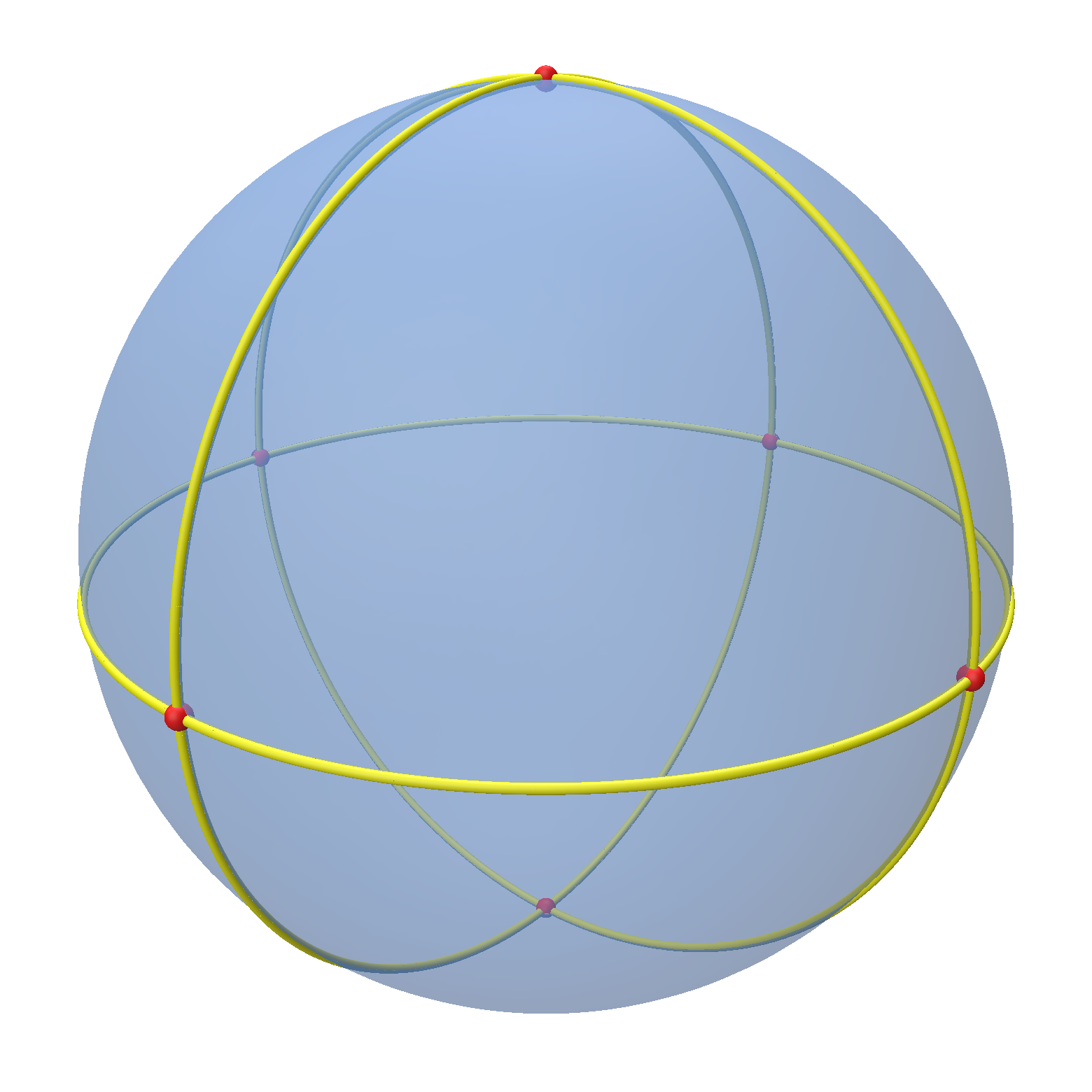} &
    \includegraphics[width=0.22\linewidth]{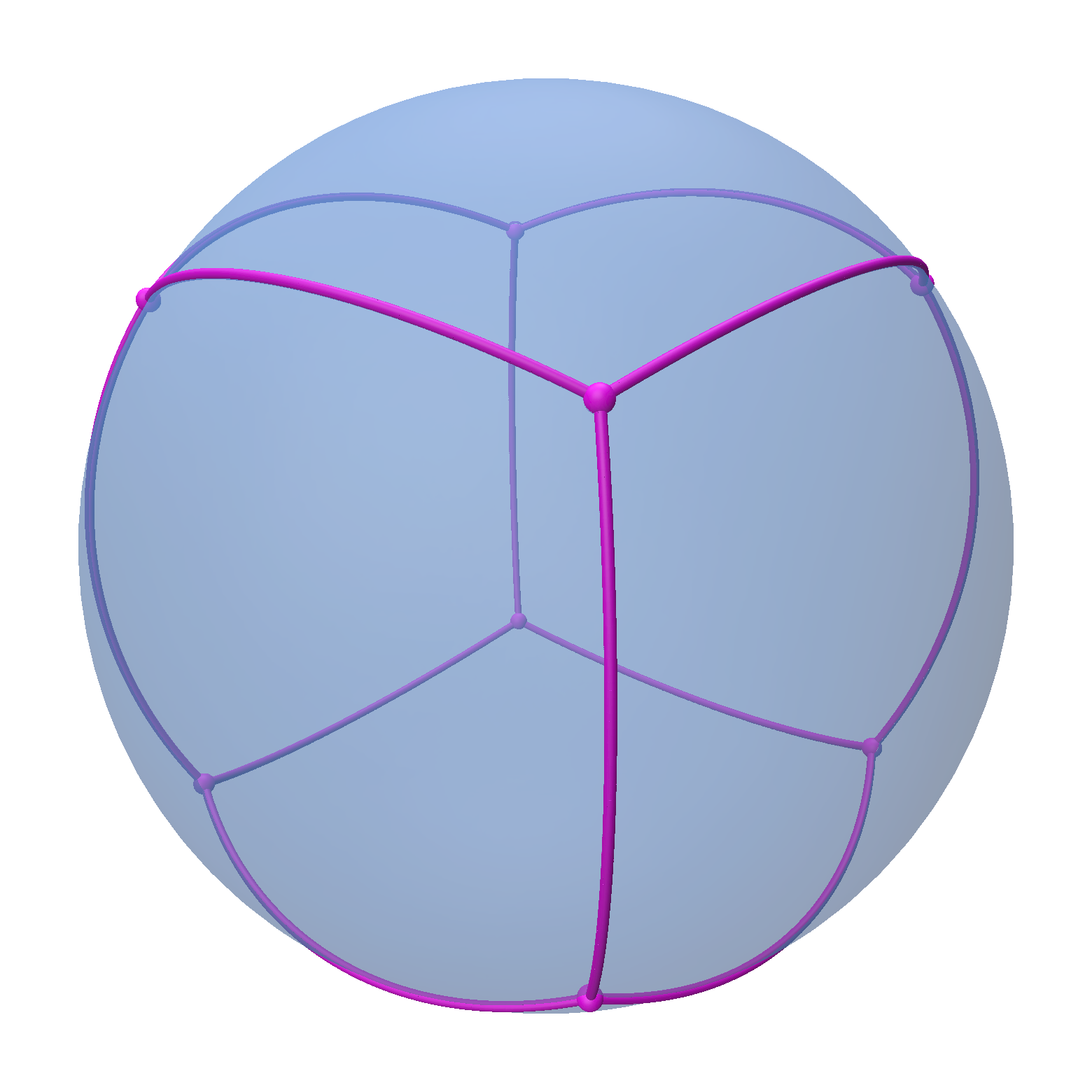} \\[-2pt]
    {\footnotesize (a) Input directions} &
    {\footnotesize (b) Candidate $G$ and its dual $G^*$} &
    {\footnotesize (c) Dual realization $Q^*$} &
    {\footnotesize (d) Primal SQS $Q$}
  \end{tabular}}
  \caption{Pipeline overview for $n=6$ input branch directions.
    \textbf{(a)}~The branch directions define the point set $P\subset\mathbb{S}^2$ (red points).
    \textbf{(b)}~A candidate simple quadrangulation $G$ (black) and its plane dual $G^*$ (red) enumerated by \texttt{plantri}~\cite{BMV24}.
    \textbf{(c)}~The candidate dual $G^*$ is realized on~$\mathbb{S}^2$ as $Q^*$, with vertices assigned to $P$ and edges drawn as minor arcs (yellow).
    \textbf{(d)}~The primal SQS $Q$ (magenta) partitions $\mathbb{S}^2$ into six quadrilateral faces.}
  \label{fig:sqs_pipeline}
\end{figure*}

At a skeletal joint with $n$ incident branches, consider the unit sphere $\mathbb{S}^2$ centered at the joint. 
For each branch $i$, the ray from the joint along its unit direction $\vec{d}_i$ intersects the sphere at a unique point $p_i$.
Let $P=\{p_0,\ldots,p_{n-1}\}$ denote the resulting point set.
Our goal is to construct a simple quadrangulation $Q$ of the sphere (SQS) such that each of its $n$ faces contains exactly one point of $P$ in its strict interior~\cite{PRW*18,PMA21}.

We solve the problem in the dual. 
For a candidate dual plane graph $G^*$, we seek a bijective assignment of its vertices to $P$ that realizes the dual edges as minor arcs on~$\mathbb{S}^2$ with no crossings between arcs that share no endpoint. 
The pipeline has three stages. 
We first enumerate simple quadrangulations with $n$ faces using \texttt{plantri}~\cite{BMV24} and use their plane duals as candidates (\S\ref{subsec:sqs_enumeration}).
We then solve, for each candidate, a conflict-constrained assignment problem, a variant of the Assignment Problem with Conflicts (APC)~\cite{OSAA19}, in which conflicts encode arc crossings (\S\ref{subsec:dual_realization}).
Finally, we construct the SQS from the dual realization (\S\ref{subsec:primal_construction}). 
Figure~\ref{fig:sqs_pipeline} illustrates the construction for $n=6$. 
\subsection{Preliminaries}
\label{subsec:preliminaries}

\paragraph*{Simple Quadrangulation of the Sphere (SQS).}\leavevmode
An SQS $Q=(V,E,F)$ is a finite simple graph embedded on $\mathbb{S}^2$ whose faces are quadrilaterals, i.e., each face boundary has four edges.
We consider SQS with $n \geq 3$.
Since the connectivity of an SQS is either $2$ or $3$~\cite{BGG*05}, it is in particular $2$-connected.
By a standard fact on $2$-connected plane graphs~\cite{Die25}, every face boundary is a cycle.
Combined with the four-edge boundary condition above, every face boundary of an SQS is therefore a $4$-cycle.
Each face has four boundary edges and each edge is incident to exactly two distinct faces, so $4n = 2|E|$ and $|E| = 2n$.
Euler's formula $|V| - |E| + |F| = 2$ then yields $|V| = n + 2$.

\paragraph*{Dual Graph $Q^*$.}\leavevmode
Let $Q^*=(V^*,E^*,F^*)$ be the dual of $Q$, with $|V^*|=n$, $|E^*|=2n$ and $|F^*|=n+2$.
Since every face of $Q$ has degree $4$, $Q^*$ is $4$-regular. 
Because SQSs with $n\ge 3$ are $2$-connected, $Q$ is bridgeless, hence $Q^*$ has no loops~\cite{BGG*05}. 
Moreover, for $n>2$, any two faces of $Q$ share at most two adjacent edges~\cite{PMA21}, so the edge multiplicity in $Q^*$ is at most $2$. 
We call multiplicity-$2$ edges \emph{double edges} and denote their number by $d$. 
Each double edge bounds a digon, so $Q^*$ contains exactly $d$ digons, dual to the degree-$2$ vertices of $Q$.

\paragraph*{Geodesic Convexity.}
For a simple spherical polygon whose edges are minor arcs, geodesic convexity is equivalent to all interior angles being strictly less than~$\pi$~\cite{DCM*18}.
For any geodesic embedding of $Q$, the faces incident to each vertex partition a neighborhood of that vertex. 
Their interior angles in the tangent plane therefore sum to~$2\pi$. 
At a degree-$2$ primal vertex, at least one incident face has an interior angle of at least~$\pi$ and is not geodesically convex. 
Thus $d=0$ is necessary for all primal faces to be geodesically convex.
\subsection{Enumeration of simple quadrangulations}
\label{subsec:sqs_enumeration}
Let $\mathcal{G}_n$ be the set of isomorphism classes of simple quadrangulations of $\mathbb{S}^2$ with $n$ faces, where two quadrangulations are equivalent under an embedding-preserving isomorphism (possibly orientation-reversing)~\cite{BGG*05}.
Let $\mathcal{G}_n^*$ be the corresponding dual classes. 
For $n \ge 3$, $\mathcal{G}_n^*$ coincides with the isomorphism classes of $4$-regular, $4$-edge-connected plane multigraphs on $n$ vertices~\cite{BGG*05}.

\texttt{plantri}~\cite{BMV24,BM07} uses McKay's canonical construction path method~\cite{McK98} together with the simple quadrangulation expansions of Brinkmann et al.~\cite{BGG*05} to generate exactly one embedded representative of each class in $\mathcal{G}_n$.
By plane duality, this yields one embedded representative for each class in $\mathcal{G}_n^*$~\cite{Die25,BGG*05}.
Verhetsel et al.~\cite{VPR19} use the same enumeration paradigm for a different downstream task: given a spherical quadrangulation as boundary input, they search for combinatorial hexahedral meshes and explicitly ignore the geometrical embedding problem. 
In contrast, we use the duals of these quadrangulations as candidate dual plane graphs for realization.

Each representative $G^*$ of a class in $\mathcal{G}_n^*$ is combinatorially encoded by the labeled vertex set $\mathcal{V}^* = \{0, \ldots, n-1\}$ and a clockwise rotation system $\rho^* = \{\rho^*_j\}_{j \in \mathcal{V}^*}$, where $\rho^*_j$ lists the neighbors of vertex~$j$ in cyclic order in the fixed plane embedding. 
When parallel edges are present, the corresponding neighbor label is repeated; since dual edge multiplicity is at most~$2$, each label appears at most twice. 
Hence the underlying simple edge set $\mathcal{E}^*$ and the multiplicity function $m \colon \mathcal{E}^* \to \{1,2\}$ are read off directly from~$\rho^*$. Since every double edge bounds a digon, the two occurrences of a repeated label in~$\rho^*_j$ are necessarily consecutive.

\begin{table}[t]
  \centering
  \small
  \caption{Enumeration of dual graph classes $\mathcal{G}_n^*$ and their digon-free subclass $\mathcal{G}_{n,0}^*$ using \texttt{plantri}~\cite{BMV24} (counts matching those reported in Tables~1 and~2 of Brinkmann et al.~\cite{BGG*05}).}
  \label{tab:enumeration_counts}
  \setlength{\tabcolsep}{8pt}
  \begin{tabular}{@{}rrr@{}}
    \toprule
    $n$ & $|\mathcal{G}_n^*|$ & $|\mathcal{G}_{n,0}^*|$ \\
    \midrule
     3 &                    1 &        0 \\
     4 &                    2 &        0 \\
     5 &                    3 &        0 \\
     6 &                    9 &        1 \\
     7 &                   18 &        0 \\
     8 &                   62 &        1 \\
     9 &                  198 &        1 \\
    10 &                  803 &        3 \\
    11 &              3{,}378 &        3 \\
    12 &             15{,}882 &       12 \\
    13 &             77{,}185 &       19 \\
    14 &            393{,}075 &       64 \\
    15 &        2{,}049{,}974 &      155 \\
    16 &       10{,}938{,}182 &      510 \\
    17 &       59{,}312{,}272 &  1{,}514 \\
    18 &      326{,}258{,}544 &  5{,}146 \\
    19 &  1{,}815{,}910{,}231 & 16{,}966 \\
    20 & 10{,}213{,}424{,}233 & 58{,}782 \\
    \bottomrule
  \end{tabular}
\end{table}

We first consider the digon-free subclass $\mathcal{G}_{n,0}^* \subseteq \mathcal{G}_n^*$, i.e., the classes whose representatives have no double edges. 
A dual digon forces a degree-$2$ primal vertex at which at least one incident quad has interior angle $\geq\pi$. 
Hence geodesic convexity of all primal faces requires $d=0$. 
Table~\ref{tab:enumeration_counts} shows that the total number of enumerated graph classes grows rapidly while these subclasses remain extremely sparse: at $n=20$, only $58{,}782$ of $10{,}213{,}424{,}233$ classes ($\approx 0.00058\%$) are digon-free.
\subsection{Dual graph realization}
\label{subsec:dual_realization}
Given a representative $G^*$ of a class in $\mathcal{G}^*_n$ and the point set $P$, we seek a bijection $\sigma \colon \mathcal{V}^* \to P$ that realizes each edge $\bar{e} = \{j,k\} \in \mathcal{E}^*$ as the minor arc joining $\sigma(j)$ and $\sigma(k)$. 
For this realization to define a valid geometric embedding of $G^*$ on $\mathbb{S}^2$, we require the following conditions:

\begin{enumerate}
  \item[(i)] \textbf{Bijection.} 
  $\sigma \colon \mathcal{V}^* \to P$ is bijective.

  \item[(ii)] \textbf{Valid geometric realization.} 
  (a)~For every $\bar{e} = \{j,k\} \in \mathcal{E}^*$, the points $\sigma(j)$ and $\sigma(k)$ are non-antipodal, so that the minor arc is uniquely defined. 
  (b)~The interior of each minor arc contains no other point of $P$. 
  (c)~The interiors of any two minor arcs that share no endpoint are disjoint.

  \item[(iii)] \textbf{Cyclic-order compatibility.} 
  For each $j \in \mathcal{V}^*$, the geometric clockwise cyclic order of the arcs incident to $\sigma(j)$, as seen from the exterior of the sphere, is consistent with the clockwise cyclic order encoded by $\rho^*_j$.
\end{enumerate}

We formulate dual graph realization as a variant of APC, which seeks a minimum-cost perfect matching in a complete bipartite graph without selecting both edges of any prescribed conflicting pair~\cite{OSAA19}. 
In our variant, the conflicts are pairs of crossing candidate minor arcs on~$\mathbb{S}^2$ and depend only on~$P$
(\S\ref{subsubsec:precomputation}). 
The conflict structure is therefore computed once per point set and reused across candidate graphs. We solve each model with CP-SAT, a constraint programming solver in Google OR-Tools that operates on integer variables and uses Boolean satisfiability (SAT) methods~\cite{PD26}.
We encode conditions~(i)--(ii) as static constraints (\S\ref{subsubsec:crossing_free_formulation}) and enforce condition~(iii) through a lazy solve--verify--separate loop
(\S\ref{subsubsec:cyclic_order_separation}).
\subsubsection{Precomputation of candidate arcs and conflicts}
\label{subsubsec:precomputation}
The minor arcs and conflicts depend only on~$P$, so we precompute them once.  
We index the unordered point pairs by $\ell \in \mathcal{L} = \{0, \ldots, \binom{n}{2}-1\}$, with $\alpha_\ell, \beta_\ell$ the two point indices of pair~$\ell$. 
When $p_{\alpha_\ell}$ and $p_{\beta_\ell}$ are non-antipodal, their minor arc is a \emph{candidate arc} $a_\ell$ of length $\lambda_\ell$.

\begin{figure}[t]
  \centering
  \setlength{\tabcolsep}{4pt}
  \begin{tabular}{ccc}
    \includegraphics[width=0.30\linewidth,trim=200 200 200 200,clip]{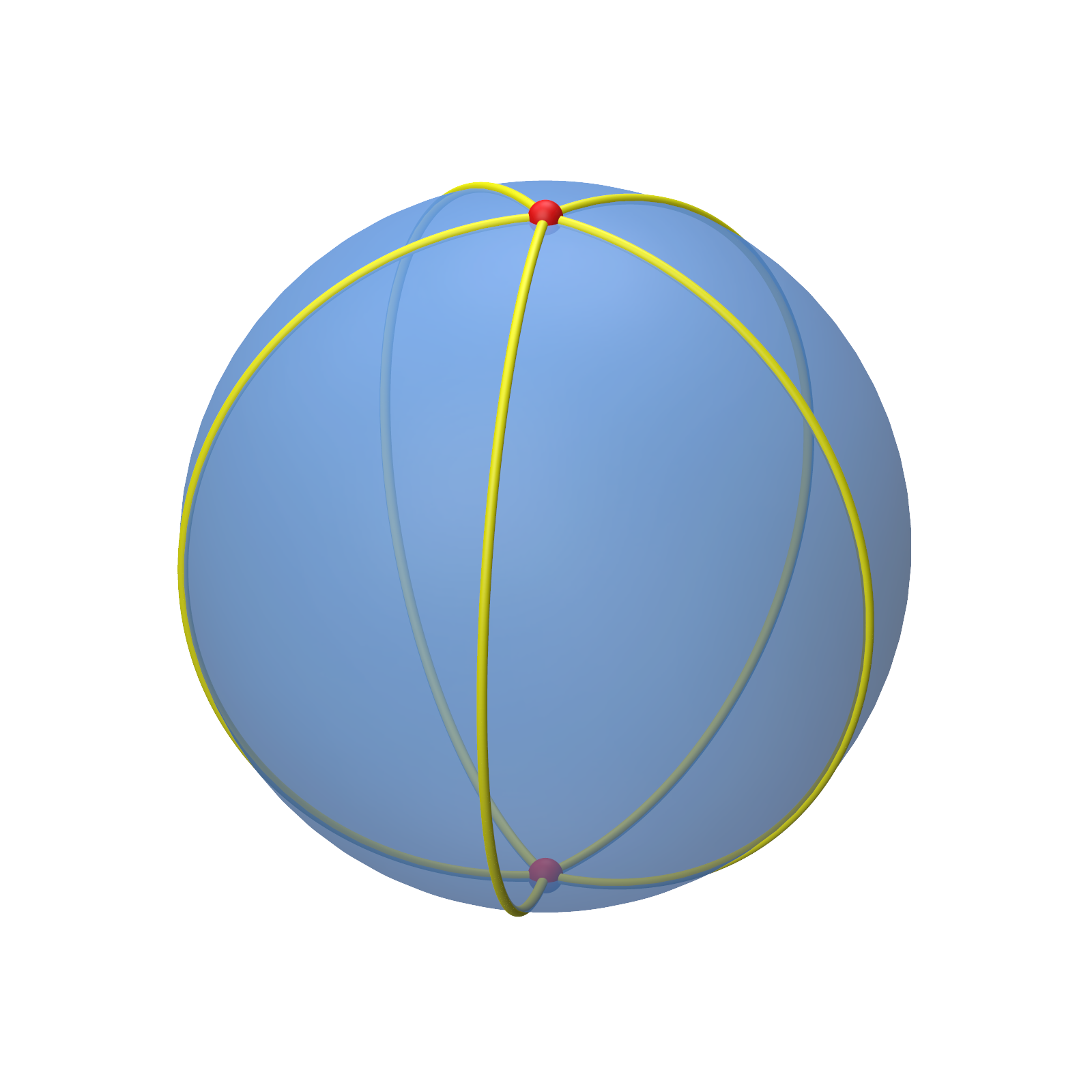} &
    \raisebox{-15pt}{\includegraphics[width=0.30\linewidth,trim=200 200 200 200,clip]{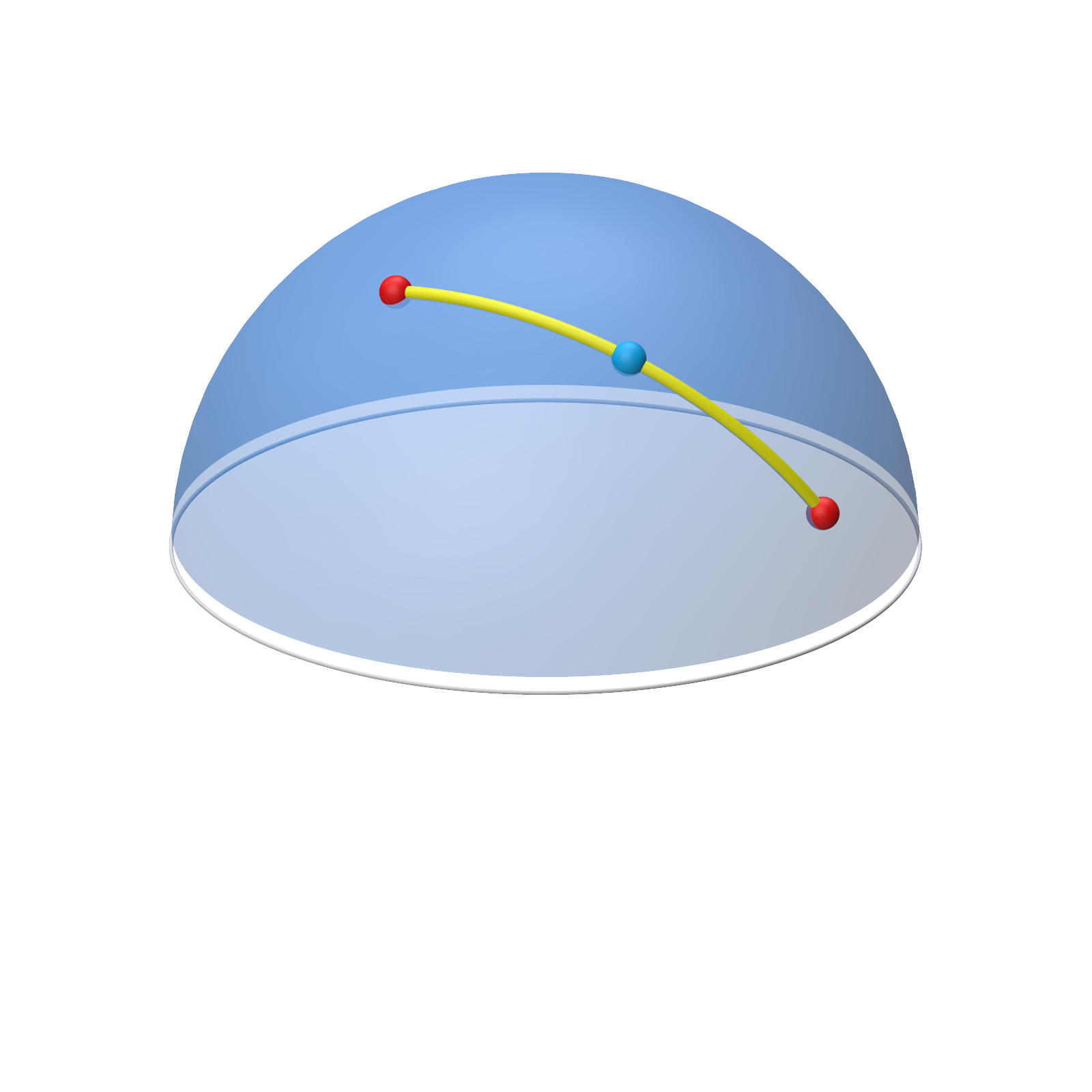}} &
    \raisebox{-15pt}{\includegraphics[width=0.30\linewidth,trim=200 200 200 200,clip]{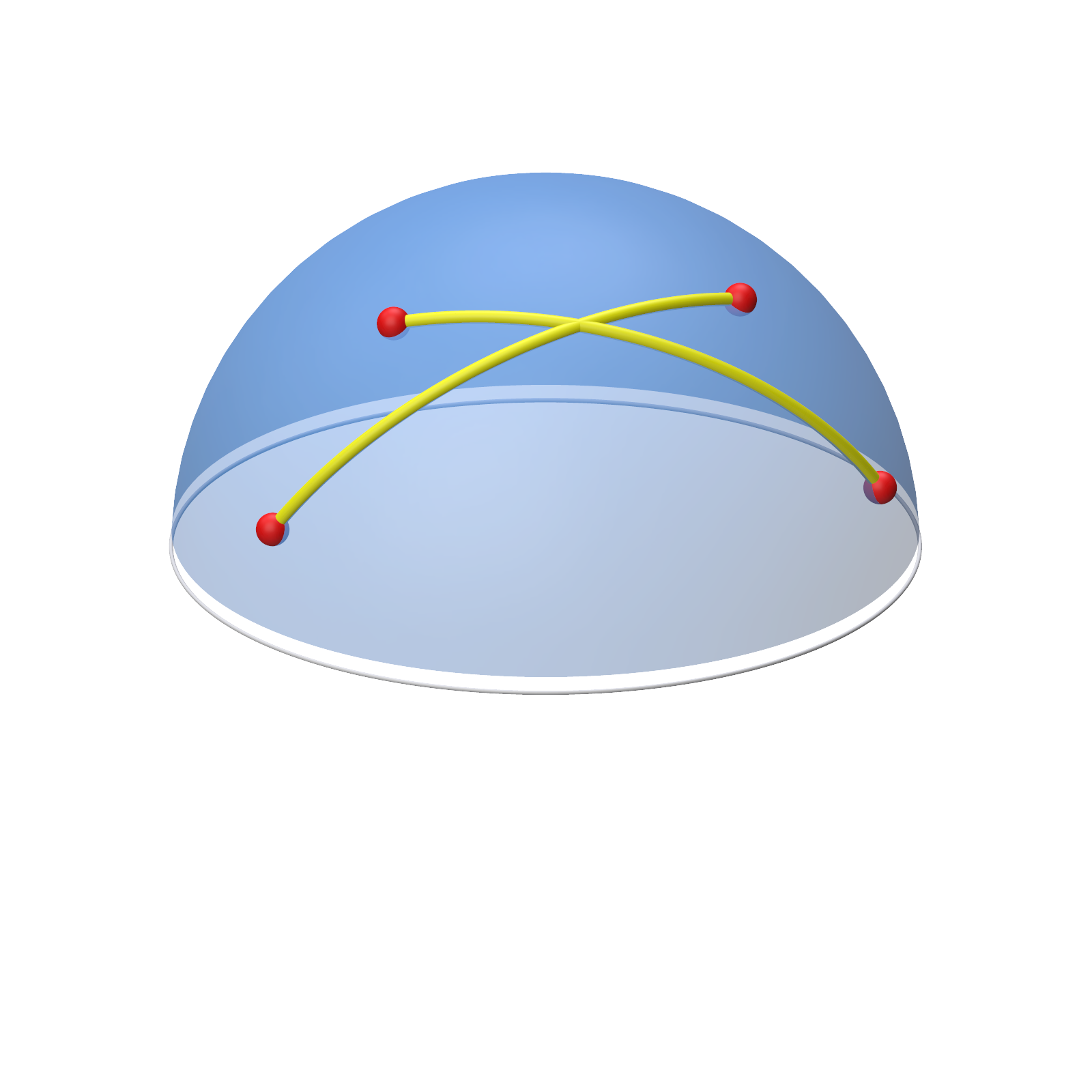}} \\[-6pt]
    {\footnotesize (a) Antipodal} &
    {\footnotesize (b) Third point on arc} &
    {\footnotesize (c) Crossing}
  \end{tabular}
  \caption{Violations of condition~(ii). 
  \textbf{(a)}~The minor arc is not uniquely defined. 
  \textbf{(b)}~A third point of~$P$ (blue) lies on the minor arc. 
  \textbf{(c)}~Two arcs with disjoint endpoint sets cross.}
  \label{fig:forbidden_pairs_and_conflicts}
\end{figure}

The \emph{forbidden} set $\mathcal{L}_{\text{forbid}} \subseteq \mathcal{L}$ collects the indices (Figure~\ref{fig:forbidden_pairs_and_conflicts}(a,b)) for which either (a)~the endpoints are antipodal, so there is no unique minor arc, or (b)~the interior of $a_\ell$ contains a third point of~$P$. 
The remaining indices form $\mathcal{L}_{\text{allow}} = \mathcal{L} \setminus \mathcal{L}_{\text{forbid}}$.

Two allowed candidate arcs $a_{\ell_1}$ and $a_{\ell_2}$ with disjoint endpoint sets form a \emph{conflict pair} $\{\ell_1, \ell_2\}\mathrel{\in}\mathcal{C}$ (Figure~\ref{fig:forbidden_pairs_and_conflicts}(c)) if their interiors intersect. 
Pairs sharing an endpoint need not be considered: if two such arcs overlap beyond the shared endpoint, one non-shared endpoint lies in the interior of the other arc, so at least one arc is already forbidden by condition~(ii)(b); otherwise, their interiors are disjoint. 
Algorithm~\ref{alg:precomputation} summarizes this precomputation.

\begin{algorithm}[t]
  \caption{Precomputation of candidate arcs and conflicts}
  \label{alg:precomputation}
  \begin{algorithmic}[1]
    \Require Pairwise distinct points $P=\{p_0,\dots,p_{n-1}\}\subset\mathbb{S}^2$
    \Ensure Allowed arc indices $\mathcal{L}_{\mathrm{allow}}$, conflicts $\mathcal{C}$, arc lengths $(\lambda_\ell)$

    \State $\mathcal{L}\gets$ all unordered pairs of $P$; pair $\ell$ has indices $(\alpha_\ell,\beta_\ell)$
    \State $\mathcal{L}_{\mathrm{allow}},\,\mathcal{C}\gets\emptyset$

    \ForAll{$\ell\in\mathcal{L}$ with non-antipodal endpoints} \Comment{(ii)(a)}
    \State $a_\ell\gets$ minor arc $\widehat{p_{\alpha_\ell}p_{\beta_\ell}}$
    \If{no third point of $P$ in its interior} \Comment{(ii)(b)}
    \State $\lambda_\ell\gets$ length of $a_\ell$
    \State add $\ell$ to $\mathcal{L}_{\mathrm{allow}}$
    \EndIf
    \EndFor

    \ForAll{$\{\ell_1,\ell_2\}\mathrel{\subseteq}\mathcal{L}_{\mathrm{allow}}$ with disjoint endpoint sets}
    \If{interiors of $a_{\ell_1}$ and $a_{\ell_2}$ intersect} \Comment{(ii)(c)}
    \State add $\{\ell_1,\ell_2\}$ to $\mathcal{C}$
    \EndIf
    \EndFor

    \State \Return $\mathcal{L}_{\mathrm{allow}},\,\mathcal{C},\,(\lambda_\ell)$
  \end{algorithmic}
\end{algorithm}
\subsubsection{Crossing-free assignment formulation}
\label{subsubsec:crossing_free_formulation}
Using the output of Algorithm~\ref{alg:precomputation}, we formulate the assignment problem for a candidate dual plane graph~$G^*$ as a CP-SAT model.

\paragraph*{Variables.}
The CP-SAT model uses two families of binary variables. 
For each dual vertex $j \in \mathcal{V}^*$ and each input point $p_i \in P$, the \emph{vertex assignment variable} $x_{j,i} \in \{0,1\}$ indicates whether $\sigma(j) = p_i$. 
For each underlying dual edge $\bar{e} = \{j,k\} \in \mathcal{E}^*$ and each allowed candidate arc $a_\ell$ with $\ell \in \mathcal{L}_{\text{allow}}$, the \emph{edge mapping variable} $y_{\bar{e},\ell} \in \{0,1\}$ indicates whether $\bar{e}$ is realized by $a_\ell$. 
Variables for $\ell \in \mathcal{L}_{\text{forbid}}$ are not created, thereby excluding those mappings a priori.

\paragraph*{Constraints.}\leavevmode
Three families of constraints encode conditions~(i) and~(ii). 
First, the \emph{bijection constraints} enforce condition~(i) for the assignment $\sigma\colon\mathcal{V}^* \to P$:
\begin{equation}\label{eq:bijection}
  \begin{aligned}
    \sum_{i=0}^{n-1} x_{j,i} &= 1
    \quad \forall\, j \in \mathcal{V}^*,\\
    \sum_{j \in \mathcal{V}^*} x_{j,i} &= 1
    \quad \forall\, i \in \{0,\ldots,n-1\}.
  \end{aligned}
\end{equation}

Second, the \emph{edge mapping constraints} ensure that each underlying dual edge is realized by exactly one allowed candidate arc, and that this choice is consistent with the vertex assignment.
For each $\bar{e}=\{j,k\}\in\mathcal{E}^*$, we impose
\begin{equation}\label{eq:edge-map-unique}
  \sum_{\ell \in \mathcal{L}_{\text{allow}}}
  y_{\bar{e},\ell} = 1.
\end{equation}
Since the candidate arc $a_\ell$ is undirected, the edge $\bar{e}=\{j,k\}$ may realize it by matching $\{j,k\}$ to the two endpoints $p_{\alpha_\ell}$ and~$p_{\beta_\ell}$ in either order. 
Accordingly, we require
\begin{equation}\label{eq:edge-map-equivalence}
  y_{\bar{e},\ell}
  \;\Longleftrightarrow\;
  \bigl(x_{j,\alpha_\ell} \land x_{k,\beta_\ell}\bigr)
  \;\lor\;
  \bigl(x_{j,\beta_\ell} \land x_{k,\alpha_\ell}\bigr).
\end{equation}
Under the bijection constraints, this equivalence is enforced by the following four channeling implications for each $(\bar{e},\ell)$:
\begin{equation}\label{eq:channeling}
  \begin{aligned}
    y_{\bar{e},\ell}
    &\;\Rightarrow\;
    x_{j,\alpha_\ell}\lor x_{j,\beta_\ell},\\
    y_{\bar{e},\ell}
    &\;\Rightarrow\;
    x_{k,\alpha_\ell}\lor x_{k,\beta_\ell},\\
    x_{j,\alpha_\ell}\land x_{k,\beta_\ell}
    &\;\Rightarrow\;
    y_{\bar{e},\ell},\\
    x_{j,\beta_\ell}\land x_{k,\alpha_\ell}
    &\;\Rightarrow\;
    y_{\bar{e},\ell}.
  \end{aligned}
\end{equation}
Together with the bijection constraints, the first two forward implications ensure that, when $y_{\bar{e},\ell}=1$, vertices $j$ and $k$ occupy the two endpoints of $a_\ell$; the last two backward implications enforce the converse.

Third, the \emph{conflict avoidance constraints} forbid the simultaneous realization of two crossing candidate arcs. 
For each conflict pair $\{\ell_1, \ell_2\} \in \mathcal{C}$:
\begin{equation}\label{eq:conflict}
  \sum_{\bar{e} \in \mathcal{E}^*} y_{\bar{e},\ell_1}
  \;+\;
  \sum_{\bar{e} \in \mathcal{E}^*} y_{\bar{e},\ell_2}
  \;\leq\; 1.
\end{equation}
By bijectivity together with the channeling implications, each allowed candidate arc realizes at most one underlying dual edge, so each of the two sums above is at most~$1$.

\paragraph*{Objective.}
\leavevmode
Given a candidate $G^*$, we minimize the multiplicity-weighted total length of the realized dual edges:
\begin{equation}\label{eq:dual_length_objective}
  \min \sum_{\bar{e} \in \mathcal{E}^*}
  \sum_{\ell \in \mathcal{L}_{\mathrm{allow}}}
  m(\bar{e})\,\lambda_\ell\, y_{\bar{e},\ell},
\end{equation}
where $\lambda_\ell$ is the precomputed geodesic length of~$a_\ell$.
The objective is linear in the edge mapping variables and ranks feasible assignments by total dual edge length.
Note that the interior angles of the primal faces depend on the primal vertices constructed in \S\ref{subsec:primal_construction} and are therefore not directly represented in this objective.
This objective neither directly optimizes nor guarantees primal angle quality.
\subsubsection{Cyclic-order compatibility via lazy separation}
\label{subsubsec:cyclic_order_separation}
The formulation in \S\ref{subsubsec:crossing_free_formulation} enforces conditions~(i)--(ii), which do not imply~(iii). 
Under conditions~(i) and~(ii)(a)--(b), the geometric clockwise cyclic order of the distinct minor arcs incident to each $\sigma(j)$ is well defined and determined by the values of~$\sigma$ at~$j$ and its distinct neighbors.
We therefore enforce condition~(iii) lazily through a solve--verify--separate loop, adding constraints only when violations are detected.

At each iteration, we solve the CP-SAT model~$\mathcal{M}$.
If the solve returns an assignment~$\sigma$, we identify the set $J\subseteq\mathcal{V}^*$ of vertices at which the geometric and prescribed cyclic orders disagree. 
If $J=\emptyset$, we return~$\sigma$. 
Otherwise, for each $j\in J$, we add a local \emph{no-good cut} excluding exactly the global assignments that agree with~$\sigma$ at~$j$ and its distinct neighbors, and re-solve.
By the local dependence established above, the accumulated cuts never exclude an assignment satisfying conditions~(i)--(iii).
Since each nonterminal iteration adds a new cut, the loop terminates and returns either such an assignment or \textsc{Infeasible}.
Algorithm~\ref{alg:cyclic-order-separation} summarizes the procedure for a candidate~$G^*$.

\begin{algorithm}[t]
  \caption{Solve--verify--separate loop for a candidate dual graph}
  \label{alg:cyclic-order-separation}
  \begin{algorithmic}[1]
    \Require Pairwise distinct points $P=\{p_0,\dots,p_{n-1}\}\subset\mathbb{S}^2$
    \Require Candidate dual graph $G^*=(\mathcal{V}^*,\mathcal{E}^*,\rho^*)$, multiplicity $m$
    \Require $(\mathcal{L}_{\mathrm{allow}},\,\mathcal{C},\,(\lambda_\ell))$ from Alg.\,\ref{alg:precomputation}
    \Ensure $\sigma\colon\mathcal{V}^*\to P$ satisfying (i)--(iii), or~\textsc{Infeasible}

    \State Build the CP-SAT model $\mathcal{M}$ encoding conditions~(i)--(ii)
    \State Set the objective in Eq.~\eqref{eq:dual_length_objective}

    \Loop
      \State Solve $\mathcal{M}$
      \If{$\mathcal{M}$ is infeasible}
        \State \Return \textsc{Infeasible}
      \EndIf

      \State Extract assignment $\sigma$ from $\{x_{j,i}\}$
      \State $J\gets
        \textproc{CyclicOrderViolations}(\sigma,\rho^*)$
        \Comment{condition~(iii)}

      \If{$J=\emptyset$}
        \State \Return $\sigma$
      \EndIf

      \ForAll{$j\in J$}
        \State Add the local no-good cut induced by $\sigma$ at $j$
      \EndFor
    \EndLoop
  \end{algorithmic}
\end{algorithm}
\subsection{Primal graph construction}
\label{subsec:primal_construction}
From the dual graph $Q^* = (V^*, E^*, F^*)$ realized in \S\ref{subsec:dual_realization}, we construct the primal SQS $Q = (V, E, F)$ in two stages: (1)~deriving primal vertices from dual faces (\S\ref{subsubsec:primal_vertex}), and (2)~deriving primal edges and faces from the embedding of~$Q^*$ (\S\ref{subsubsec:primal_edge_face}).
\subsubsection{Primal vertices from dual faces}
\label{subsubsec:primal_vertex}
Each dual face $f^*_i$ of degree $k$ corresponds to one primal vertex $v_i$, where $k \ge 2$ because $Q^*$ has no loops~(\S\ref{subsec:preliminaries}).
$v_i$ is placed at an interior point of~$f^*_i$ if $k \ge 3$ and at the geodesic midpoint of the double edge if $k = 2$. 

\paragraph*{Regular dual face ($k \geq 3$).}
\leavevmode
Let $\mathbf{c}_0,\dots,\mathbf{c}_{k-1}$ be the boundary vertices of~$f^*_i$ in the cyclic order that keeps $f^*_i$ on the right when viewed from the exterior. 
We compute the boundary moment vector $\mathbf{g}$ of~$f^*_i$ and its unit vector $\hat{\mathbf{g}}$:
\begin{equation}
  \mathbf{g} = \sum_{r=0}^{k-1} \hat{\mathbf{n}}_r\,\theta_r,
  \qquad
  \hat{\mathbf{g}} = \mathbf{g} / \lVert\mathbf{g}\rVert,
\end{equation}
where $\hat{\mathbf{n}}_r = (\mathbf{c}_r \times \mathbf{c}_{r+1}) /\lVert\mathbf{c}_r \times \mathbf{c}_{r+1}\rVert$ is the great-circle pole and $\theta_r$ is the arc length of the $r$-th edge. 
This boundary-moment vector is based on Brock's spherical-triangle centroid formula~\cite{Bro75}, and our implementation is based on the spherical-polygon centroid computation in d3-geo~\cite{Bos24}.
With this orientation, $\mathbf{g}$ points opposite to the centroid vector of~$f^*_i$, so $-\hat{\mathbf{g}}$ is its radial projection and no sign search is needed. 
We accept $v_i=-\hat{\mathbf{g}}$ only if it lies in the interior of~$f^*_i$, as determined by the point in spherical polygon algorithm~\cite{KRB22}.  
For a non-convex face, the current assignment is rejected if the projected centroid lies outside the face.
We then add a no-good cut excluding only this bijection and re-solve the model for the same candidate~$G^*$.
If the model is proved infeasible, we proceed to the next candidate.

\paragraph*{Digon dual face ($k = 2$).}
Under the minor-arc realization, a digon's two boundary arcs coincide, so it has no interior. 
Instead, we place its primal vertex at the geodesic midpoint of the double edge. 
This choice is symmetric in the two endpoints and independent of boundary orientation.
\subsubsection{Primal edges and faces}
\label{subsubsec:primal_edge_face}
We now derive the primal edges and faces from the realized dual $Q^*$ by plane duality.

\paragraph*{Primal edges.}
Let $\bar e \in \mathcal{E}^*$ be an underlying dual edge.
If $m(\bar e) = 1$, then $\bar e$ has two distinct incident dual faces $f_L^*$ and $f_R^*$: since $Q$ is simple, it has no loops, so $Q^*$ has no bridges. 
This yields the primal edge $(v_{f_L^*}, v_{f_R^*})$. If $m(\bar e) = 2$, its two parallel edges bound a digon face $f_D^*$; on their opposite sides, the two edges are incident to distinct dual faces $f_L^*$ and $f_R^*$. 
In the primal graph, this produces the two-edge chain $(v_{f_L^*}, v_{f_D^*})$, $(v_{f_D^*}, v_{f_R^*})$ through the degree-$2$ primal
vertex $v_{f_D^*}$ associated with the digon (\S\ref{subsubsec:primal_vertex}). 
Summing over $\mathcal{E}^*$, with $s$ and $d$ denoting the numbers of edges with $m(\bar e) = 1$ and $m(\bar e) = 2$ respectively, $\lvert E\rvert = s + 2d = \lvert E^*\rvert = 2n$.
Plane duality does not ensure that the minor-arc realization of the primal graph is crossing-free.
We realize each primal edge as the minor arc between its endpoints and reject the corresponding assignment if any two primal edges that do not share an endpoint intersect.

\paragraph*{Primal faces.}
Each dual vertex~$v_j^*$ corresponds to one primal face~$f_j$.
Since $Q^*$ is $4$-regular, reading the four dual faces encountered around $v_j^*$ in the cyclic order of the embedding yields four primal vertices. 
Because $Q^*$ is the dual of a simple quadrangulation~(\S\ref{subsec:preliminaries}),  these four vertices are pairwise distinct, so $f_j$ is the quadrilateral $(v_a,v_b,v_c,v_d)$. 
If a double edge is incident to~$v_j^*$, the corresponding digon appears once in this cyclic list and contributes its degree-$2$ primal vertex as a single corner of~$f_j$.
\section{Experimental results}
\label{sec:results}
\subsection{Experimental setup}
\label{subsec:setup}
All experiments ran on a machine with an Intel Core i7-13700F CPU and 64\,GB of RAM. Candidate quadrangulation graphs were generated with \texttt{plantri}~v5.5~\cite{BMV24}, and the resulting assignment problems (\S\ref{subsec:dual_realization}) were solved with Google OR-Tools CP-SAT~v9.15.6755~\cite{PD26}. 
The pipeline was implemented in Python~3.11.

Each realized SQS was classified according to the geometry of its primal faces. 
Using the convention introduced in \S\ref{subsec:preliminaries}, an SQS was labeled \textsc{Convex} if every primal face is geodesically convex. 
An SQS was labeled \textsc{Degenerate} if the largest interior angle over all primal faces equals~$\pi$, and \textsc{Concave} if it exceeds~$\pi$.
\subsection{SQS construction}
\label{subsec:sqs_construction}

For a candidate dual plane graph $G^*$, let $\boldsymbol{k}^{\downarrow}=(k_1,\ldots,k_{n+2})$ denote the degrees of its faces in non-increasing order. 
By duality, this is also the non-increasing sequence of primal vertex valences. 
Let $\theta_{\min}$ denote the minimum interior angle over all primal faces. 
The angle sum at a primal vertex of valence $k_1$ yields $\theta_{\min}\le 2\pi/k_1$. 
We process candidates in ascending lexicographic order of $(d,\boldsymbol{k}^{\downarrow})$, breaking remaining ties by their original order in the corresponding \texttt{plantri} output. 
Since $d$ is the primary ordering key and $d=0$ is necessary for geodesic convexity, all digon-free candidates precede candidates with double edges.
Within a fixed $d$, however, an earlier candidate need not admit a geodesically convex realization or one with a larger $\theta_{\min}$.

When the digon-free subclass $\mathcal{G}_{n,0}^*$ is nonempty, we process its candidates in the order above and stop at the first valid geodesically convex SQS.
If the search exhausts this subclass without finding such an SQS, we select among the valid constructions found in the subclass, preferring \textsc{Degenerate} to \textsc{Concave} and breaking ties by shorter total primal edge length. 
For $n\in\{3,4,5,7\}$, the subclass $\mathcal{G}_{n,0}^*$ is empty (Table~\ref{tab:enumeration_counts}), so we evaluate all candidates in $\mathcal{G}_n^*$. 
The valid constructions found are then ranked first by geometry class, with \textsc{Degenerate} preferred to \textsc{Concave}, next by smaller $d$, and finally by shorter total primal edge length.

\begin{figure*}[t]
  \centering
  \setlength{\tabcolsep}{1pt}
  \begin{tabular}{@{}cccccc@{}}
    \includegraphics[width=0.160\linewidth]{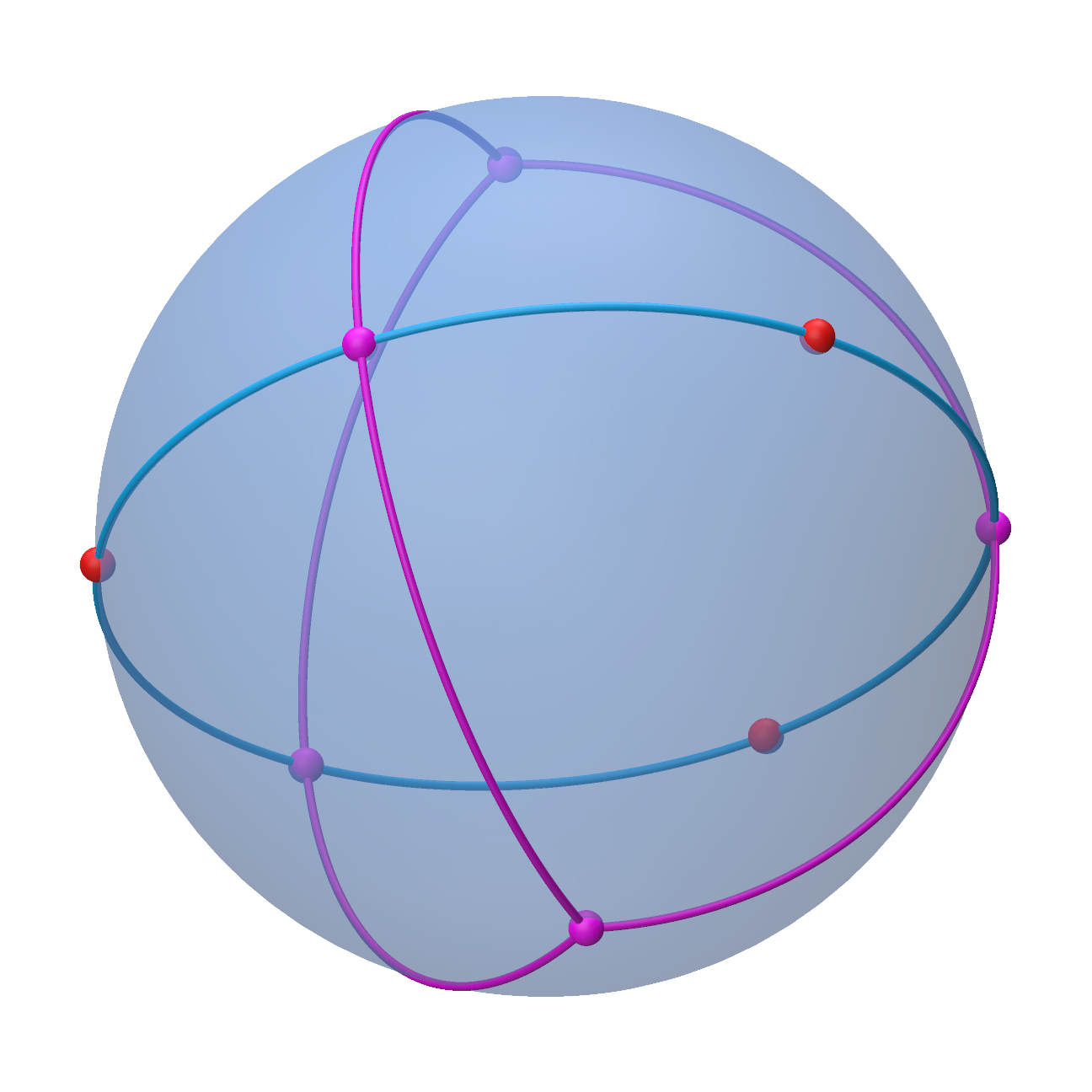} &
    \includegraphics[width=0.160\linewidth]{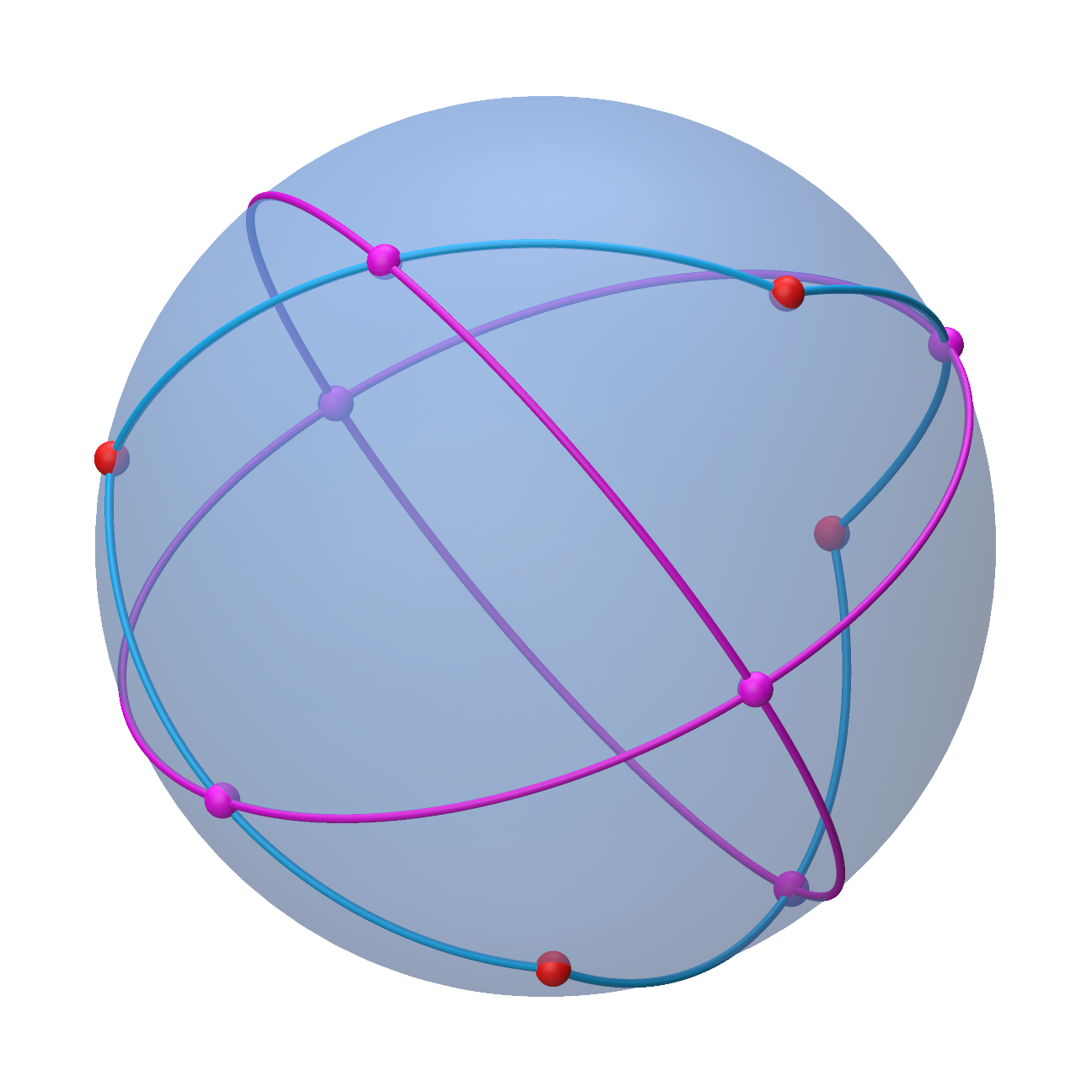} &
    \includegraphics[width=0.160\linewidth]{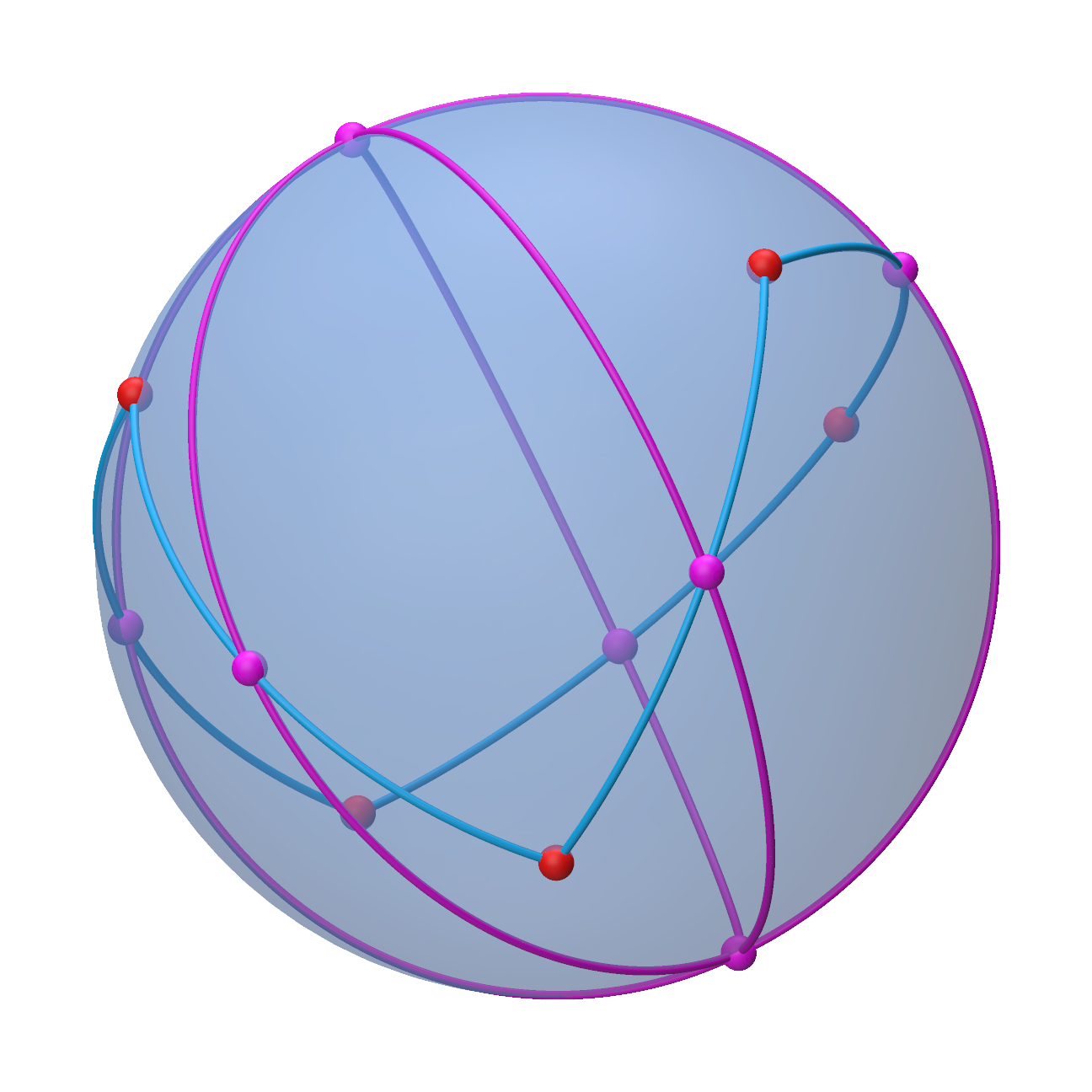} &
    \includegraphics[width=0.160\linewidth]{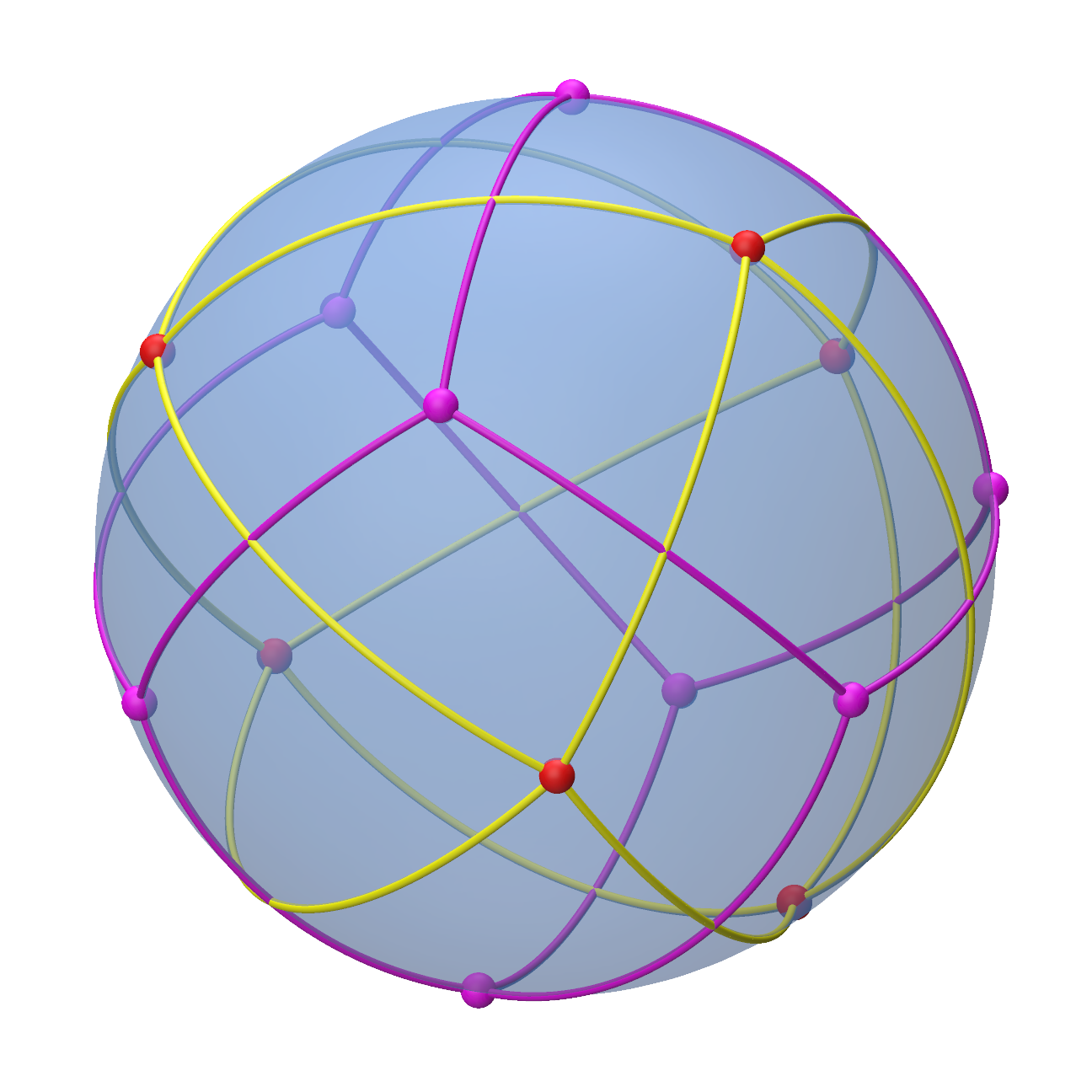} &
    \includegraphics[width=0.160\linewidth]{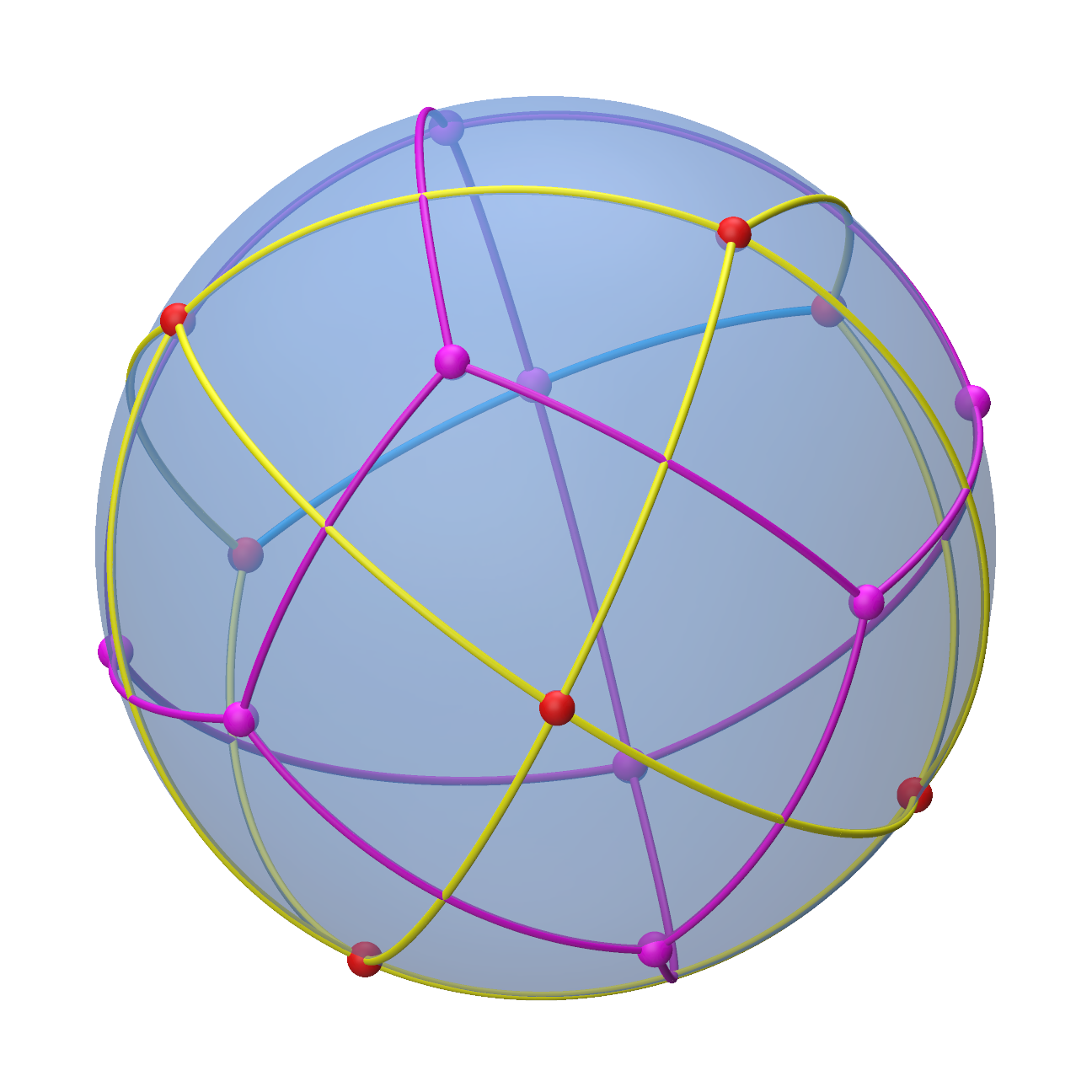} &
    \includegraphics[width=0.160\linewidth]{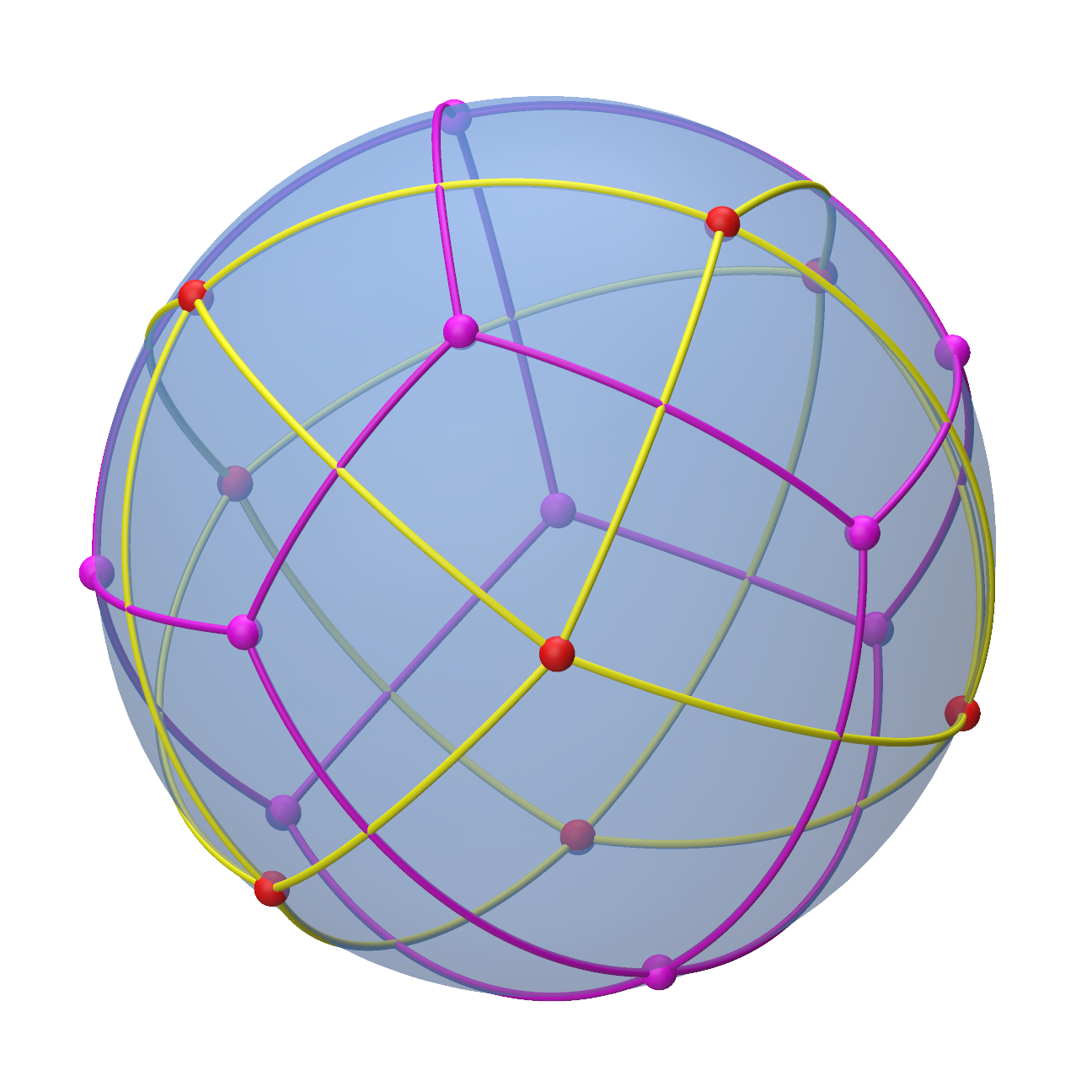} \\[-7pt]
    {\scriptsize $n{=}3$} &
    {\scriptsize $n{=}4$} &
    {\scriptsize $n{=}5$} &
    {\scriptsize $n{=}6$} &
    {\scriptsize $n{=}7$} &
    {\scriptsize $n{=}8$} \\[1pt]
    \includegraphics[width=0.160\linewidth]{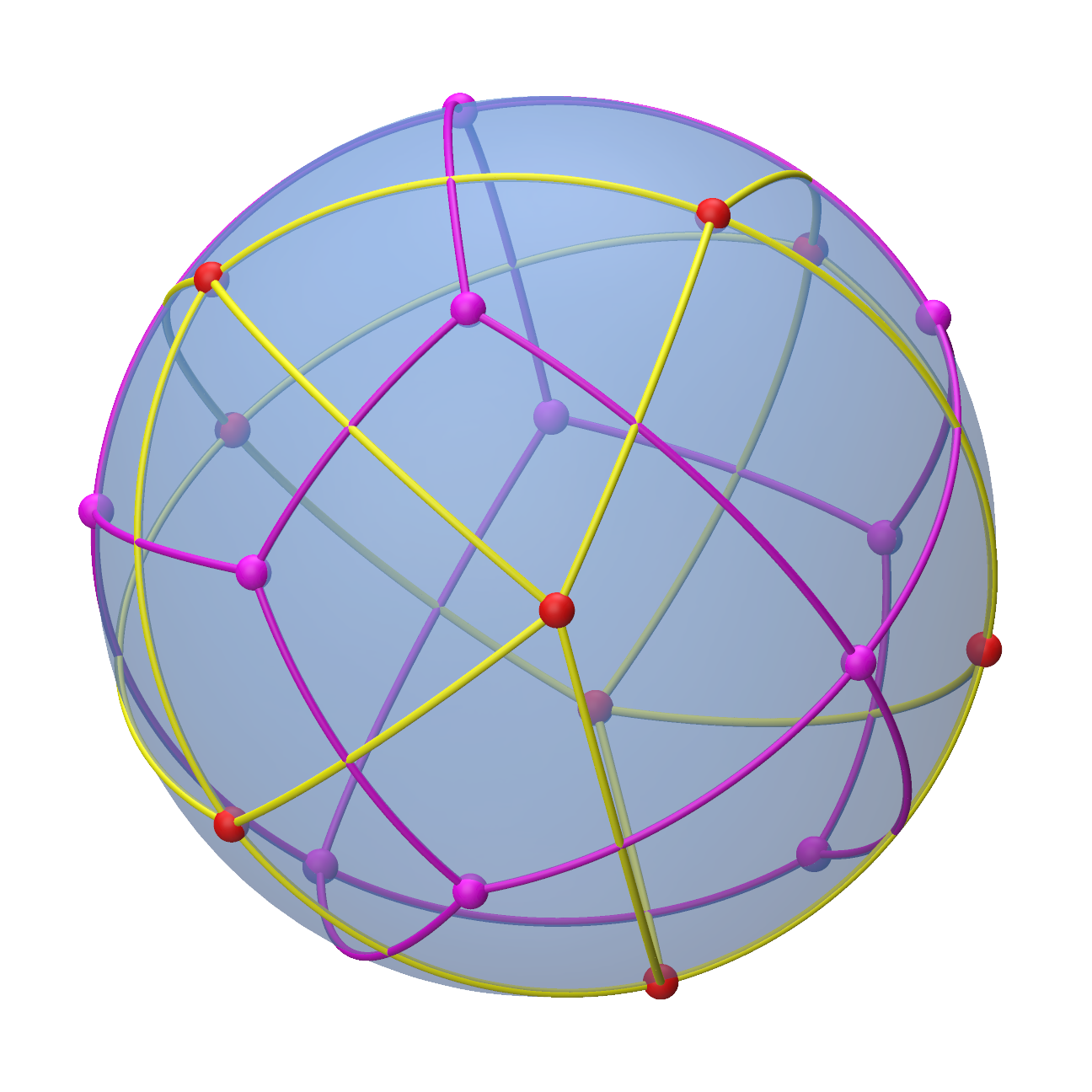} &
    \includegraphics[width=0.160\linewidth]{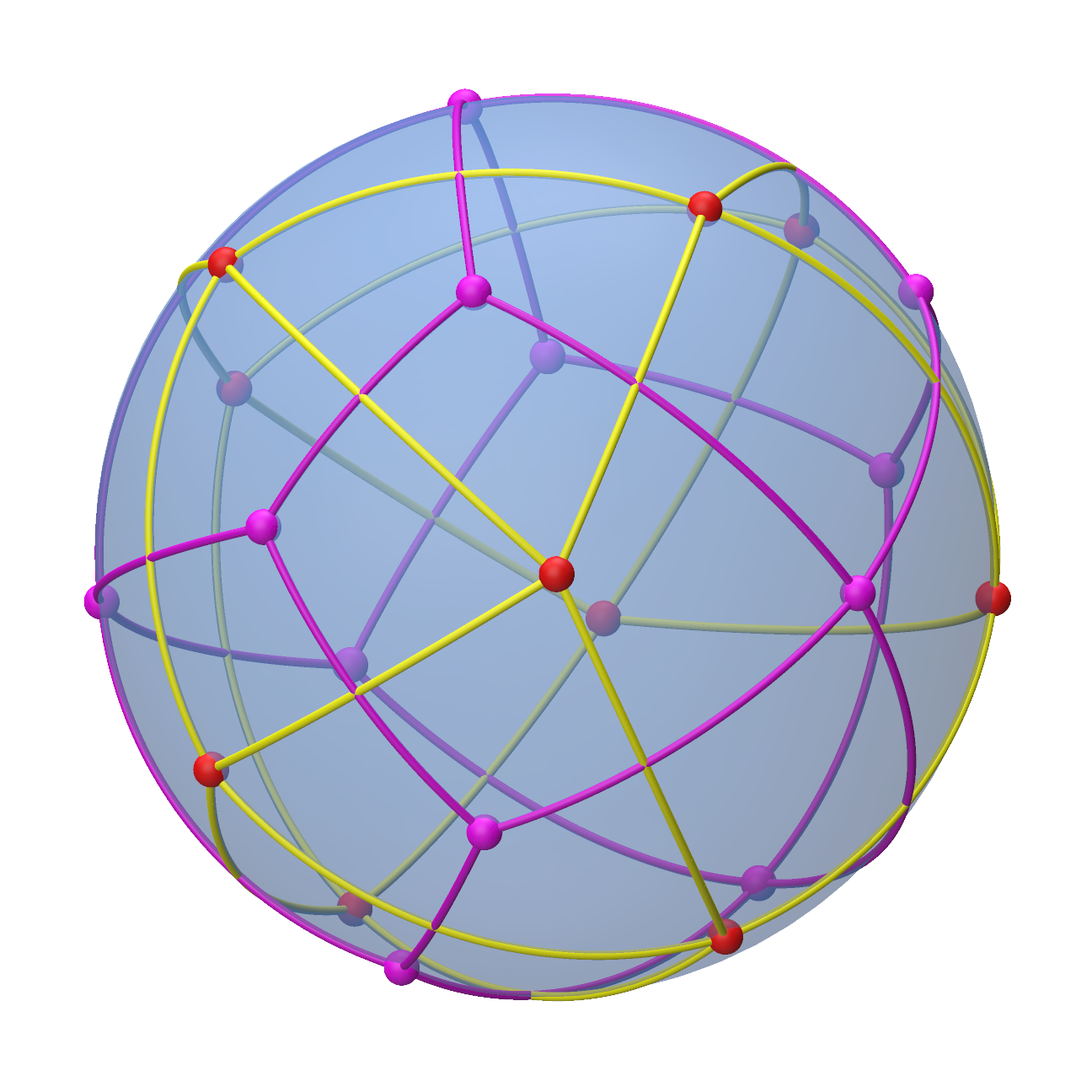} &
    \includegraphics[width=0.160\linewidth]{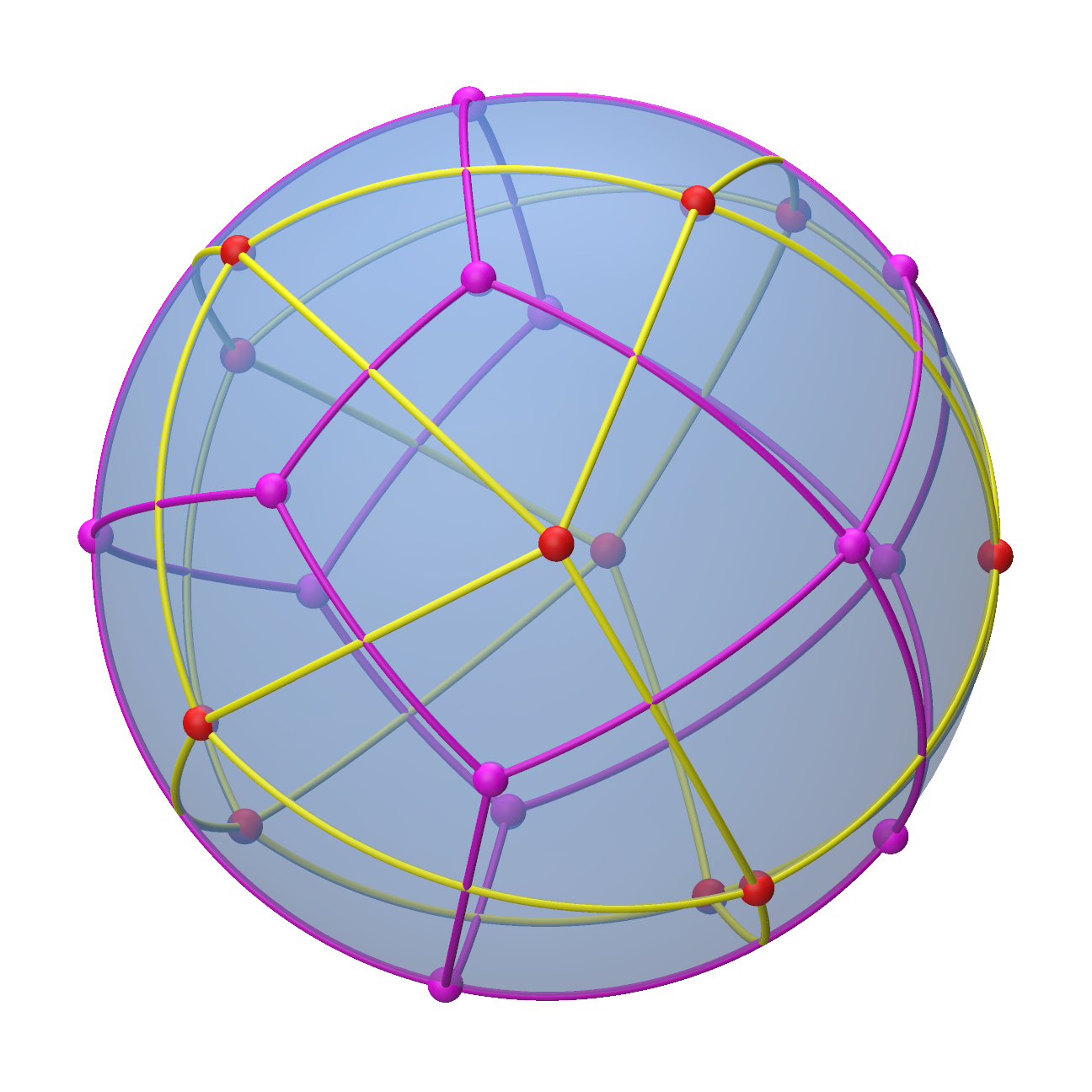} &
    \includegraphics[width=0.160\linewidth]{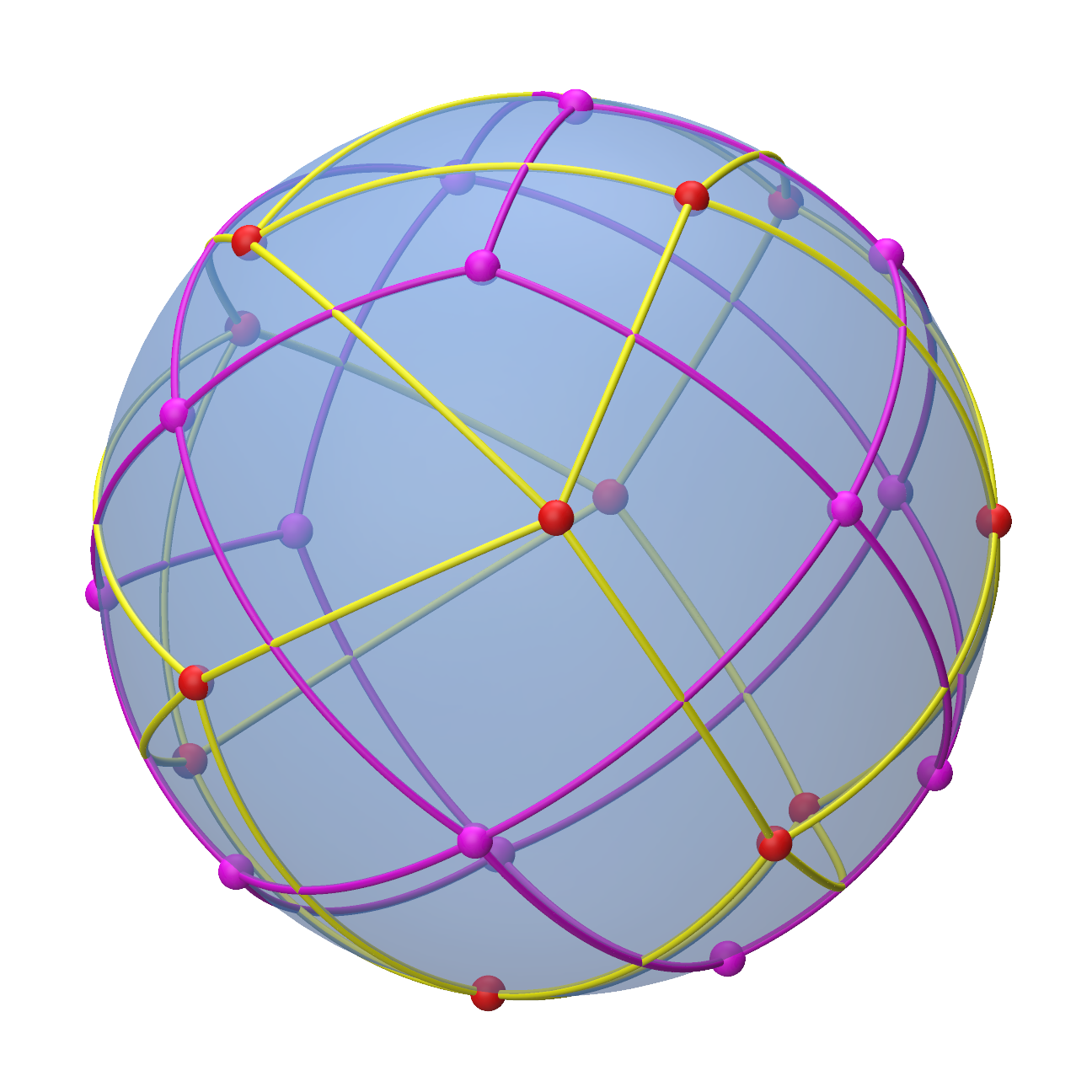} &
    \includegraphics[width=0.160\linewidth]{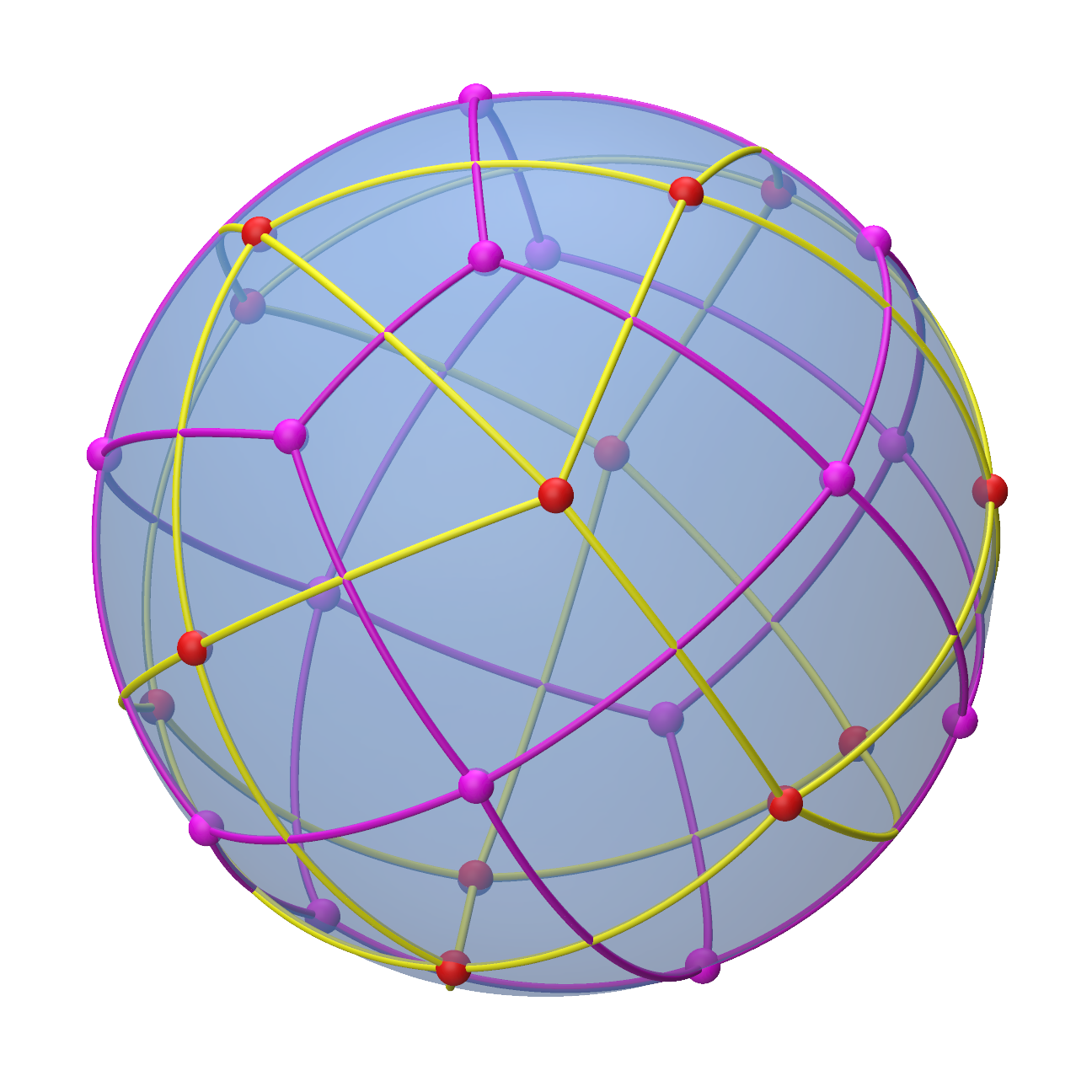} &
    \includegraphics[width=0.160\linewidth]{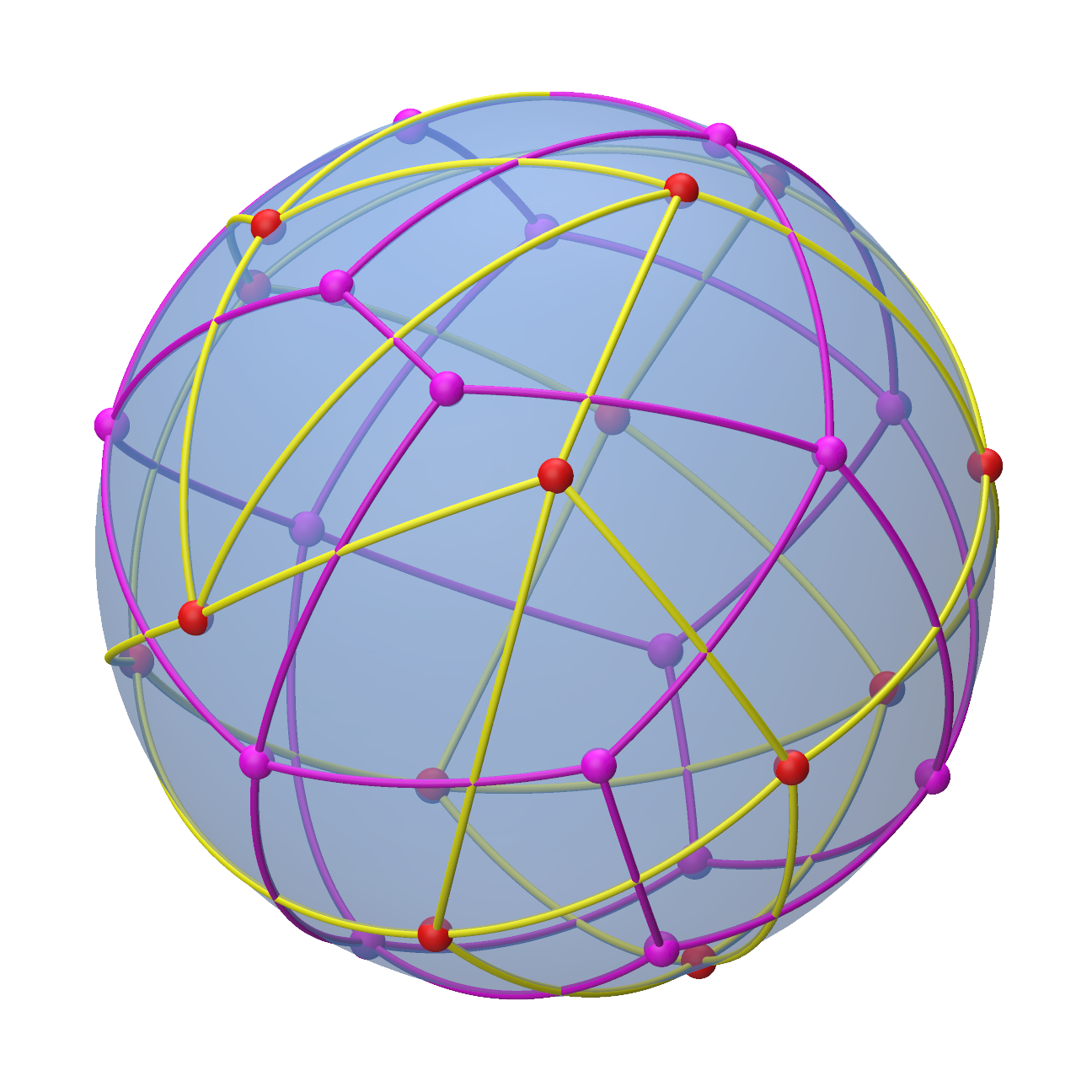} \\[-7pt]
    {\scriptsize $n{=}9$} &
    {\scriptsize $n{=}10$} &
    {\scriptsize $n{=}11$} &
    {\scriptsize $n{=}12$} &
    {\scriptsize $n{=}13$} &
    {\scriptsize $n{=}14$} \\[1pt]
    \includegraphics[width=0.160\linewidth]{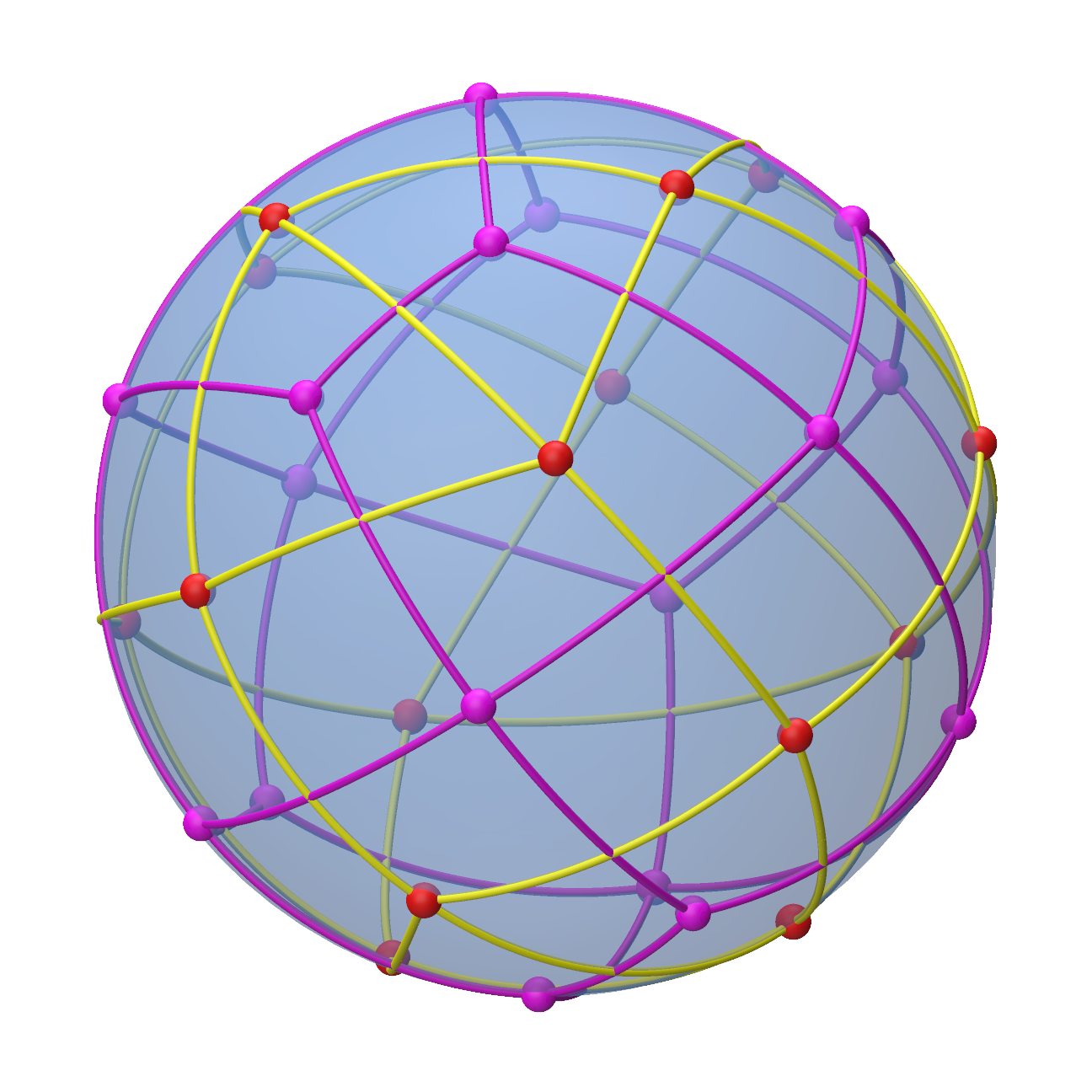} &
    \includegraphics[width=0.160\linewidth]{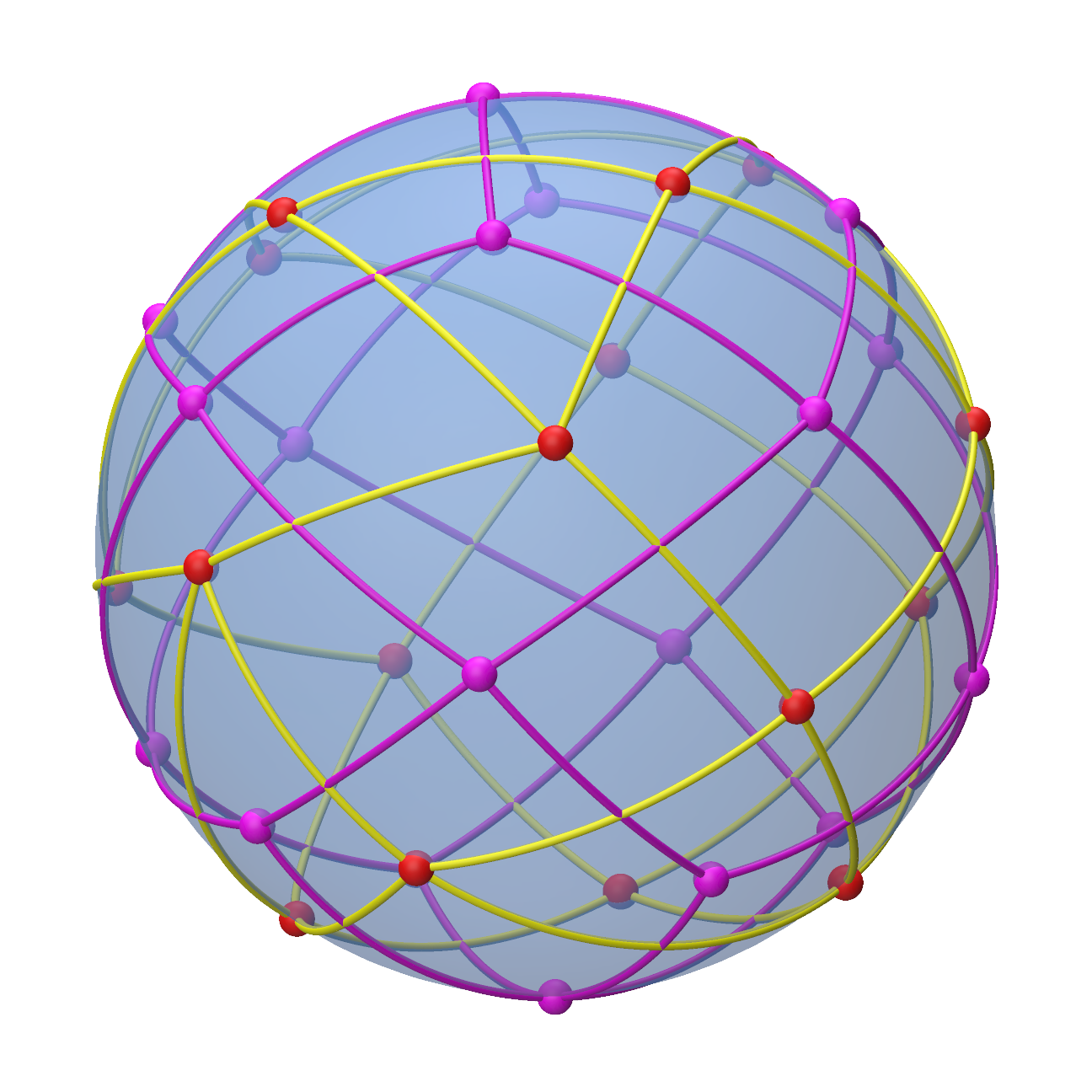} &
    \includegraphics[width=0.160\linewidth]{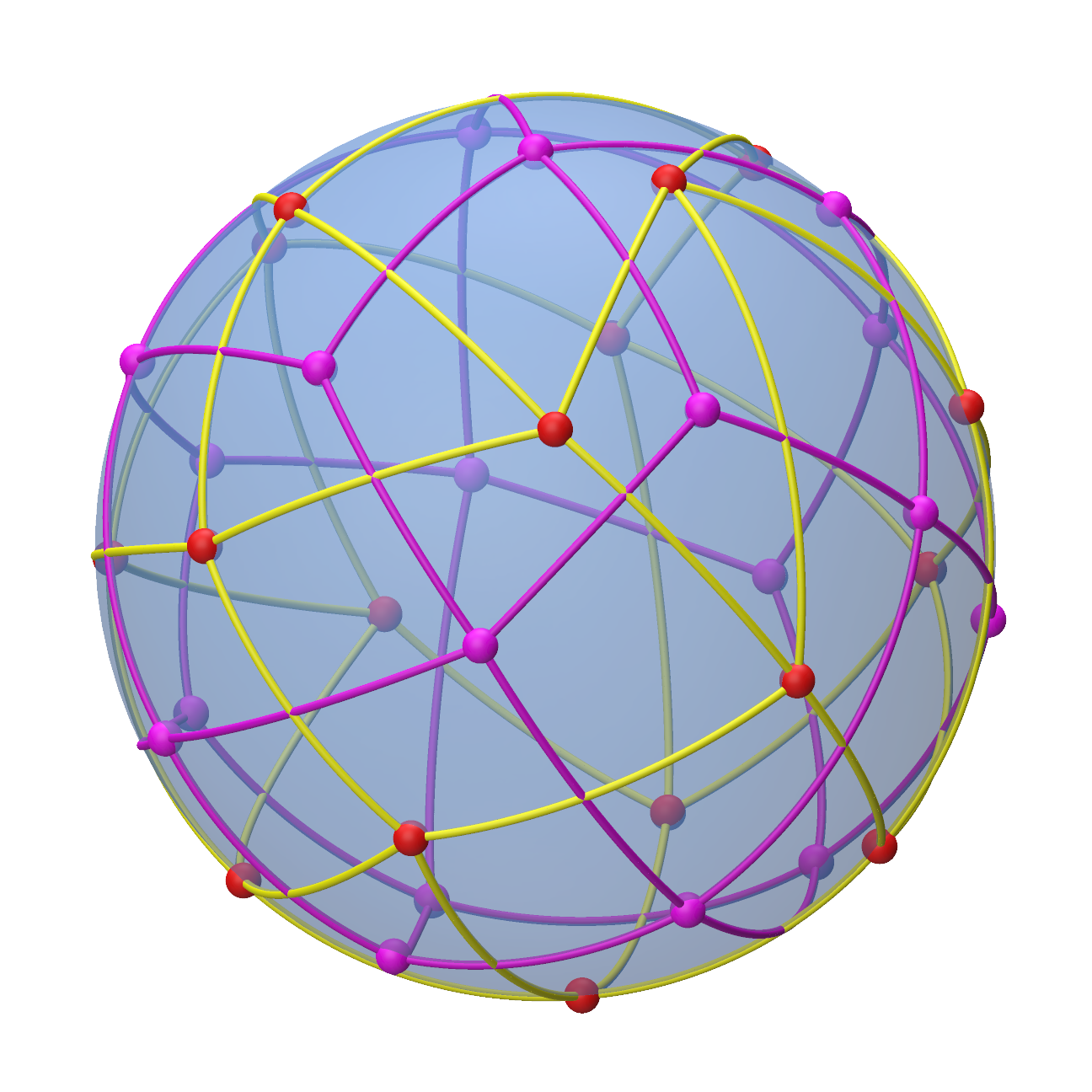} &
    \includegraphics[width=0.160\linewidth]{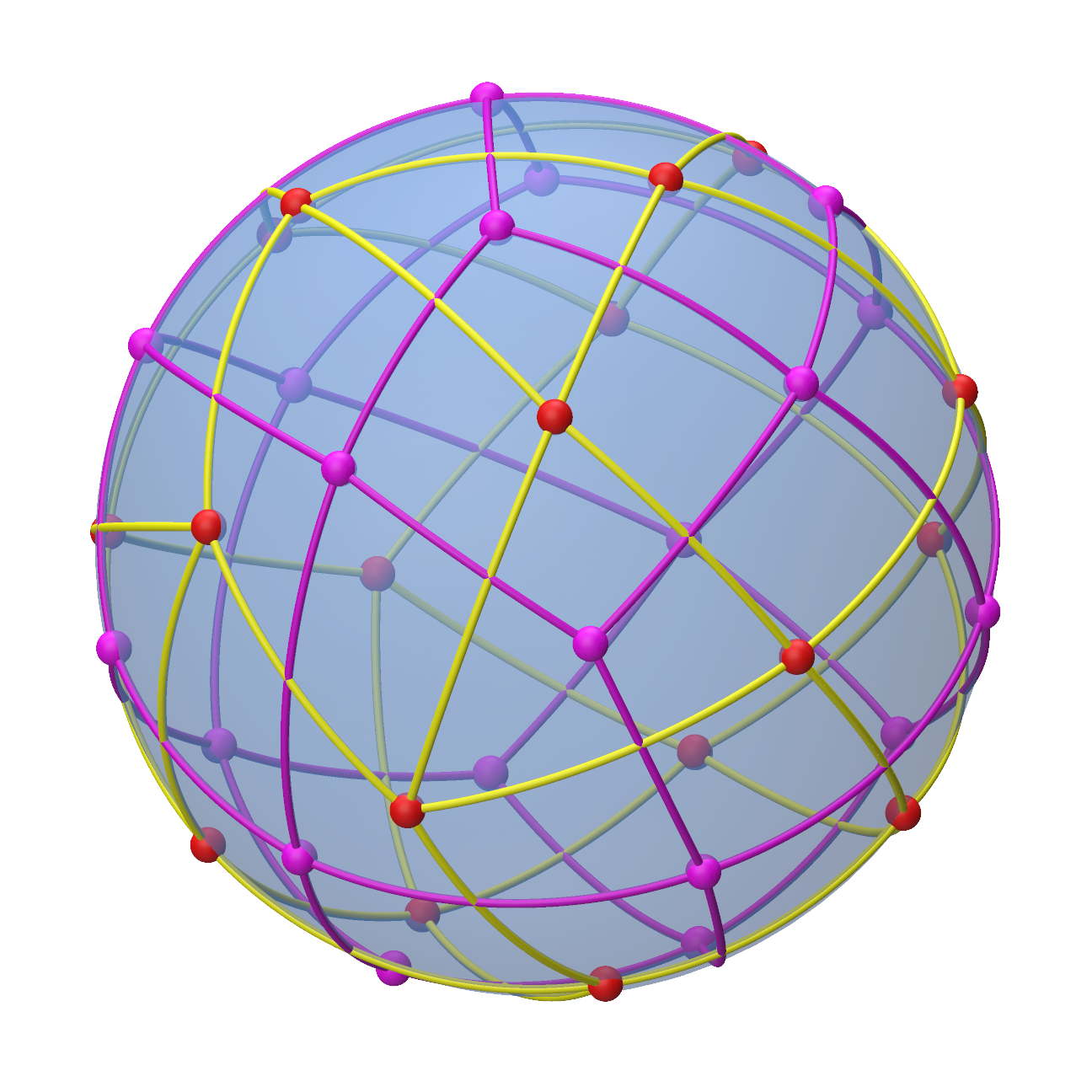} &
    \includegraphics[width=0.160\linewidth]{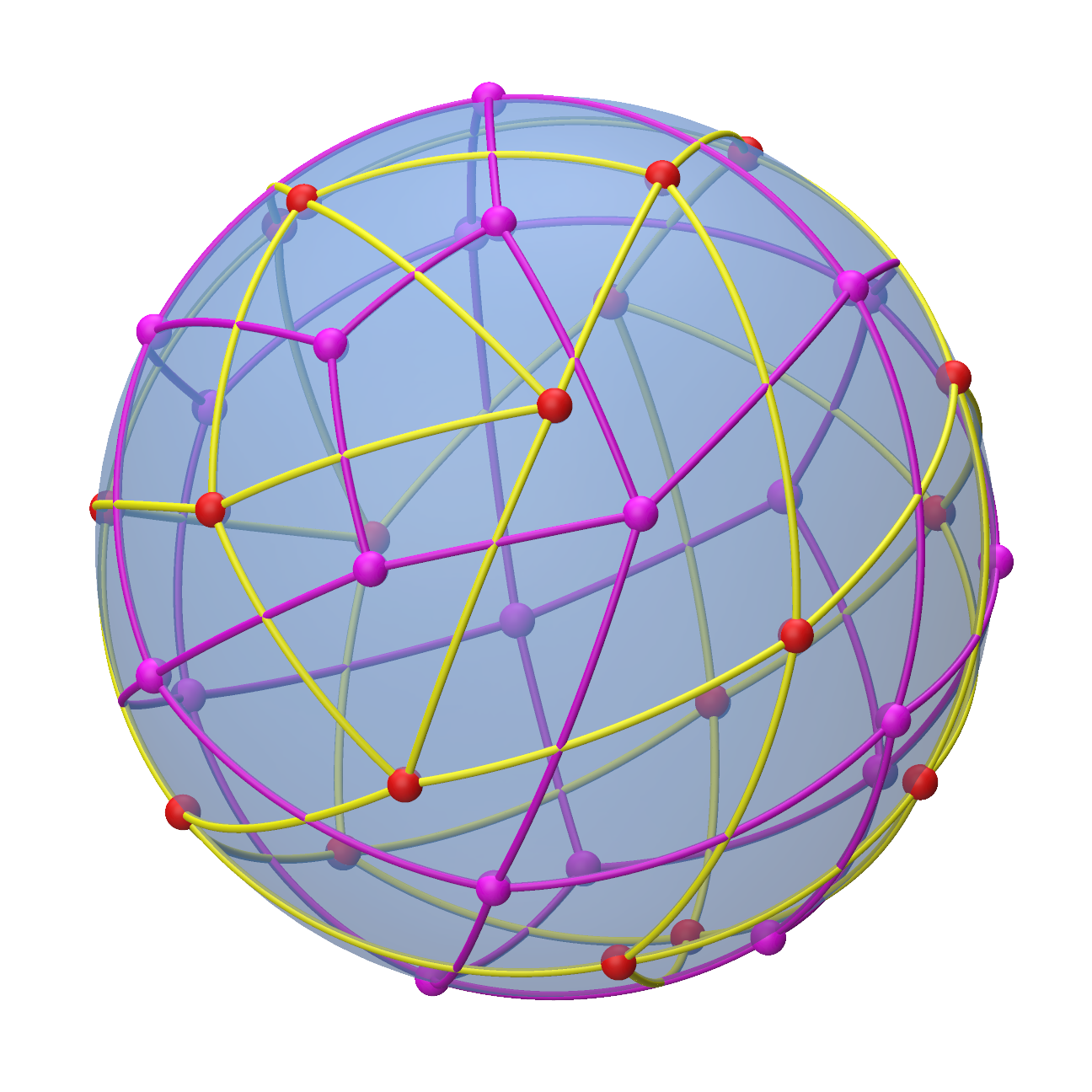} &
    \includegraphics[width=0.160\linewidth]{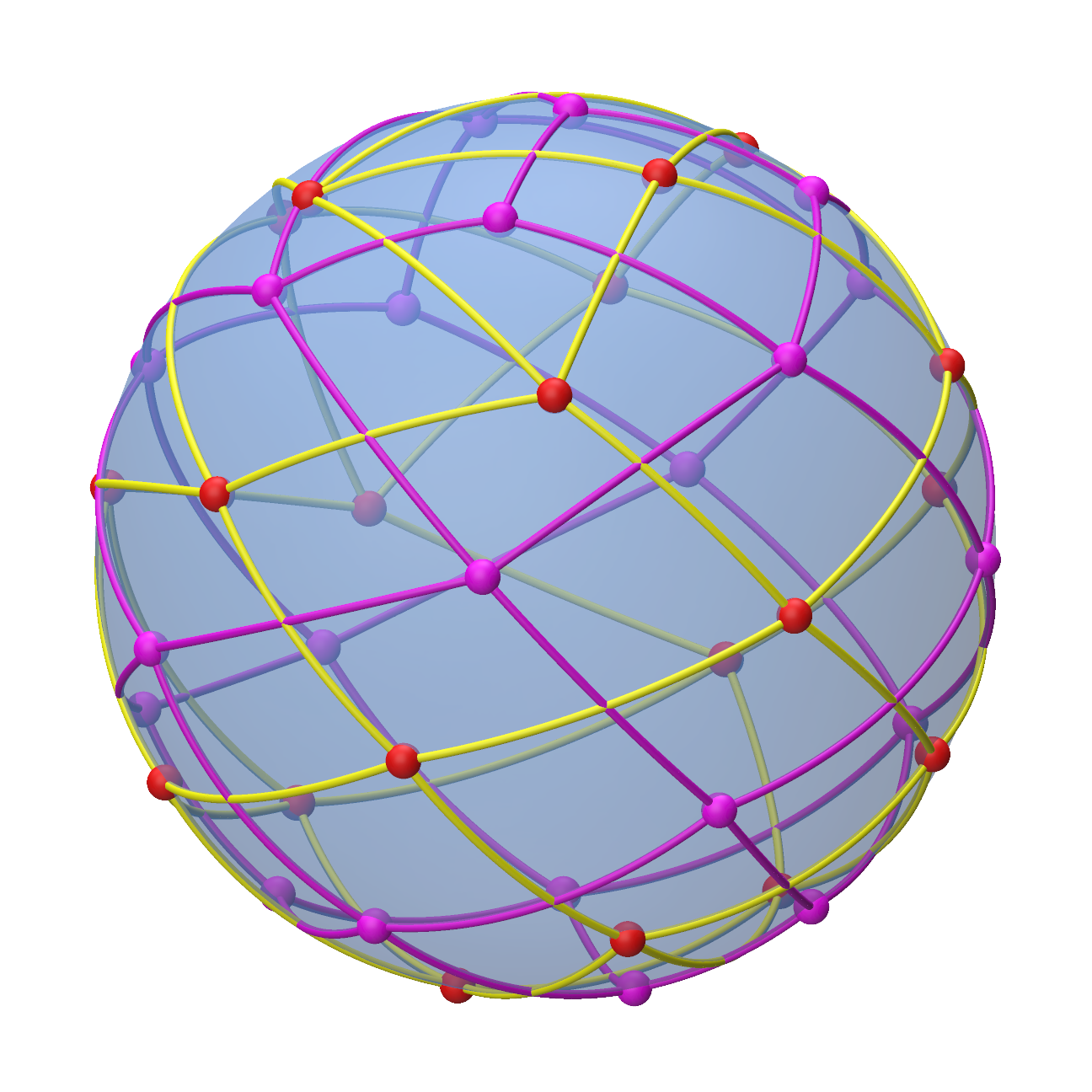} \\[-7pt]
    {\scriptsize $n{=}15$} &
    {\scriptsize $n{=}16$} &
    {\scriptsize $n{=}17$} &
    {\scriptsize $n{=}18$} &
    {\scriptsize $n{=}19$} &
    {\scriptsize $n{=}20$}
  \end{tabular}
  \caption{SQS constructions from Fibonacci direction sets for $n=3$--$20$. 
  Blue arcs denote double dual edges; yellow arcs denote single dual edges; 
  magenta arcs denote primal edges; magenta dots denote primal vertices; 
  red dots denote dual vertices.}
  \label{fig:fibonacci_sqs_results}
\end{figure*}

\begin{table}[t]
  \centering
  \small
  \caption{Dual realization performance and SQS geometry for Fibonacci inputs ($n=3$--$20$). 
    $N_{\mathrm{eval}}$: number of candidate dual graphs evaluated under the stopping policy; 
    $t_{\mathrm{solve}}$: cumulative CP-SAT solve time over all evaluated candidates;    
    $d$: double-edge count of the reported dual realization~$Q^*$; 
    $\theta_{\min}$, $\theta_{\max}$: minimum and maximum of the interior angles of all primal faces; 
    Status: classification of the reported SQS (\S\ref{subsec:setup}).}
  \label{tab:fibonacci_sqs_results}
  \resizebox{\columnwidth}{!}{%
  \begin{tabular}{@{}r r r r r r l@{}}
    \toprule
    & \multicolumn{2}{c}{Evaluation} & \multicolumn{4}{c}{Reported SQS} \\
    \cmidrule(lr){2-3}\cmidrule(lr){4-7}
    $n$ & $N_{\mathrm{eval}}$ & $t_{\mathrm{solve}}$ (ms) & $d$ & $\theta_{\min}$\,($^\circ$) & $\theta_{\max}$\,($^\circ$) & Status \\
    \midrule
    3  &  1 &        12 & 3 & 118.3 & 180.0 & \textsc{Degenerate} \\
    4  &  2 &        20 & 4 &  81.3 & 180.0 & \textsc{Degenerate} \\
    5  &  3 &        44 & 5 &  63.1 & 180.0 & \textsc{Degenerate} \\
    6  &  1 &        23 & 0 & 111.5 & 131.1 & \textsc{Convex}     \\
    7  & 18 &   1{,}600 & 1 &  83.7 & 180.0 & \textsc{Degenerate} \\
    8  &  1 &        45 & 0 &  86.9 & 125.2 & \textsc{Convex}     \\
    9  &  1 &        56 & 0 &  82.7 & 143.0 & \textsc{Convex}     \\
    10 &  1 &       298 & 0 &  84.8 & 135.6 & \textsc{Convex}     \\
    11 &  1 &       430 & 0 &  80.4 & 130.5 & \textsc{Convex}     \\
    12 &  1 &   1{,}047 & 0 &  70.3 & 146.3 & \textsc{Convex}     \\
    13 &  1 &       891 & 0 &  81.8 & 130.8 & \textsc{Convex}     \\
    14 &  1 &   1{,}701 & 0 &  67.5 & 145.2 & \textsc{Convex}     \\
    15 &  1 &       748 & 0 &  75.3 & 132.5 & \textsc{Convex}     \\
    16 &  1 &   1{,}525 & 0 &  67.7 & 163.0 & \textsc{Convex}     \\
    17 &  1 &   6{,}388 & 0 &  70.1 & 155.7 & \textsc{Convex}     \\
    18 &  1 &   2{,}565 & 0 &  73.7 & 149.2 & \textsc{Convex}     \\
    19 &  1 &  14{,}427 & 0 &  60.9 & 169.9 & \textsc{Convex}     \\
    20 &  1 &  40{,}098 & 0 &  57.0 & 171.7 & \textsc{Convex}     \\
    \bottomrule
  \end{tabular}
  } 
\end{table}

For each $n=3,\ldots,20$, we evaluate our method on one direction set generated by spherical Fibonacci sampling on~$\mathbb{S}^2$~\cite{Gon10,KISS15}.
Figure~\ref{fig:fibonacci_sqs_results} shows the reported SQS for each~$n$ under the policy above, and Table~\ref{tab:fibonacci_sqs_results} summarizes the corresponding candidate evaluations, CP-SAT solve time, and SQS geometry.
For all tested values of~$n$ with $\mathcal{G}_{n,0}^*\neq\emptyset$, the first candidate evaluated yielded a convex SQS. 
For $n\in\{3,4,5,7\}$, we evaluate every candidate in $\mathcal{G}_n^*$ as $\mathcal{G}_{n,0}^*=\emptyset$, 
and report a degenerate construction that the policy above ranks highest among the valid constructions found.
Across the tested instances, $t_{\mathrm{solve}}$ ranges from $12$\,ms to $40.1$\,s and varies non-monotonically with~$n$. 
For $n\in\{4,5,7\}$, $t_{\mathrm{solve}}$ accumulates over multiple evaluated candidates, with $N_{\mathrm{eval}}=18$ for $n=7$.

\begin{figure}[t]
  \centering
  \begin{tabular}{@{}c@{\hspace{10pt}}c@{}}
    \includegraphics[width=0.44\linewidth]{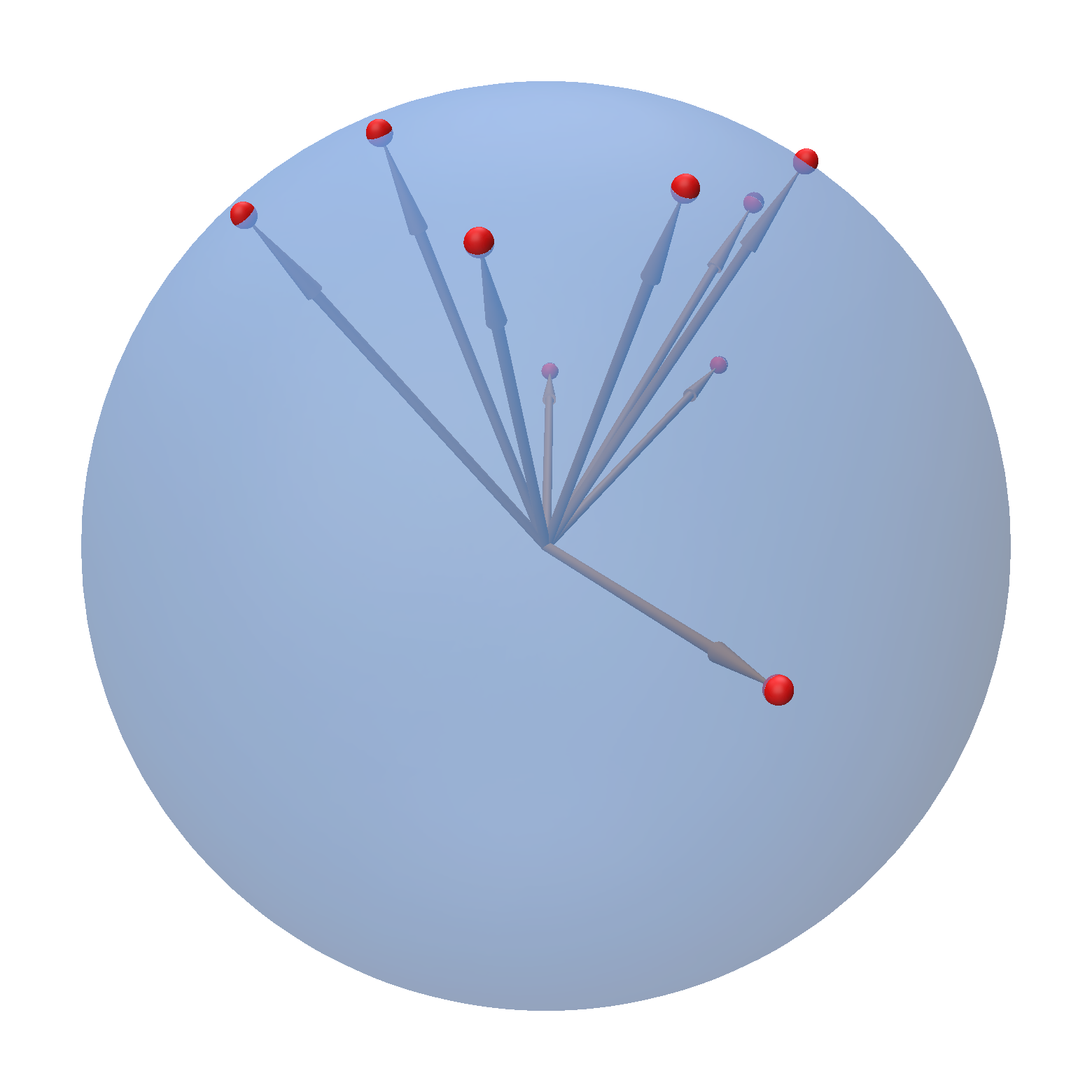} &
    \includegraphics[width=0.44\linewidth]{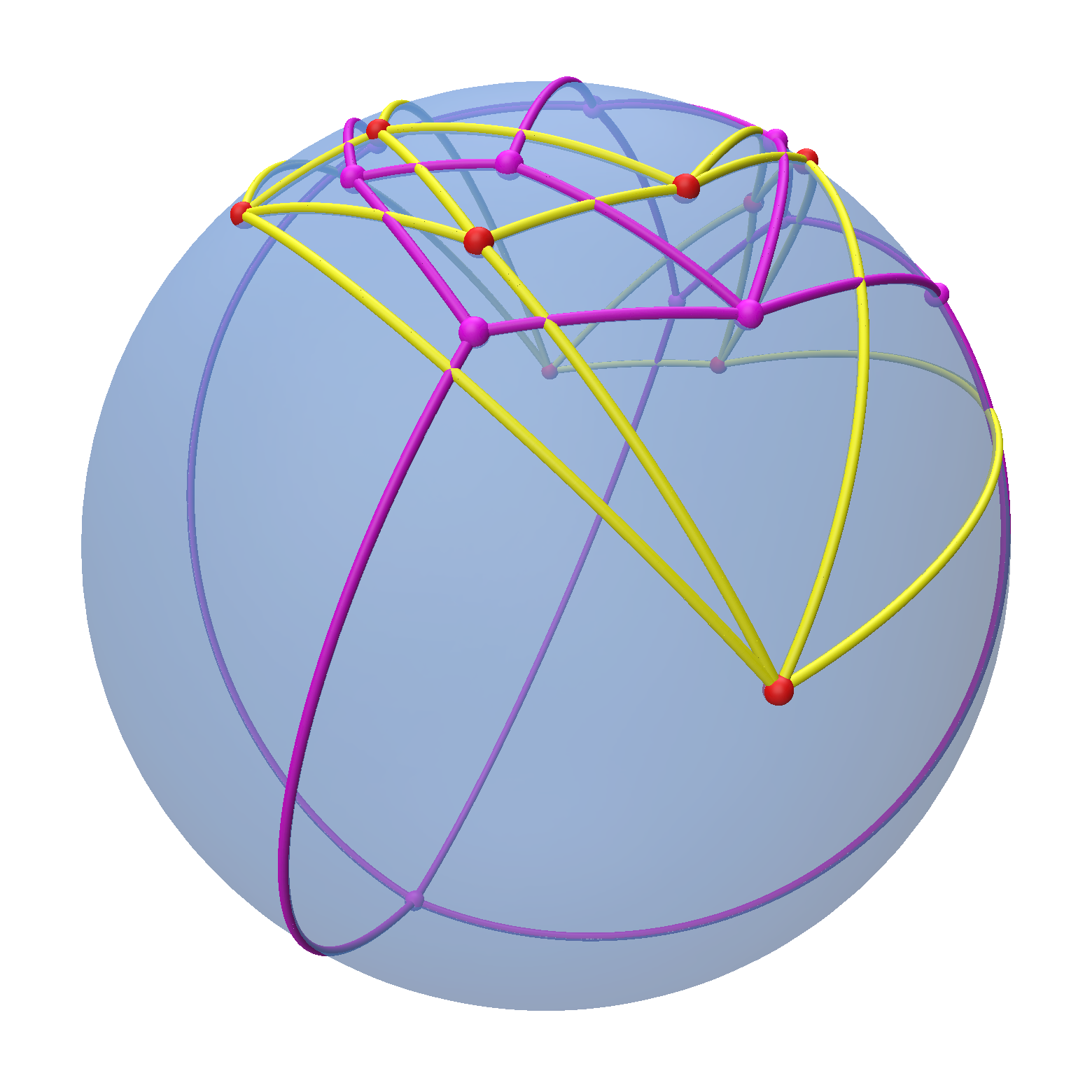} \\[0pt]
    \footnotesize (a) Hemisphere-contained input ($n{=}9$) &
    \footnotesize (b) Representative concave SQS
  \end{tabular}
  \caption{For the digon-free candidate dual graph~$G^*$ used in the $n{=}9$ Fibonacci result in Figure~\ref{fig:fibonacci_sqs_results} and Table~\ref{tab:fibonacci_sqs_results}, the input directions can affect whether our construction yields a convex SQS.}
  \label{fig:direction_dependence}
\end{figure}

For $n=9$, the digon-free subclass $\mathcal{G}_{9,0}^*$ contains a single dual graph class (Table~\ref{tab:enumeration_counts}).
Using a representative $G^*$ of this class, we obtain a convex SQS for the Fibonacci direction set, as shown in Figure~\ref{fig:fibonacci_sqs_results} and reported in Table~\ref{tab:fibonacci_sqs_results}.
For the hemisphere-contained input in Figure~\ref{fig:direction_dependence}(a), exhaustive evaluation of the $9!$ bijections finds valid SQSs, all of which are concave.
Since $d=0$ is necessary for geodesic convexity, no $n=9$ candidate yields a convex realization for this input under our construction.
Thus, the input directions can affect convex realizability even for the same candidate topology.
\subsection{Comparison with prior work}
\label{subsec:comparison_prior_work}

\begin{figure}[t]
  \centering
  \setlength{\tabcolsep}{2pt}
  \begin{tabular}{cc}
    \begin{tikzpicture}
      \node[anchor=south west,inner sep=0] (vimg) at (0,0)
        {\includegraphics[width=0.46\linewidth]{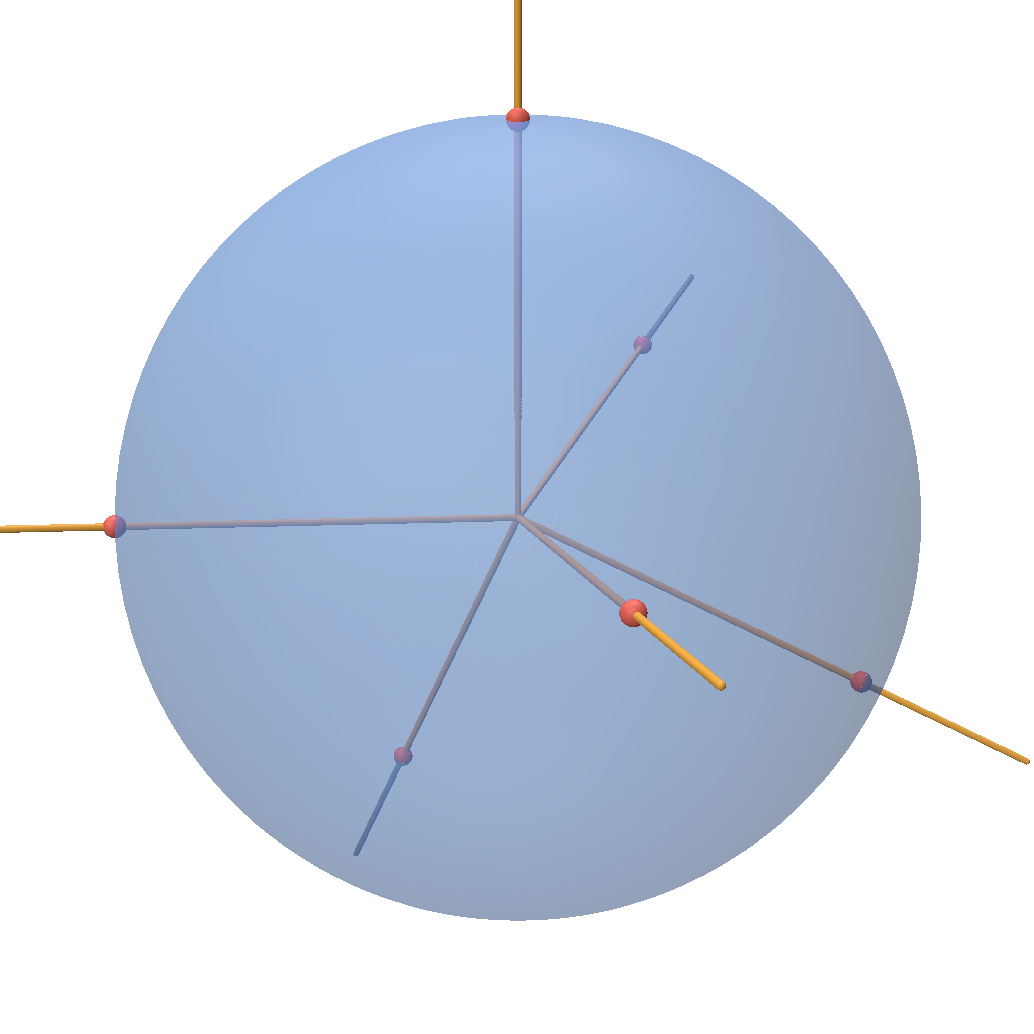}};
      \begin{scope}[x={(vimg.south east)},y={(vimg.north west)},
            every node/.append style={font=\footnotesize}]
        \node at (0.57,0.92) {$\vec{d}_0$};
        \node at (0.08,0.53) {$\vec{d}_1$};
        \node at (0.70,0.70) {$\vec{d}_2$};
        \node at (0.59,0.35) {$\vec{d}_3$};
        \node at (0.32,0.22) {$\vec{d}_4$};
        \node at (0.91,0.37) {$\vec{d}_5$};
      \end{scope}
    \end{tikzpicture} &
    \includegraphics[width=0.46\linewidth]{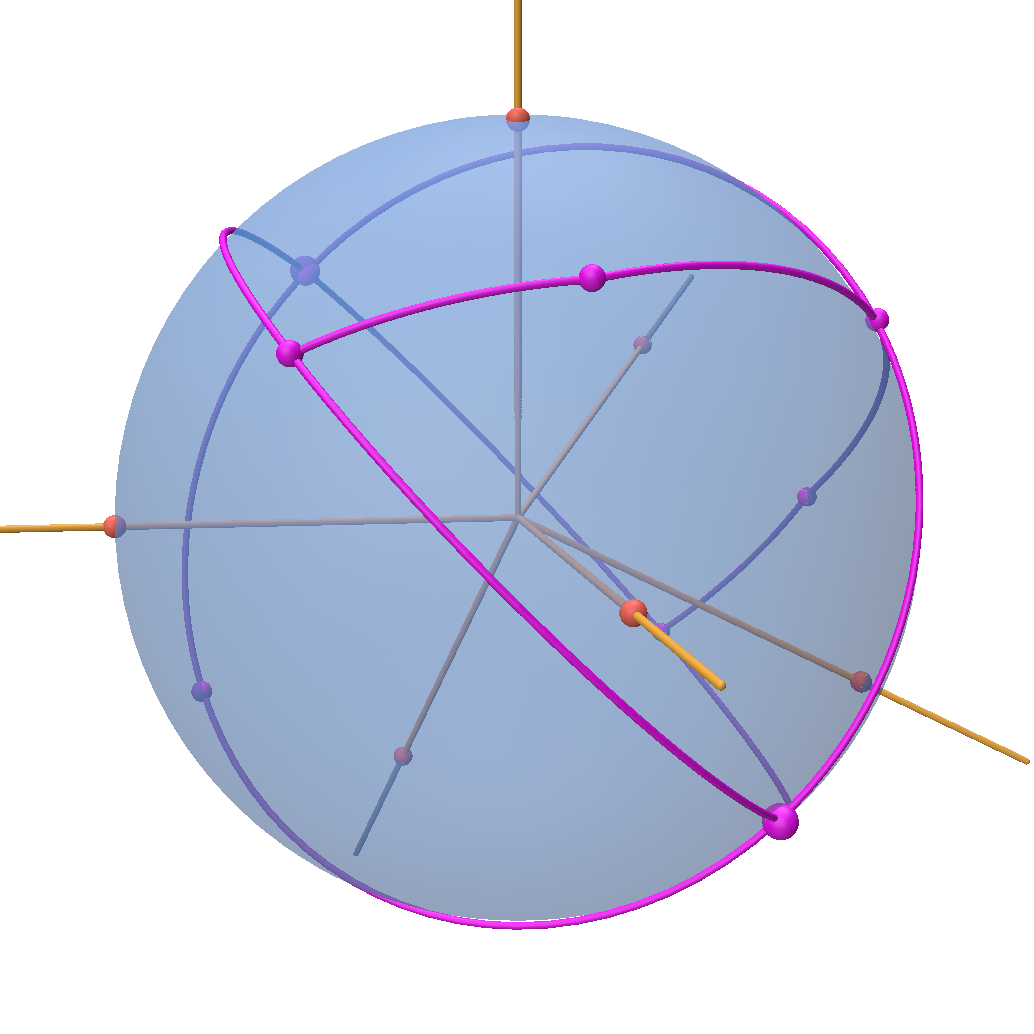} \\[-6pt]
    {\footnotesize (a) Input directions ($n{=}6$)} &
    {\footnotesize (b) Panotopoulou et al.~\cite{PRW*18}} \\[6pt]
    \includegraphics[width=0.46\linewidth]{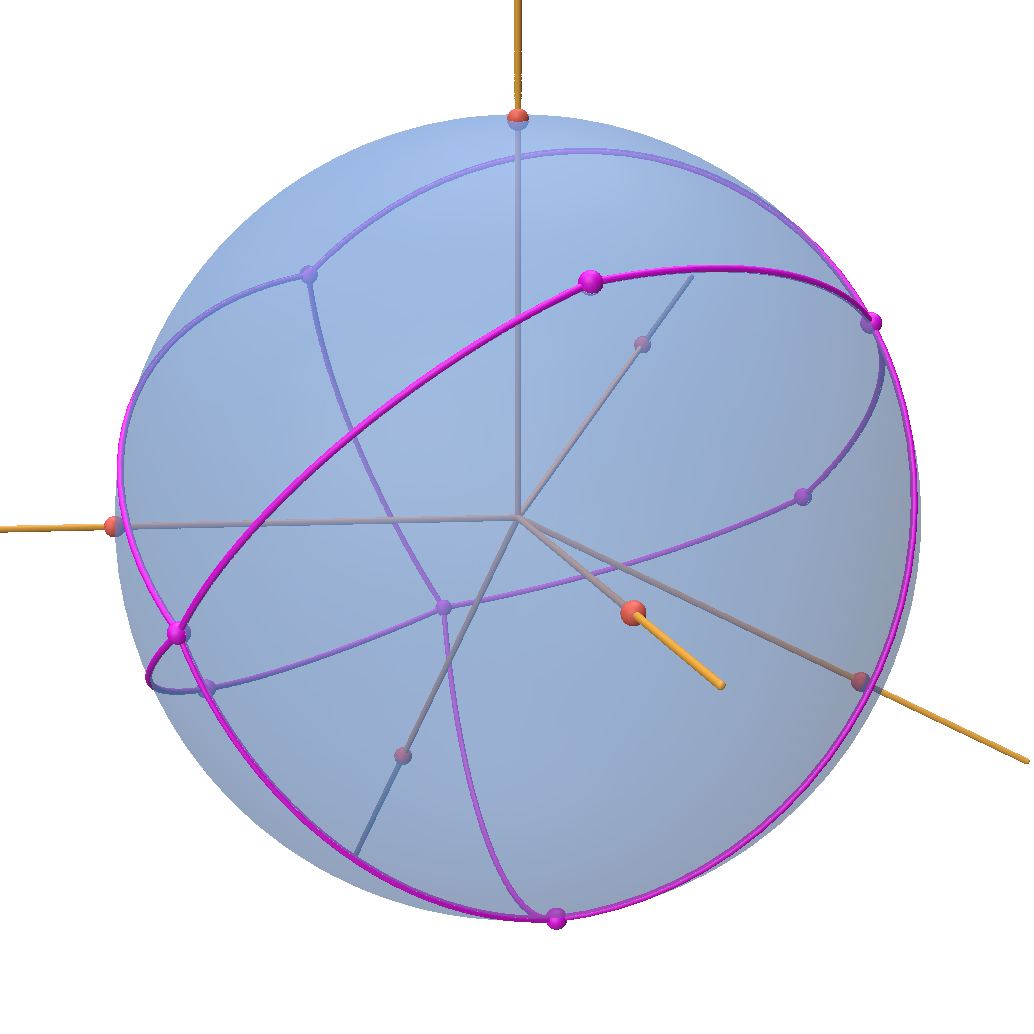} &
    \includegraphics[width=0.46\linewidth]{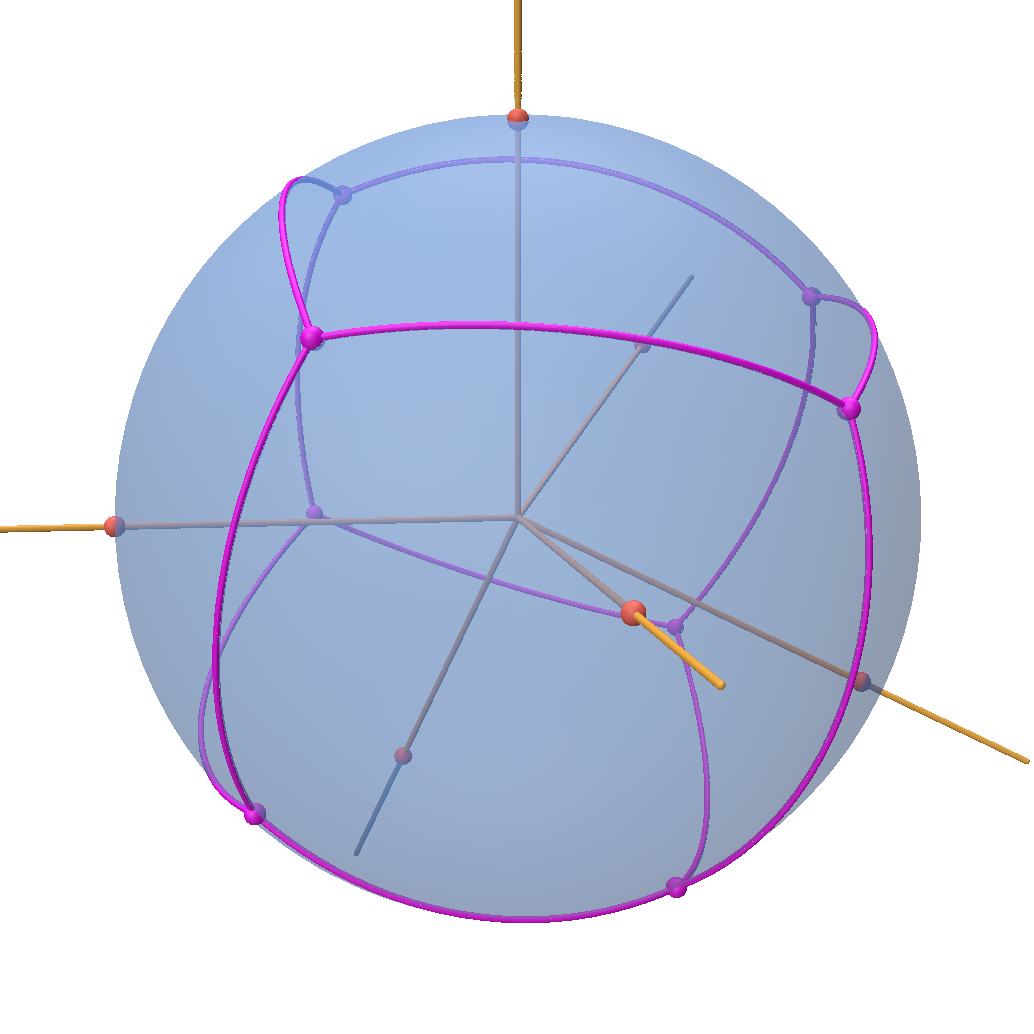} \\[-6pt]
    {\footnotesize (c) Peltier et al.~\cite{PMA21}} &
    {\footnotesize (d) Ours}
  \end{tabular}
  \caption{Joint-sphere partition comparison on the same six directions. 
    \textbf{(a)}~Input $\vec{d}_0,\dots,\vec{d}_5$ are numbered in insertion order. 
    \textbf{(b)}~Panotopoulou et al.~\cite{PRW*18} and \textbf{(c)}~Peltier et al.~\cite{PMA21} build the partition incrementally and are order-dependent; both are non-convex on these directions, whereas \textbf{(d)}~ours treats them as an unordered set and returns a crossing-free convex partition.}
  \label{fig:partition_comparison}
\end{figure}

We compare the two incremental methods and ours at the common joint-sphere partition stage, before scaffold or volume construction (Figure~\ref{fig:partition_comparison}).
Panotopoulou et al.~\cite{PRW*18} build a one-quad-per-branch partition incrementally and note that the result is not unique and depends on the branch order. 
Peltier et al.~\cite{PMA21} likewise generate their junction SQS by an ordered sequence of openings that affects both topology and geometry, and leave a canonical, order-independent SQS as future work. 
Both baselines produce topologically valid one-quad-per-branch layouts that are non-convex on this input, for different reasons. 
For~(b) convexity is structurally impossible, as the quad-splitting of~\cite{PRW*18} always produces at least two quads sharing two adjacent edges~\cite{PMA21}, that is, a dual digon, dual to a degree-$2$ primal vertex that forces a non-convex face for any insertion order (\S\ref{subsec:preliminaries}). 
For~(c) the openings of~\cite{PMA21} can in principle realize any SQS topology, but in practice the construction follows a single insertion sequence and local choices rather than searching over alternative orders or candidate dual graphs. 
For~(d), our method instead treats the directions as an unordered set, searches over candidate dual graphs and assignments, and returns a crossing-free convex partition for this input.

\begin{figure}[t]
  \centering
  \setlength{\tabcolsep}{8pt}
  \begin{tabular}{cc}
    \multicolumn{2}{c}{\includegraphics[width=0.42\linewidth]{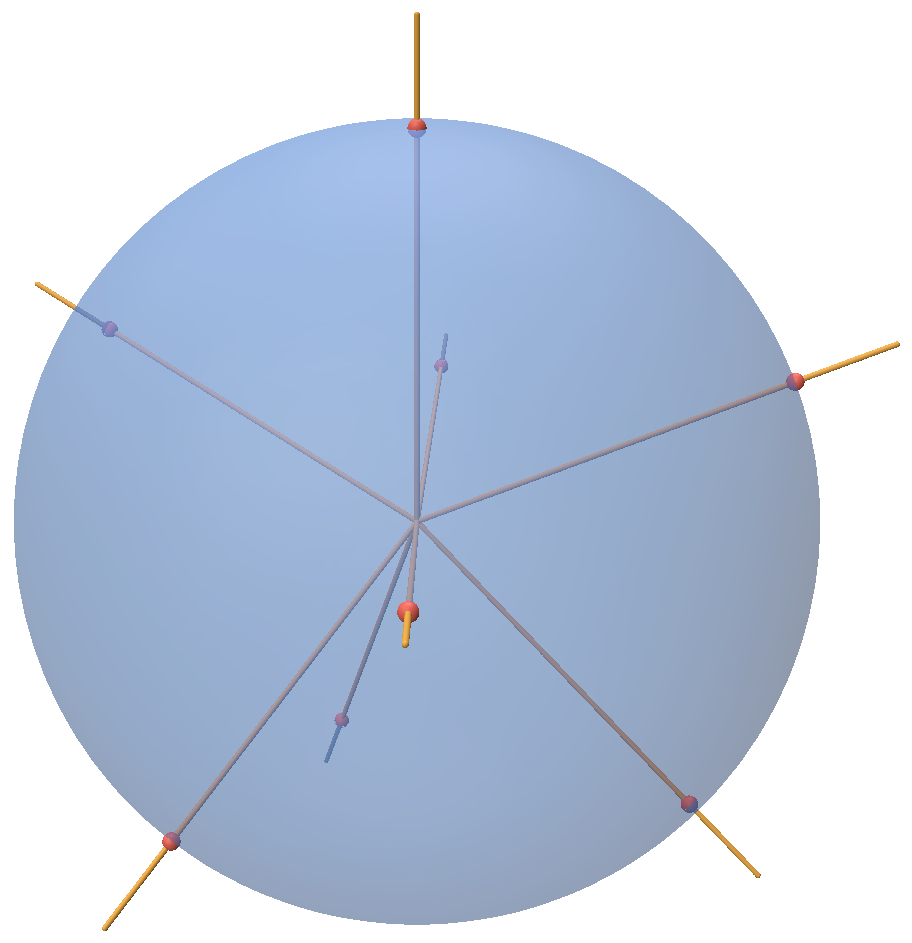}} \\[0.5em]
    \multicolumn{2}{c}{\footnotesize (a) Input directions ($n{=}8$)} \\[8pt]
    \includegraphics[width=0.42\linewidth]{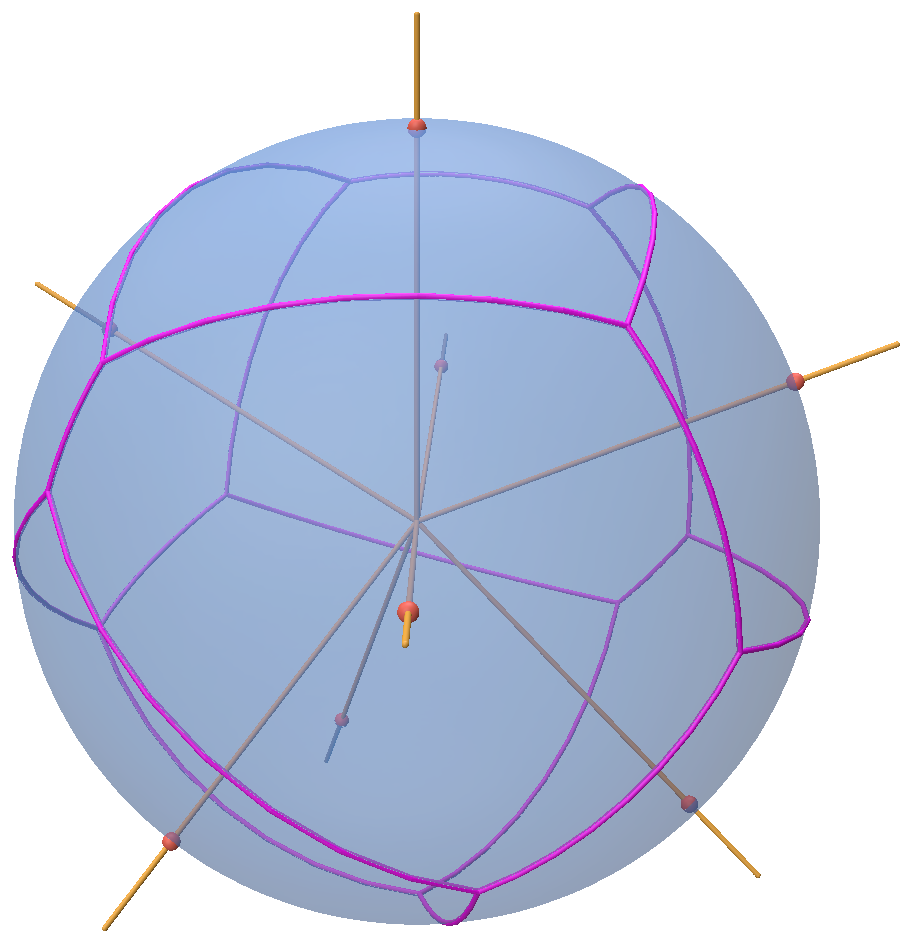} &
    \includegraphics[width=0.42\linewidth]{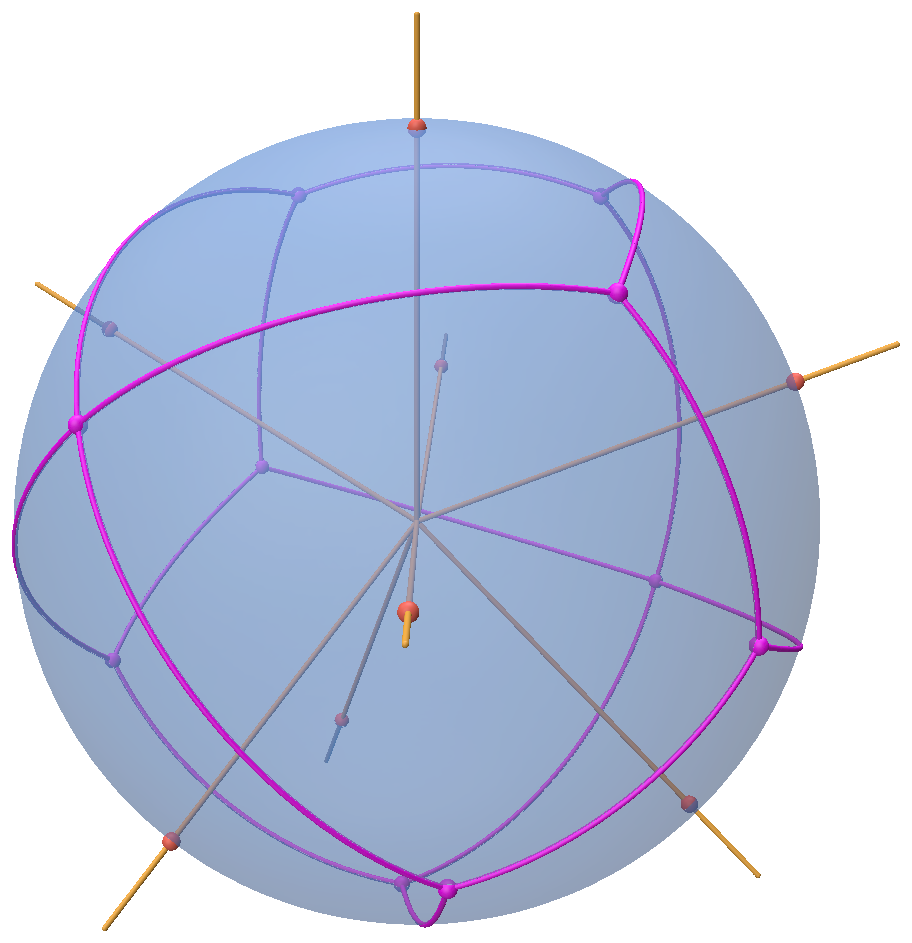} \\[0.5em]
    {\footnotesize \shortstack{(b) Fuentes Su{\'a}rez\\ and Hubert~\cite{FSH18}}} &
    {\footnotesize \shortstack{(c) Ours\\\strut}}
  \end{tabular}
  \caption{Joint-sphere partitions for the same eight branch directions.
  \textbf{(a)}~Input directions.
  \textbf{(b)}~Fuentes Su{\'a}rez and Hubert~\cite{FSH18} partition the sphere by the spherical Voronoi diagram, whose regions are general spherical polygons, not necessarily quads. 
  \textbf{(c)}~Ours is a simple quadrangulation, one quad per branch.}
  \label{fig:voronoi_comparison}
\end{figure}

Fuentes Su{\'a}rez and Hubert~\cite{FSH18} are also order-independent, with a construction feasible for any skeleton topology. 
Although their final scaffold is a quad mesh, the joint-sphere partition that defines it is the spherical Voronoi diagram of the incident directions, so each branch region is a general spherical polygon that need not be a quad, with its topology inherited from the input geometry rather than chosen (Figure~\ref{fig:voronoi_comparison}). 
Our method instead selects an SQS by enumeration and assignment, keeping one quadrilateral face per branch.
\section{Application to construction of parametric volumes of skeletal joints}
\label{sec:sqs_to_volume}

\begin{figure*}[t]
  \centering
  \setlength{\tabcolsep}{2pt}
  \begin{tabular}{ccc}
    \includegraphics[width=0.24\linewidth]{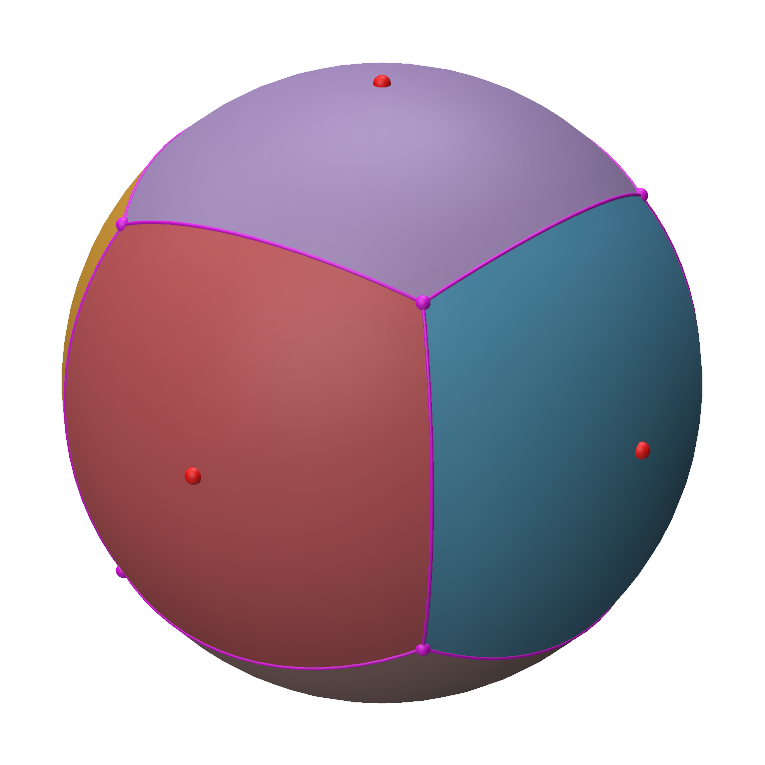} &
    \includegraphics[width=0.24\linewidth]{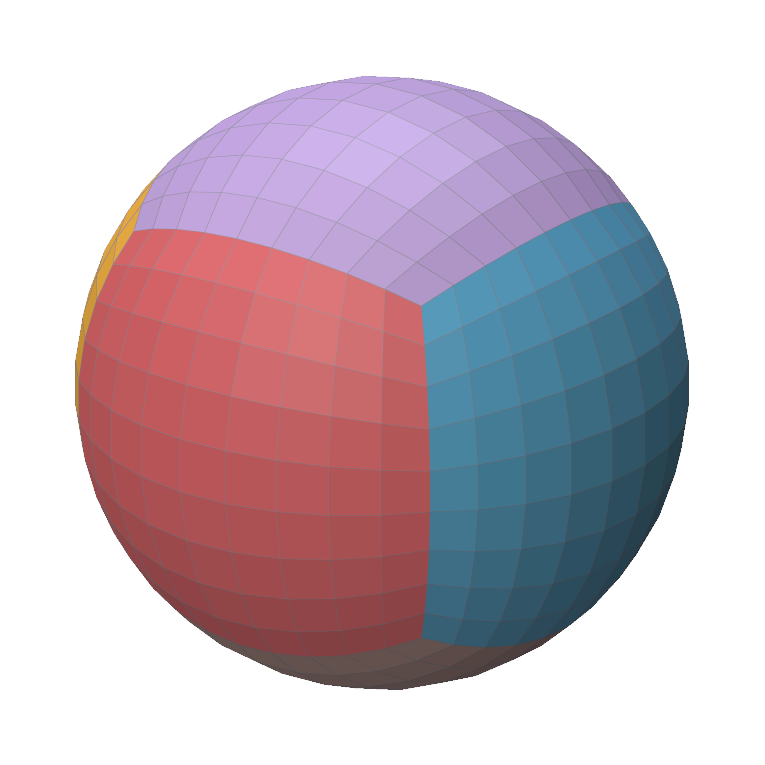} &
    \includegraphics[width=0.24\linewidth]{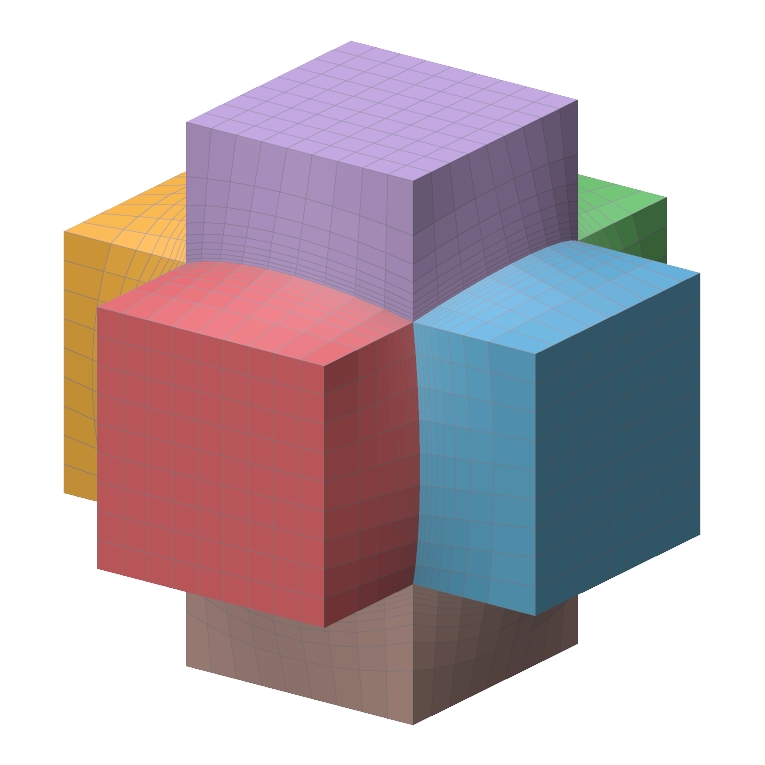} \\[-6pt]
    {\footnotesize (a) Face patches} &
    {\footnotesize (b) Shells} &
    {\footnotesize (c) Shells + columns}
  \end{tabular}
  \caption{Three-stage volume construction for $n = 6$, starting from the SQS of Figure~\ref{fig:sqs_pipeline}. 
  \textbf{(a)}~Biquartic rational B\'{e}zier face patches on~$\mathbb{S}^2$. 
  \textbf{(b)}~Radial shell volumes of thickness~$\delta$. 
  \textbf{(c)}~Branch-aligned column volumes appended to the outer shell.}
  \label{fig:sqs_volume}
\end{figure*}

\subsection{Skeletal joint construction}
We demonstrate the application of the proposed quadrangulation method by constructing parametric volumetric models of skeletal joints given the branch directions.
The SQS $Q = (V, E, F)$ constructed in \S\ref{subsec:primal_construction} is converted into a parametric joint volume in three stages (Figure~\ref{fig:sqs_volume}): we first obtain rational B\'{e}zier boundary primitives for edges and faces, construct radial shell volumes, and append column volumes extruded along the branch directions.

As each primal edge is the minor arc between its distinct, non-antipodal endpoints, the edge is represented exactly by a degree-$2$ rational B\'{e}zier curve and degree-elevated to~$4$ for compatibility with the face patches~\cite{Far01}. 
From four edge curves, we use a stereographic construction to obtain an exact biquartic $(4,4)$ rational B\'{e}zier patch $\mathbf{s}(u,v)$ on $\mathbb{S}^2$ with $\lVert\mathbf{s}(u,v)\rVert = 1$ whose four boundary curves are exactly the incident primal edges~\cite{Cob88, DCM*18}.

We then construct a radial shell volume from each face patch $\mathbf{s}(u,v)$. 
Let $r_{\mathrm{out}}=1$ and $r_{\mathrm{in}}=1-\delta$ with $0<\delta<1$. 
We construct another spherical patch $r\,\mathbf{s}(u,v)$ by scaling the spatial coordinates of the projective control points of $\mathbf{s}$ by a factor $r$. 
The degree-$(4,4,1)$ rational B\'{e}zier volume is generated for each face patch $\mathbf{s}(u,v)$ by ruling between $\mathbf{s}_{\mathrm{out}}=r_{\mathrm{out}}\mathbf{s}$ and $\mathbf{s}_{\mathrm{in}}=r_{\mathrm{in}}\mathbf{s}$ as follows:
\begin{equation}
  \mathbf{V}(u,v,\tau)=
  \bigl[(1-\tau)r_{\mathrm{out}}+\tau r_{\mathrm{in}}\bigr]
  \,\mathbf{s}(u,v).
\end{equation}
Since $r_{\mathrm{in}}=0$ would collapse each inner face to the origin and violate the non-vanishing Jacobian requirement of~\S\ref{sec:intro}, the joint is a hollow shell of thickness~$\delta$ rather than a filled ball.

Finally, the column volume is appended to join the outer shell patch to a branch-facing patch in the neighboring joint. 
Let $\mathbf{s}_{\mathrm{out}}(u,v)$ be the outer shell patch of the primal face $f_i$ and $\vec{d}_i$ be the branch direction. 
Its four parametric corners $\mathbf{s}_{\mathrm{out}}(u_c,v_c)$, $(u_c,v_c) \in \{0,1\}^2$, are first projected along $\vec{d}_i$ onto the plane $\mathbf{x} \cdot \vec{d}_i = r_{\mathrm{out}} + h$ with the prescribed interface
offset $h$, yielding four coplanar points that define a bilinear tensor-product patch $\mathbf{q}_i(u,v)$. 
The patch $\mathbf{q}_i$ is degree-elevated to $(4,4)$ for compatibility with $\mathbf{s}_{\mathrm{out}}$. 
The column volume is constructed by ruling between $\mathbf{q}_i$ and $\mathbf{s}_{\mathrm{out}}(u,v)$. Note that for each fixed $(u,v)$, the corresponding fiber is a straight segment from $\mathbf{s}_{\mathrm{out}}(u,v)$ to $\mathbf{q}_i(u,v)$.
\subsection{Auxiliary direction vectors for convex realizations}
\label{subsec:aux_directions}
As shown in \S\ref{subsec:sqs_construction}, a convex realization is impossible for $n \leq 5$ and $n = 7$ because every candidate dual graph contains at least one double edge. 
To overcome this limitation, we augment the original~$n$ direction vectors with auxiliary vectors so that the augmented count~$n'$  admits a digon-free candidate (cf.~Table~\ref{tab:enumeration_counts}). 
In practice, we target $n' \geq 8$.

The auxiliary vectors are placed by minimizing the Coulomb energy $\sum_{i<j} \|x_i - x_j\|^{-1}$ of the complete $n'$-point configuration, with the original $n$ positions fixed and the remaining $n' - n$ positions free on~$\mathbb{S}^2$~\cite{SK97}. 
This drives the auxiliary vectors away from the fixed ones and from each other, producing a well-separated configuration.

\begin{figure}[t]
  \centering
  \setlength{\tabcolsep}{2pt}
  \begin{tabular}{cc}
    \includegraphics[width=0.48\linewidth]{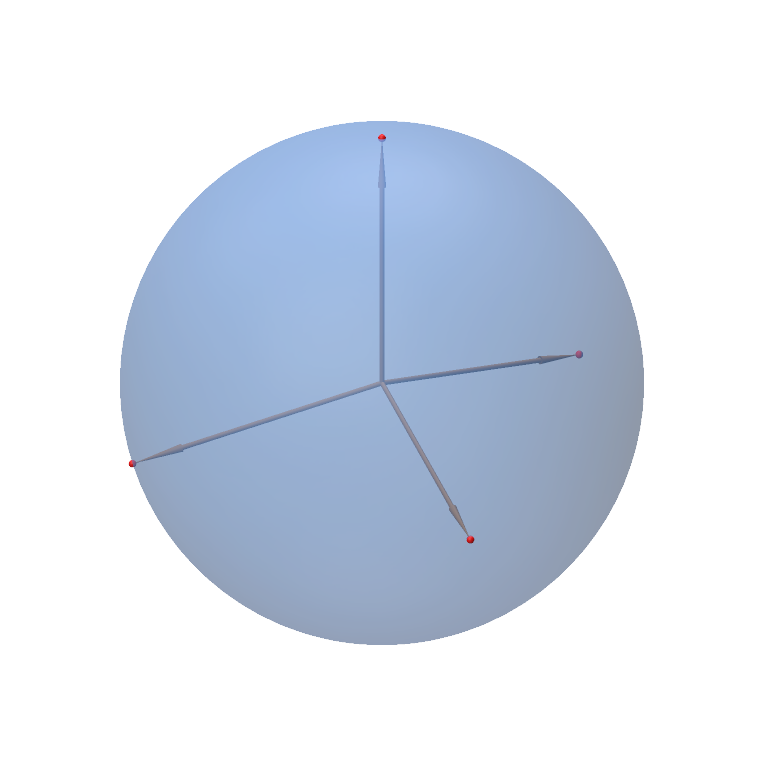} &
    \includegraphics[width=0.48\linewidth]{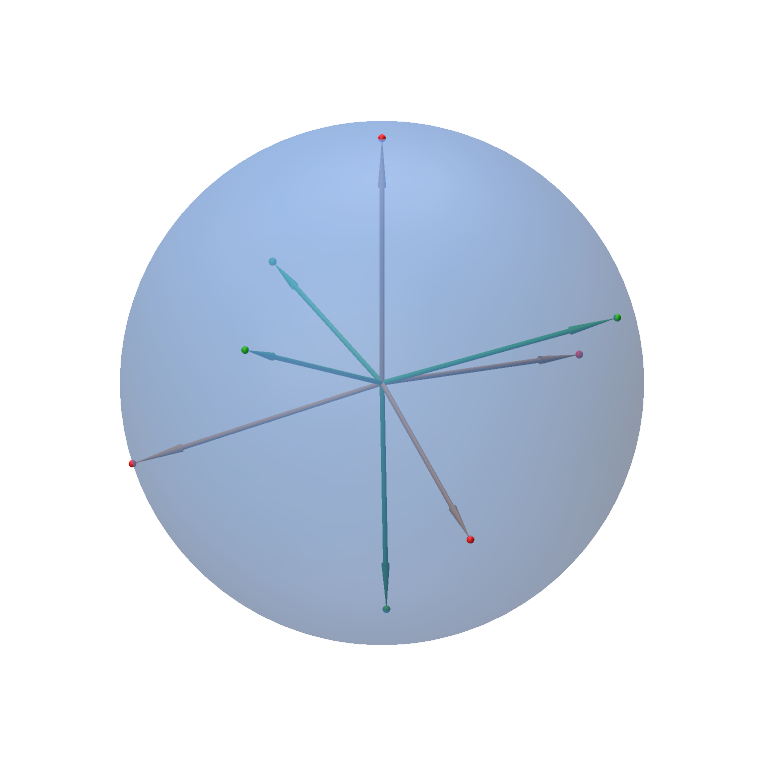} \\[-12pt]
    {\footnotesize (a) Input ($n{=}4$)} &
    {\footnotesize (b) Augmented ($n'{=}8$)} \\[12pt]
    \includegraphics[width=0.48\linewidth]{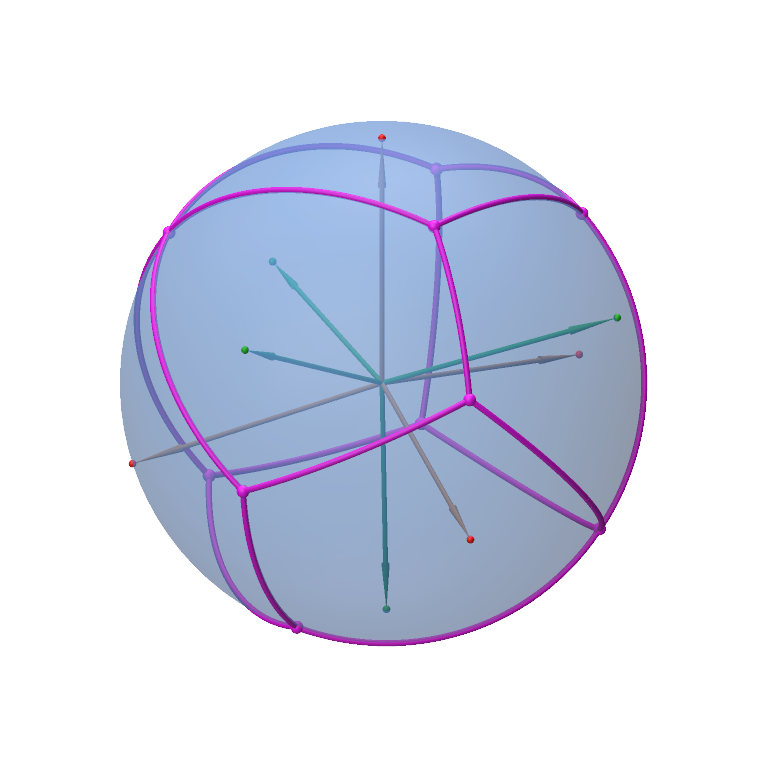} &
    \includegraphics[width=0.48\linewidth]{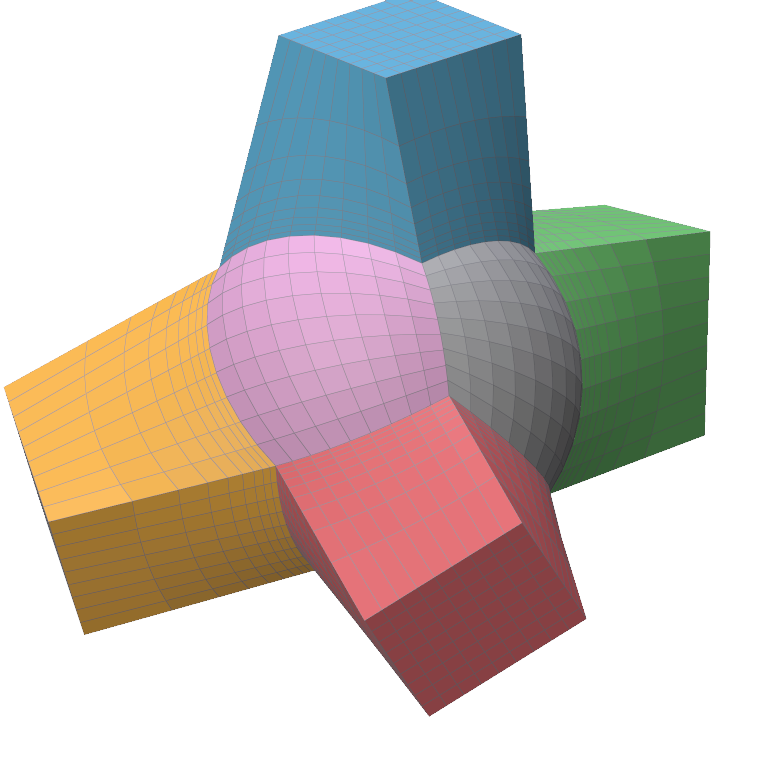} \\[-8pt]
    {\footnotesize (c) SQS} &
    {\footnotesize (d) Shell + columns}
  \end{tabular}
  \caption{Direction vector augmentation for $n = 4$. 
  \textbf{(a)}~Four initial direction vectors (red) on~$\mathbb{S}^2$. 
  \textbf{(b)}~Four auxiliary vectors (green) obtained by minimizing Coulomb energy, yielding $n'{=}8$. 
  \textbf{(c)}~Resulting SQS primal graph (magenta). 
  \textbf{(d)}~Shell volumes on all eight faces; column volumes on the four initial directions only.}
  \label{fig:aux_directions}
\end{figure}

Figure~\ref{fig:aux_directions} illustrates this strategy for a concave case, $n = 4$. 
Four auxiliary vectors (green dots in Figure~\ref{fig:aux_directions}(b)) are added to yield $n' = 8$, for which a digon-free candidate exists. 
The resulting SQS (Figure~\ref{fig:aux_directions}(c)) has all eight primal faces geodesically convex. 
In the volume generation stage (Figure~\ref{fig:aux_directions}(d)), shells are constructed for all eight faces, whereas columns are extruded only along the four original branch directions. 
Faces corresponding to auxiliary vectors carry no skeletal branches and are covered by shells alone. 
The augmentation thus enables a convex realization while preserving the original skeletal connectivity.
\subsection{Multi-center assembly of junction volumes}
\label{subsec:multi_center}
We extend the single-center pipeline of \S\ref{sec:method} to fill an entire domain~$\Omega$ by placing $N_c$ centers. 
At each center~$c_k$, we assign one direction vector toward each neighboring center (determined by lattice connectivity or Delaunay adjacency~\cite{Aur91}), and, for boundary centers, an additional surface projection direction, yielding $n_k$ direction vectors. 
SQS construction and shell/column generation then run independently at each center. 
The branch interface plane between adjacent centers $c_k$ and~$c_l$ is the perpendicular bisector plane of~$\overline{c_k c_l}$. 
Opposing face corners are paired by cyclic alignment of their projections onto this plane, and the line joining each matched pair is intersected with the plane to produce a shared top-surface corner used by both sides.

\begin{figure}[t]
  \centering
  \setlength{\tabcolsep}{2pt}
  \begin{tabular}{@{}cc@{}}
    {\footnotesize Shells} &
    {\footnotesize Shells + columns} \\[-2pt]
    \includegraphics[width=0.46\linewidth]{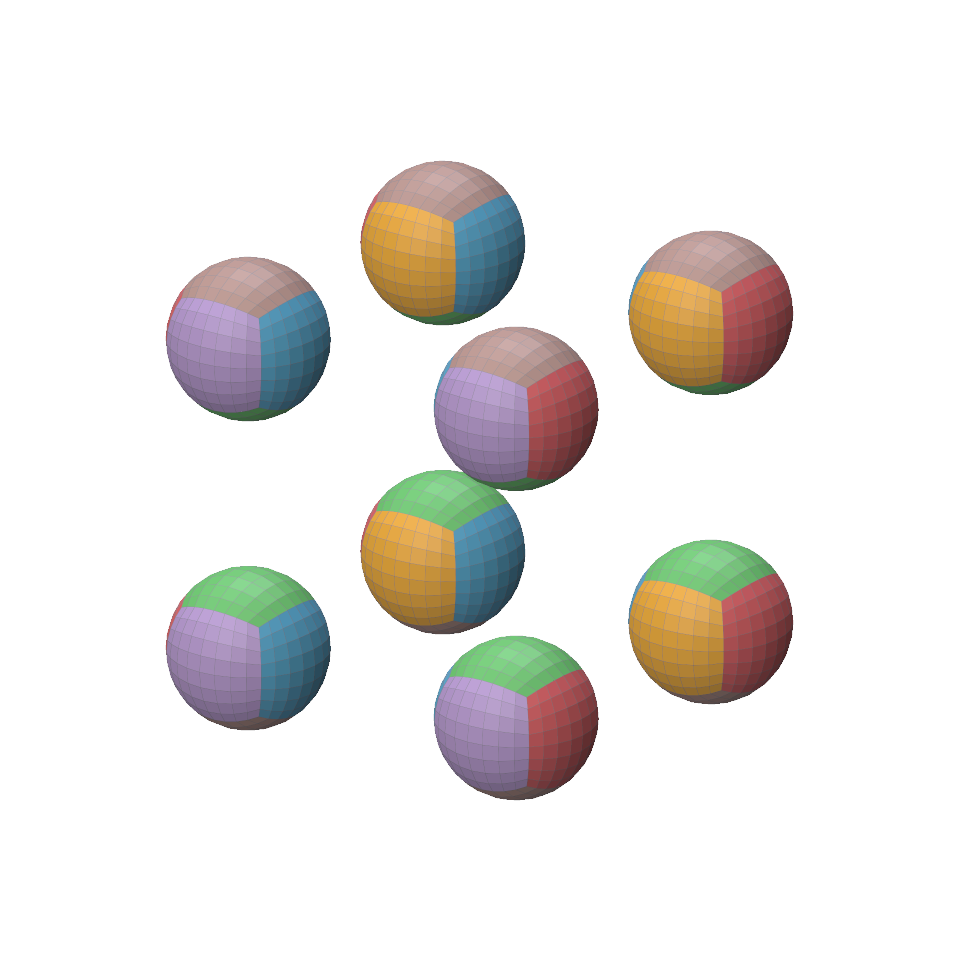} &
    \includegraphics[width=0.46\linewidth]{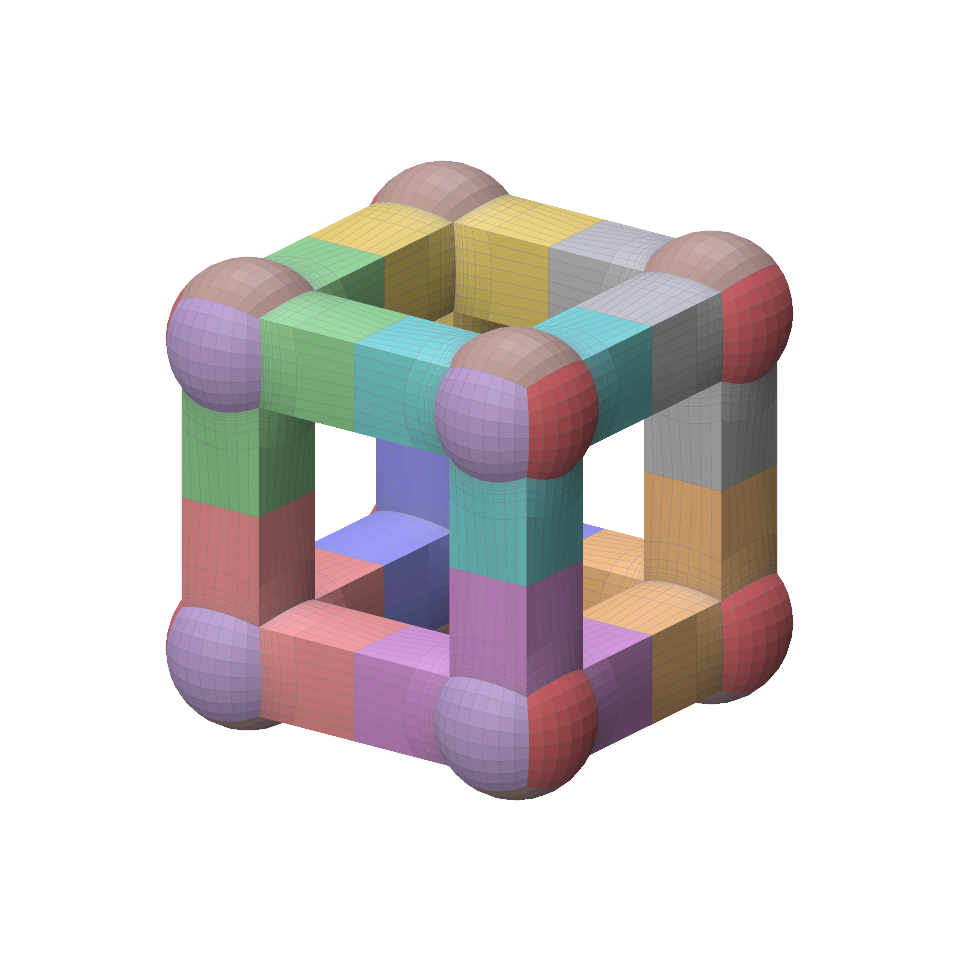} \\[-12pt]
    \multicolumn{2}{c}{\footnotesize (a) SC lattice} \\[2pt]
    \includegraphics[width=0.46\linewidth]{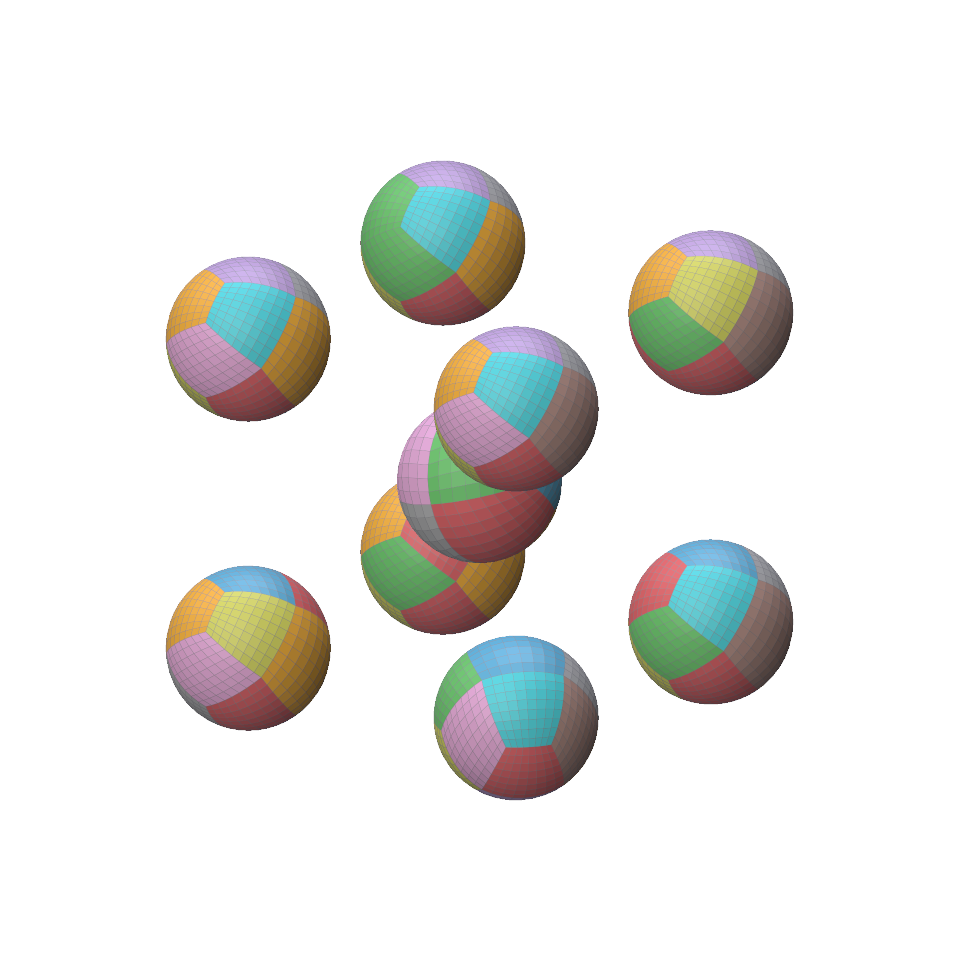} &
    \includegraphics[width=0.46\linewidth]{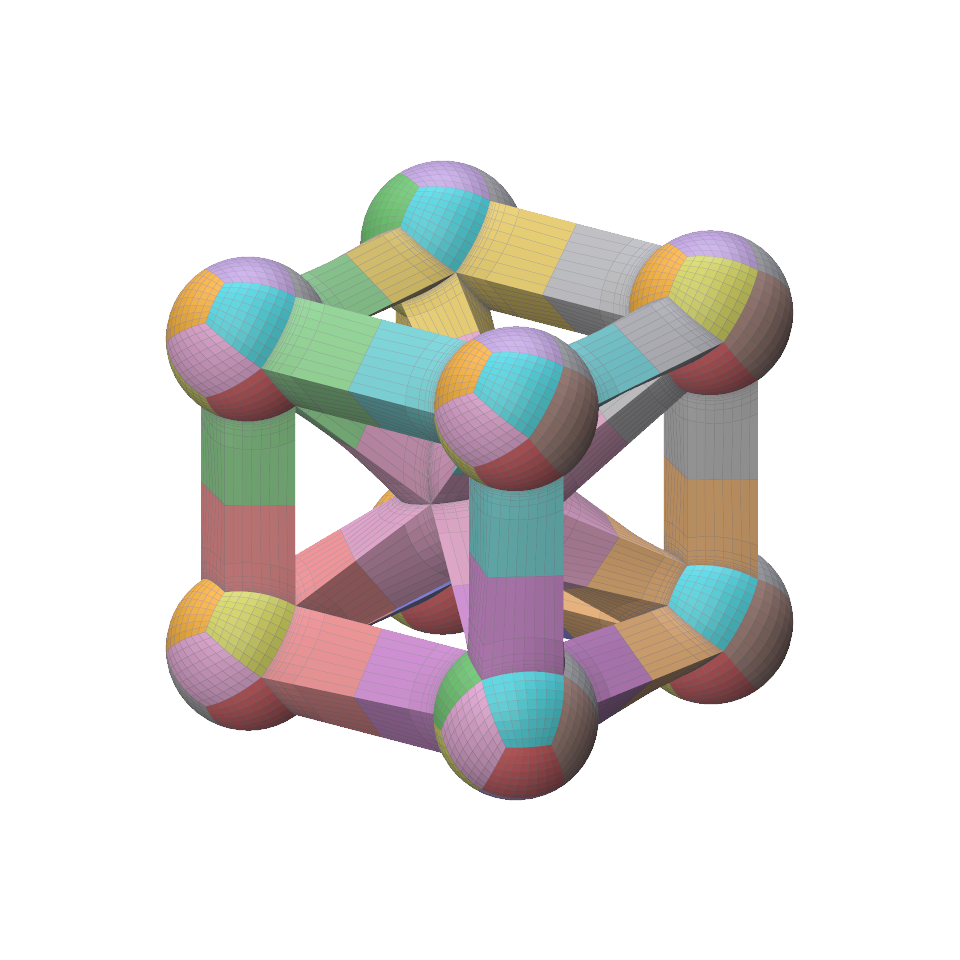} \\[-12pt]
    \multicolumn{2}{c}{\footnotesize (b) SC--BCC lattice}
  \end{tabular}
  \caption{Multi-center SQS volumes on SC and SC--BCC lattices. 
  \textbf{(a)}~SC with $N_c{=}8$ centers and $n{=}6$ directions per center. 
  \textbf{(b)}~SC--BCC with $N_c{=}9$ centers, $n_k{=}14$ at the corners and $n_k{=}8$ at the body center.}
  \label{fig:lattice}
\end{figure}

Figure~\ref{fig:lattice} shows results for two lattice layouts. 
In the simple cubic (SC) $2{\times}2{\times}2$ layout (Figure~\ref{fig:lattice}(a), $N_c = 8$), cubic symmetry ensures that all joints share the same SQS with $n = 6$. 
The SC--BCC layout (Figure~\ref{fig:lattice}(b), $N_c = 9$) comprises eight corner centers, each with $n_k = 14$ directions (six axial and eight body-diagonal directions), and one body center with $n_k = 8$ directions toward the cube corners. 
Only directions paired with another center in the displayed cell generate columns.
All joints admit convex SQS realizations, and columns from adjacent joints meet at the branch interfaces, forming the truss-like volumetric assemblies shown in Figure~\ref{fig:lattice}.

\begin{figure*}[t]
  \centering
  \setlength{\tabcolsep}{2pt}
  \begin{tabular}{@{}cc@{}}
    \includegraphics[width=0.34\linewidth]{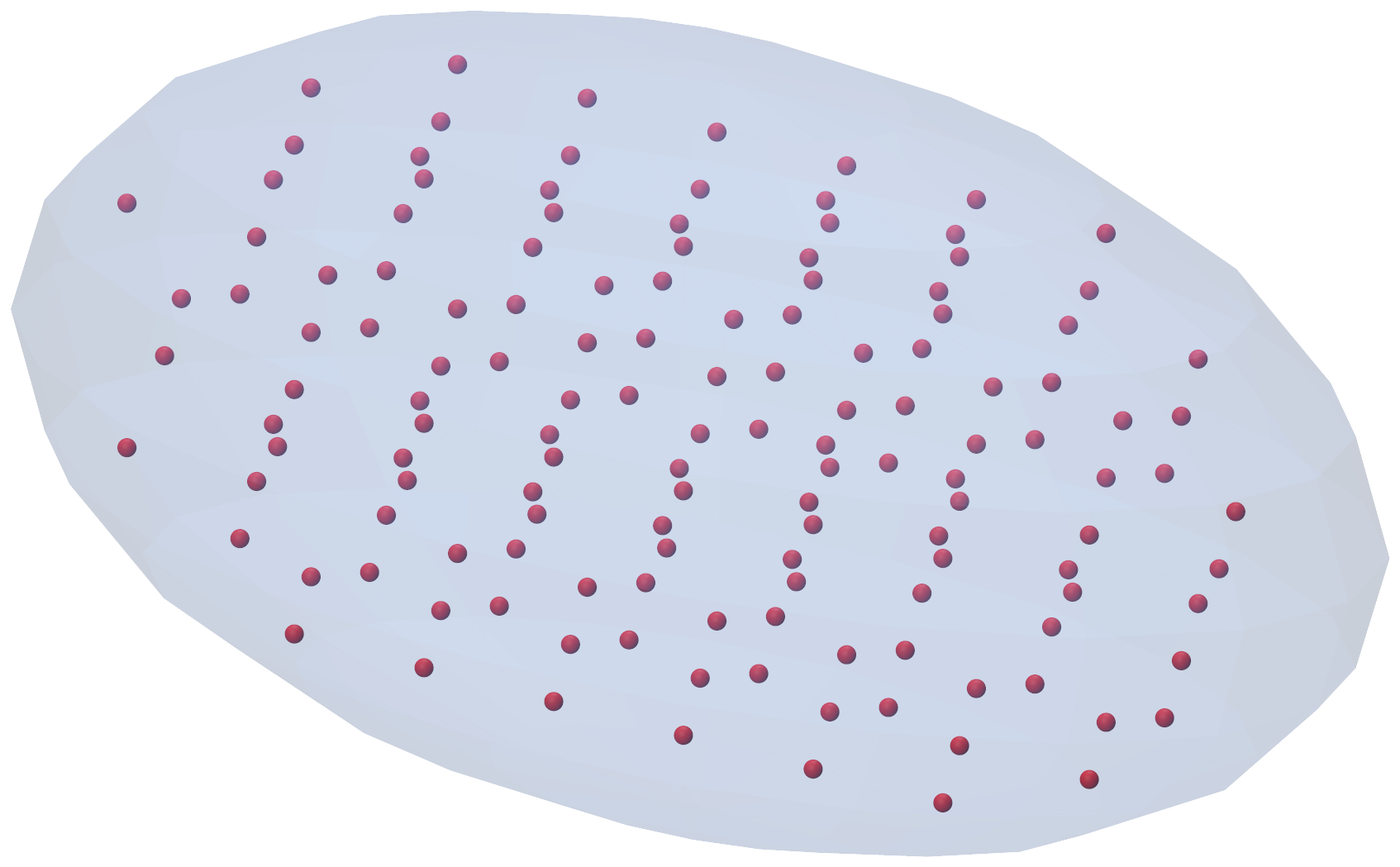} &
    \includegraphics[width=0.34\linewidth]{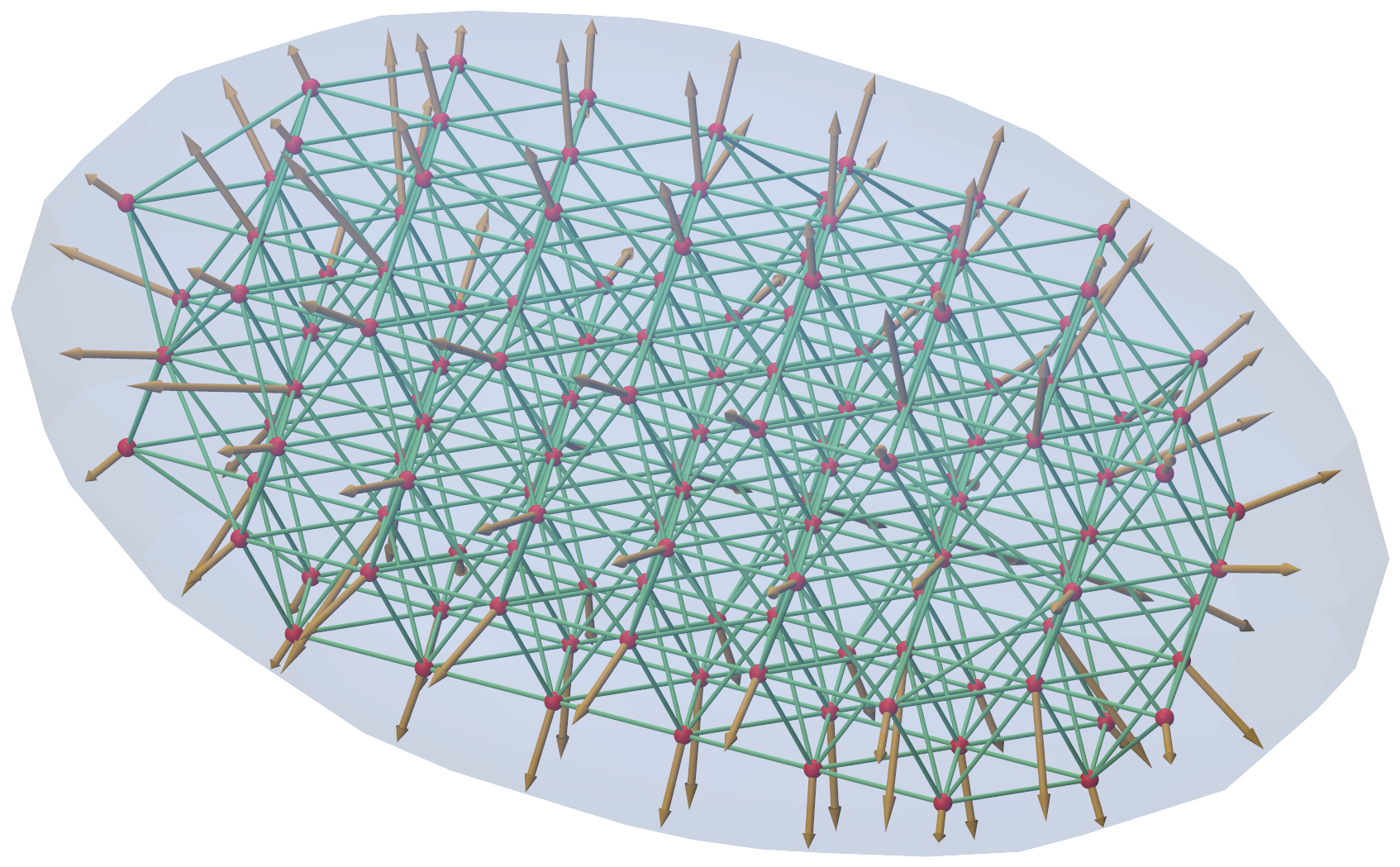} \\[2pt]
    {\footnotesize (a) Centers} &
    {\footnotesize (b) Neighbors + projections} \\[8pt]
    \includegraphics[width=0.44\linewidth]{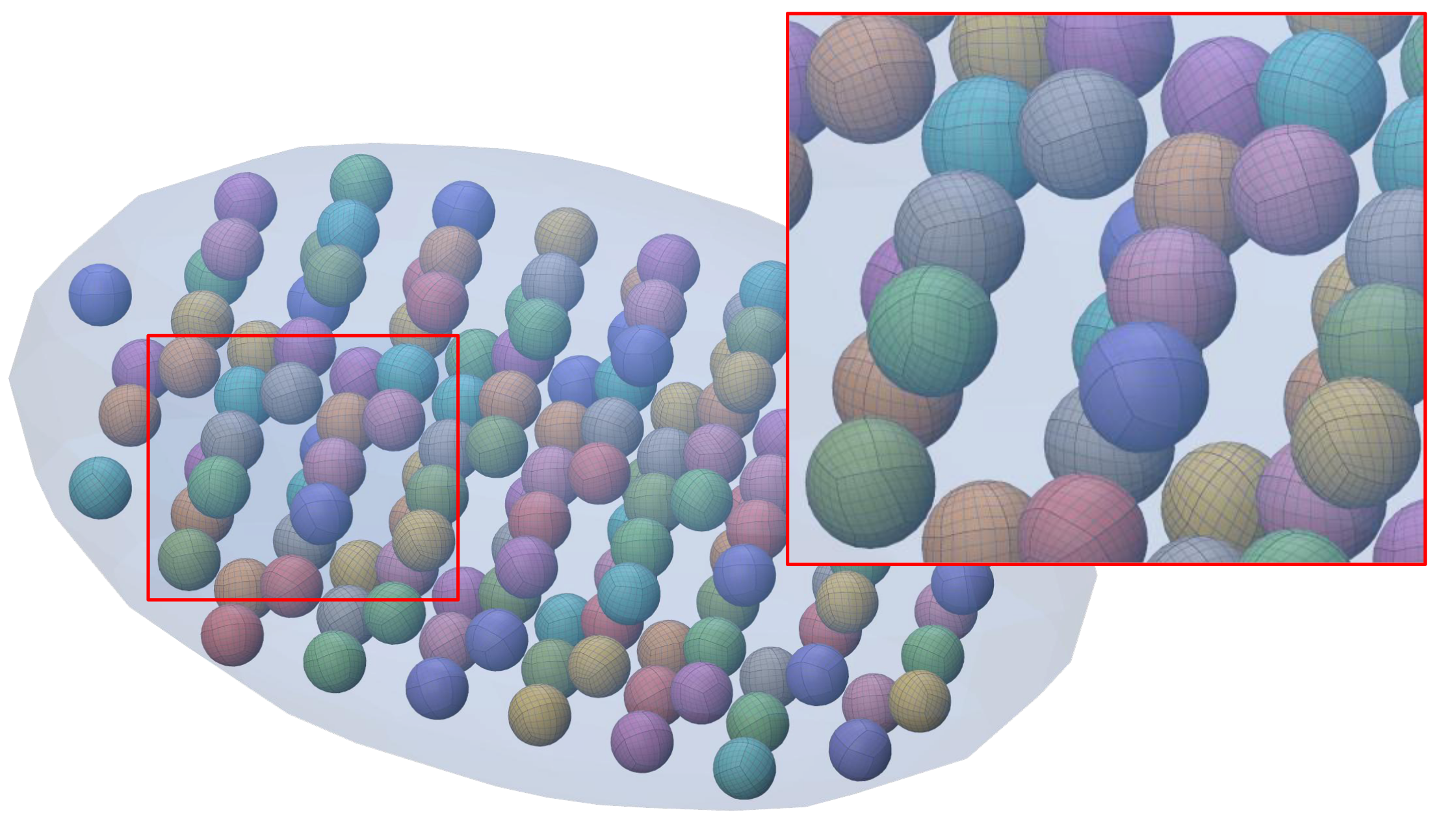} &
    \includegraphics[width=0.44\linewidth]{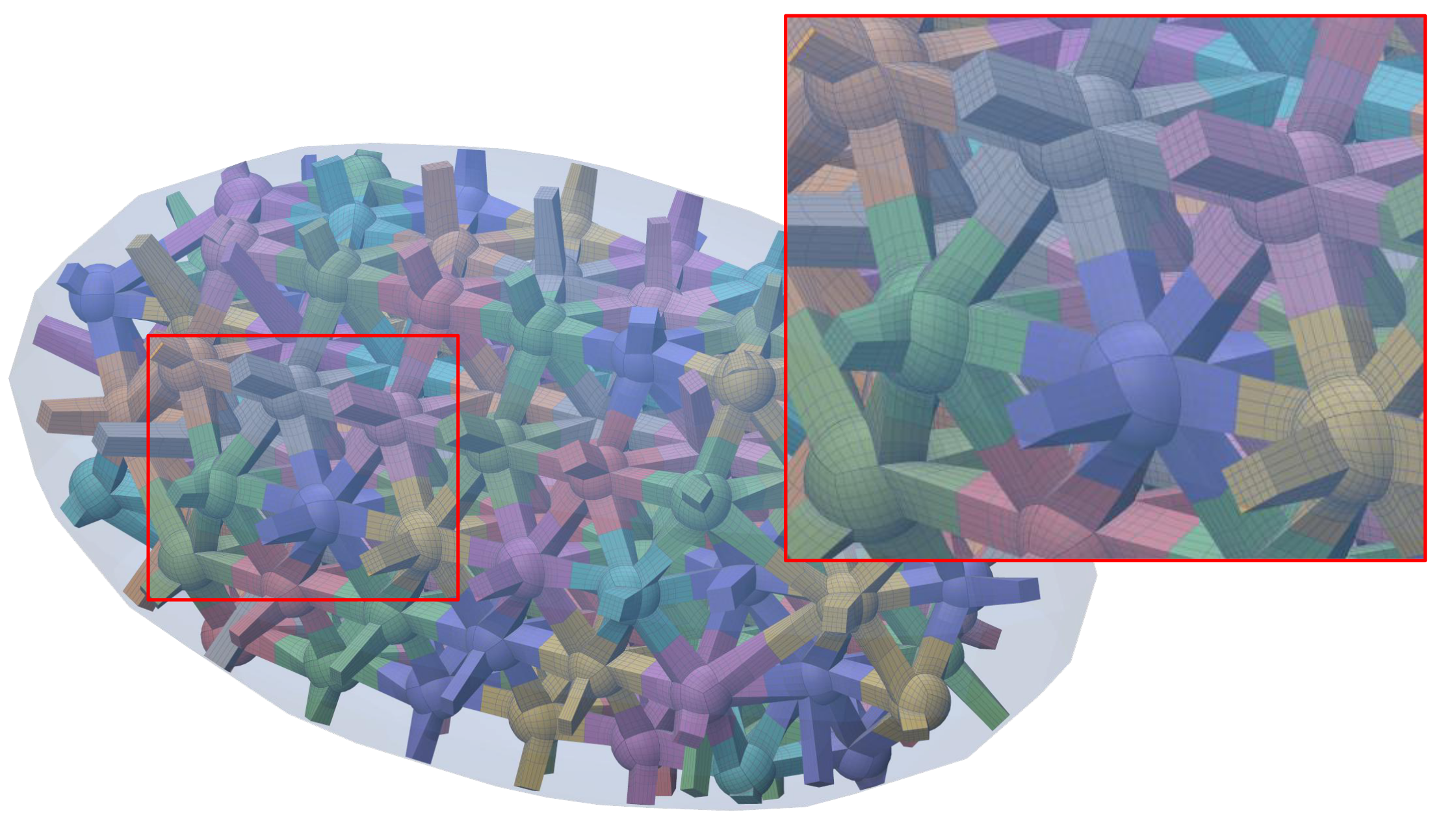} \\[2pt]
    {\footnotesize (c) Shells} &
    {\footnotesize (d) Volumes}
  \end{tabular}
  \caption{Multi-center SQS volumes inside an ellipsoidal domain~$\Omega$ with $N_c{=}137$ centers. 
  Each center is augmented when needed to reach $n_k \geq 10$ direction vectors. 
  \textbf{(a)}~Red dots: $137$ centers distributed inside~$\Omega$. 
  \textbf{(b)}~Green edges: Delaunay neighbors. 
  Orange arrows: surface projection directions from boundary centers toward~$\partial\Omega$. 
  \textbf{(c)}~SQS shells at each center (inset: detail). 
  \textbf{(d)}~Adjacent columns meet at branch interfaces with boundary columns extended to~$\partial\Omega$ (inset: detail).}
  \label{fig:ellipsoid_pipeline}
\end{figure*}

Figure~\ref{fig:ellipsoid_pipeline} demonstrates the pipeline on an unstructured domain. 
$N_c = 137$ centers are distributed inside an ellipsoid (Figure~\ref{fig:ellipsoid_pipeline}(a)), with neighbors determined by Delaunay adjacency (Figure~\ref{fig:ellipsoid_pipeline}(b)). 
For this experiment we target $n_k \geq 10$ at every center, augmenting those with fewer Delaunay neighbors following \S\ref{subsec:aux_directions}. 
Figure~\ref{fig:ellipsoid_pipeline}(c,d) shows the SQS shells at each center and the assembled result with columns connecting adjacent centers and extending to~$\partial\Omega$. 
In this experiment, convex SQS realizations are obtained at all 137 joints, demonstrating that the pipeline handles the non-uniform conditions of this instance where~$n_k$ and the direction-vector configuration vary across centers.

\begin{figure*}[t]
  \centering
  \setlength{\tabcolsep}{15pt}
  \begin{tabular}{@{}cc@{}}
    \includegraphics[width=0.40\linewidth]{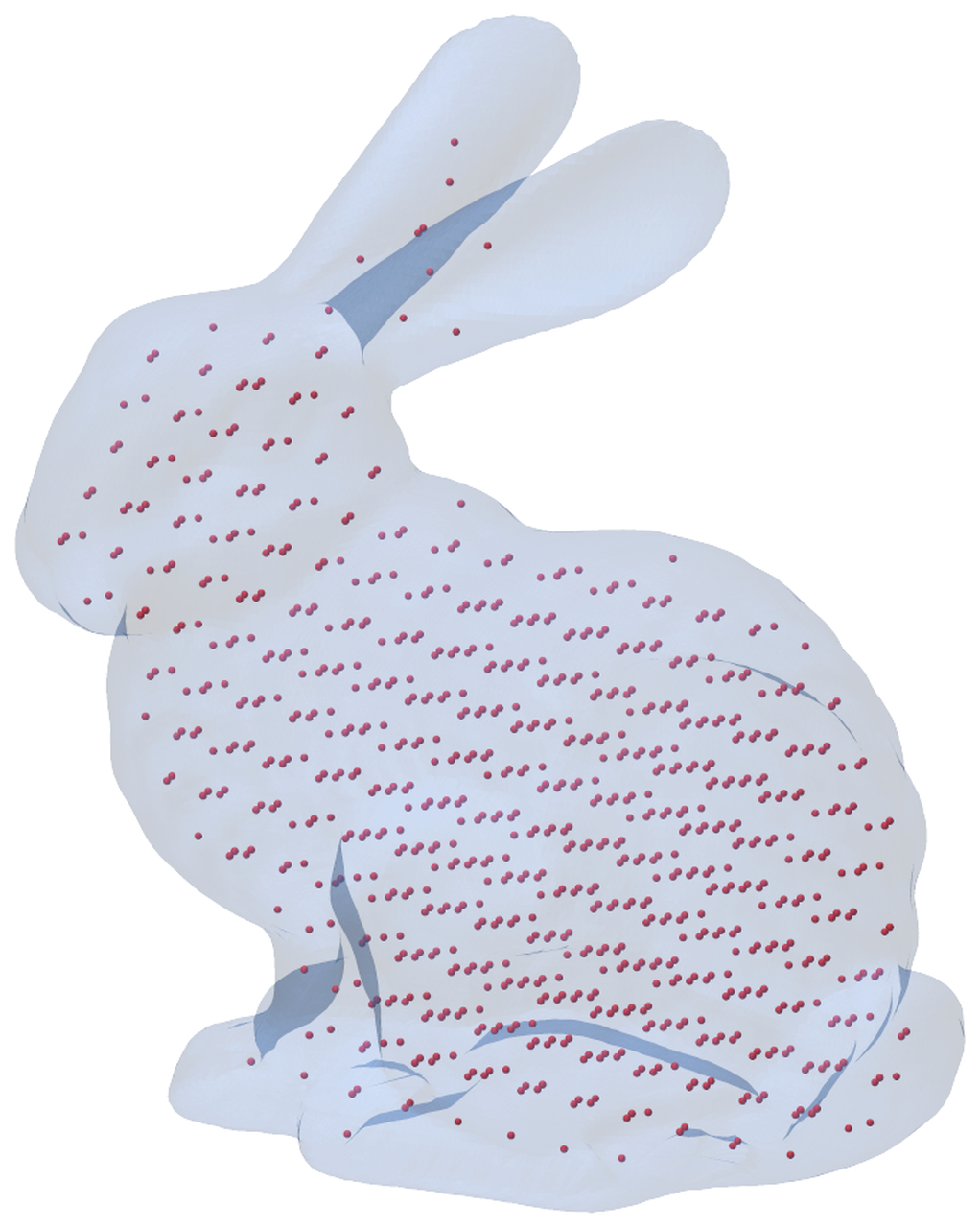} &
    \includegraphics[width=0.40\linewidth]{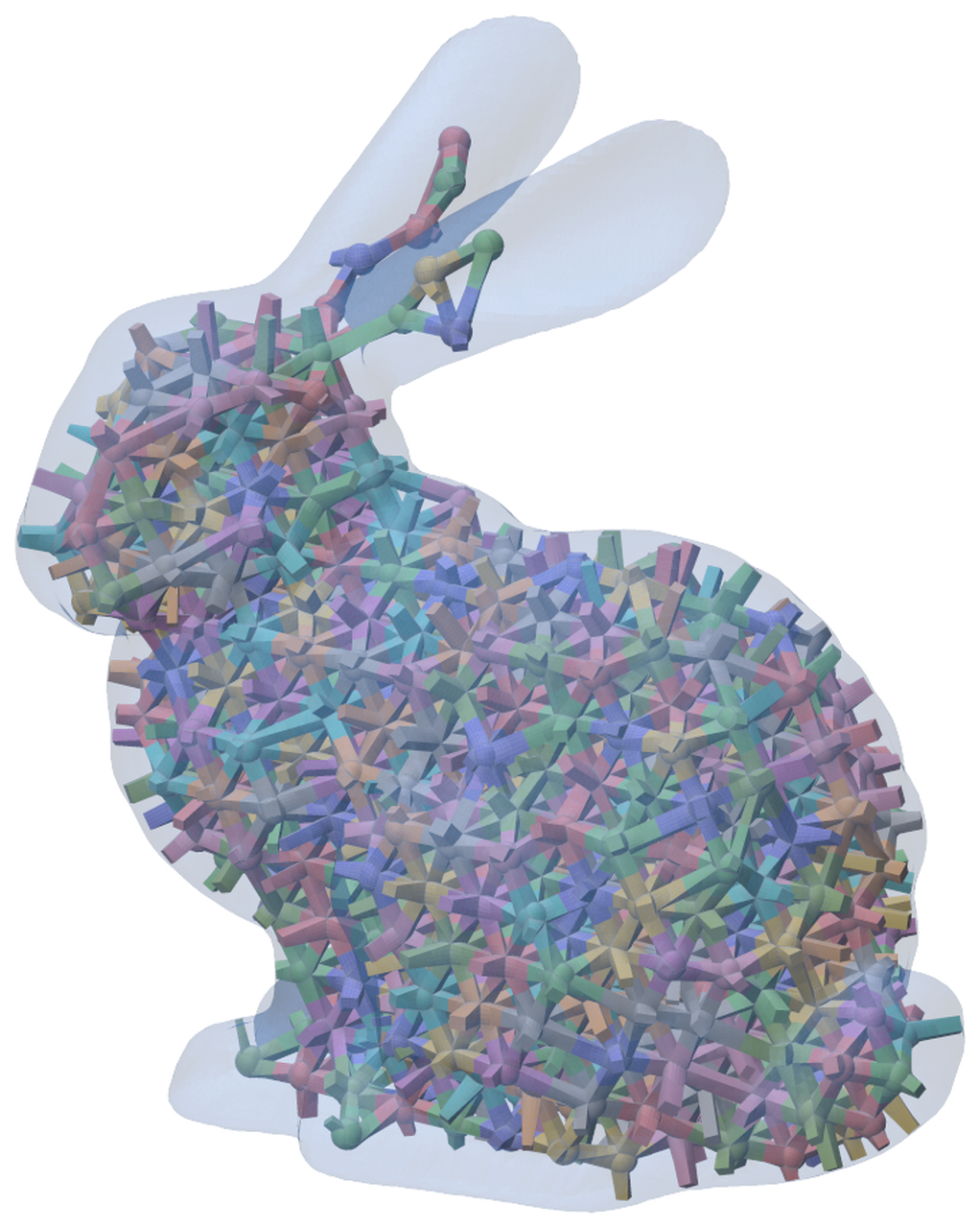} \\[-2pt]
    {\footnotesize (a) Centers} &
    {\footnotesize (b) Volumes}
  \end{tabular}
  \caption{Multi-center SQS volumes inside a bunny-shaped domain~$\Omega$ with $N_c{=}861$ centers. 
  \textbf{(a)}~Red dots: $861$ centers distributed inside~$\Omega$. 
  \textbf{(b)}~Assembled SQS shells and columns with boundary columns extended to~$\partial\Omega$.}
  \label{fig:bunny}
\end{figure*}

Figure~\ref{fig:bunny} applies the same pipeline to a bunny-shaped domain~$\Omega$ with $N_c = 861$ centers, each carrying $n_k \leq 14$ direction vectors, augmented as needed (\S\ref{subsec:aux_directions}). 
Adjacent columns meet at the branch interfaces and boundary columns extend to~$\partial\Omega$, forming the largest multi-center assembly we demonstrate.
\section{Conclusion}
\label{sec:conclusion}
We present a pipeline for constructing simple quadrangulations of the sphere (SQS) at skeletal joints.
We enumerate candidate dual plane graphs, realize a selected dual on~$\mathbb{S}^2$ via a conflict-constrained CP-SAT assignment with lazy cyclic-order separation, and recover the corresponding primal graph.
The resulting quadrilateral partition is converted into rational B\'{e}zier boundary
and volume primitives. The construction treats the input direction vectors as an unordered set, so its result is independent of branch insertion order, unlike previous incremental schemes~\cite{PRW*18, PMA21}.  
Across Fibonacci-sampled directions ($n = 3, \ldots, 20$) and multi-center assemblies of up to $861$ joints, we show that the method produces a valid SQS in every tested instance, and a convex one for the quasi-uniform inputs that admit a digon-free dual class. 

A limitation of the proposed method is scalability in the number of branch directions~$n$. 
The number of candidate dual classes to enumerate grows exponentially with~$n$~(Table~\ref{tab:enumeration_counts}), while each individual CP-SAT solve remains fast in practice~(Table~\ref{tab:fibonacci_sqs_results}). 
In this work, we report experiments for $n \leq 20$ as joints with more than 20 branches are rarely seen in practical skeletal modeling. 

The current strategy for selecting dual graph candidates does not provide a guarantee of global optimality with respect to the primal quality metrics.
One direction for future work is to incorporate a direct angle-aware objective into the assignment model.
Another direction is to optimize the placement of primal vertices directly within dual faces. 
The quality of quadrilateral elements could be further improved by replacing the current centroid-based heuristic, particularly for irregular or clustered direction distributions.
\section*{Acknowledgments}
This research was supported by the National Research Foundation of Korea (NRF) grant funded by the Korea government (MSIT) (RS-2024-00352439).
\bibliographystyle{eg-alpha-doi}
\bibliography{paper1378_CRC}
\end{document}